\documentclass[<options>]{elsarticle}
 \pdfoutput=1
 \usepackage{lipsum}
\makeatletter
\def\ps@pprintTitle{%
 \let\@oddhead\@empty
 \let\@evenhead\@empty
 \def\@oddfoot{}%
 \let\@evenfoot\@oddfoot}
\makeatother
\usepackage[utf8]{inputenc}
\usepackage{array}
\usepackage{indentfirst}
\usepackage{graphicx}
\usepackage{xcolor}
\usepackage{amsmath}
\usepackage{float}
\usepackage[T1]{fontenc}
\usepackage{adjustbox}
\graphicspath{{figures/}}
\usepackage{appendix}
\usepackage{hyperref}
\usepackage{subcaption}
\renewcommand{\arraystretch}{1}
\usepackage{cleveref}

\usepackage[a4paper, total={7.4in, 9in}]{geometry}

\begin{document}

\begin{frontmatter}

\title{Comparing Sinc and Harmonic Oscillator Basis for Bound States of a Gaussian Interaction}

\author{Mamoon Sharaf$^a$\footnote{Corresponding author: msharaf@iastate.edu.  This work was supported by the US Department of Energy under Grant Nos. DE-SC0018223 (SciDAC-4/NUCLEI) and DE-FG02-87ER40371}, Ryan McCarty$^a$, Robert A. M. Basili$^a$, James P. Vary$^a$}

\address{$^{a}$ Department of Physics and Astronomy, Iowa State University, Ames, Iowa, U.S.A, 50011}

\begin{abstract} \label{abstract}
We investigate the use of the sinc collocation and harmonic oscillator bases for solving a two-particle system bound by a Gaussian potential described by the radial Schrödinger equation.  We analyze the properties of the bound state wave functions by investigating where the basis-state wave functions break down and relate the breakdowns to the infrared and ultraviolet scales for both bases.  We propose a correction for the asymptotic infrared region, the long range tails of the wave functions.  We compare the calculated bound state eigenvalues and mean square radii obtained within the two bases.  From the trends in the numerical results, we identify the advantages and disadvantages of the two bases.  We find that the sinc basis performs better in our implementation for accurately computing both the deeply- and weakly-bound states whereas the harmonic oscillator basis is more convenient since the basis-state wave functions are orthogonal and maintain the same mathematical structure in both position and momentum space.  These mathematical properties of the harmonic oscillator basis are especially advantageous in problems where one employs both position and momentum space.  The main disadvantage of the harmonic oscillator basis as illustrated in this work is the large basis space size required to obtain accurate results simultaneously for deeply- and weakly-bound states.  The main disadvantage of the sinc basis could be the numerical challenges for its implementation in a many-body application.    
 \vspace{3mm}
 
 \textbf{Keywords}: sinc basis; harmonic oscillator basis; infrared properties; two-body system; ultraviolet properties; cutoff effects
 
\end{abstract}

\end{frontmatter}

\section{Introduction}\label{introduction}

The three-dimensional harmonic oscillator (HO) basis is widely used in nuclear structure theory, as it possesses several convenient features.  First, it retains rotational symmetries, which facilitate the implementation of conservation laws and therefore reduces computational cost.  Second, it simulates some features of a mean-field related to the phenomenologically successful nuclear shell model (see \cite{source1} and \cite{source2}).  Third, the HO basis functions are orthonormal and treat position and momentum on an equal footing \cite{source3}.  This feature is advantageous since, for example, basis space matrix elements of the kinetic energy term are more conveniently evaluated in momentum space while interaction matrix elements are often evaluated in position space.  Fourth, the HO basis provides a closed-form transformation from the basis of relative HO states to the basis of single-particle states which is convenient for solving quantum many-body problems \cite{source4}.  A primary drawback of the HO basis is the mismatch of its asymptotic behavior, Gaussian tails at large distances, with the exponential tails of bound state wave functions.

In the sinc basis, we can choose a function to use as a map based on \textit{a priori} knowledge of the asymptotic behavior of the wave function.  Based on the asymptotic behavior, we can choose a function $\theta(r)$ that maps the collocation points (points on a finite, semi-infinite, or infinite interval that is the domain of the solution) \cite{source5} in a way that gives us the desired asymptotic behavior.  Therefore, we can get an accurate approximation of the wave function that possesses the desired asymptotic behavior using a convenient basis space.  This assists in obtaining numerically accurate results, especially in states where the long-range physics is important.  With modest computational resources one may evaluate one-body and two-body matrix elements of typical nuclear operators in a sinc basis with useful bases of the order of $10^{3}$ dimensions.

The sinc and HO bases have several common features that make them advantageous in numerous physics problems as we will see.  First, they facilitate implementation of conservation laws, which is an essential feature in rendering problems computationally tractable.  Second, both bases have parameters that determine the basis truncation and coordinate scales covered.  Third, both basis approximations have a systematic way in which they break down and are no longer accurate (though they do so differently). For both bases, this breakdown scale can be moved away from the physically useful domains with increasing basis dimension.

Like all numerical methods, the HO basis approach is not without its faults.  While treating the position and momentum space on an equal footing does have its merits, it mixes long- and short-range physics. For many-body problems, this makes calculations numerically expensive because we need the oscillator basis to be large enough to accurately describe a set of bound states that cover different distance (or, equivalently, momentum) scales.  Moreover, one needs to test the calculations at different values of the oscillator spacing $\hbar \Omega$ in order to check that the results converge to a value independent of basis parameters.  Various renormalization schemes have been implemented for the purpose of reducing the computational cost of many-body calculations (\cite{source2},\cite{source6}) and the spectra of light and some strongly-bound nuclei such as He-4 have been calculated to high accuracy (\cite{source7},\cite{source8}).  

Those limitations motivate the search for basis spaces that improve on the HO basis and make it a topic of current interest.  In light of this, many alternative basis spaces have been utilized to solve various physical problems including the Woods-Saxon basis \cite{source9}, the Coulomb-Sturmian basis \cite{source10}, and the natural orbitals basis employed for halo nuclei \cite{source11}.  This work is part of the effort to investigate the utility of different basis spaces.

The sinc collocation approach has been applied numerous times before in solving the Schrödinger equation (and more generally, the Sturm-Liouville equation).  Problems tackled include the anharmonic oscillator \cite{source12}, the Woods-Saxon potential \cite{source13}, and the non-relativistic planar Coulomb Schrödinger equation \cite{source14}.  However, there have not been previous reports that compare results between the HO and sinc bases, to our knowledge.  It is our aim to provide such a comparison which highlights the advantages and limitations of each basis space which can guide which is selected depending on the contemplated application.

For this paper, we concern ourselves only with the two-body problem to avoid some of the computational limitations of the many-body problem.  Here, among other topics, we focus on breakdown scales, wave function structure, and accurate calculation of observables such as bound state energies and mean square radii.  We first outline the two basis function approaches and introduce the wave equation and parameters of a two-body central force problem with a Gaussian interaction in Section \ref{theory}.  We then compare the sinc and HO bases bound state eigenvalues, mean square radii, and wave functions in Section \ref{results and comparisons}.  We demonstrate how to remedy the breakdown scale problem by attaching an exponential tail to both the weakest- and deepest-bound states (WBS and DBS, respectively) in Section \ref{tail corrections}.  We conclude by discussing ways to improve our methods and future avenues of investigation in Section \ref{conclusion}.  We discuss additional numerical details in the appendices.

\section{Theory} \label{theory}

The equation we solve is the two-body radial Schrödinger equation whose matrix representation is

\begin{equation}  
\label{eq1}
Hv=\mu v.
\end{equation}
Here, $H$ is the Hamiltonian
\begin{equation}
\label{eq2}
H=T+V,
\end{equation}
where $T$ is the kinetic energy operator and $V$ is the Gaussian interaction in matrix form.  Using the standard reduction of a two-body problem into a one-body problem describing the relative motion between two nucleons \cite{source15}, we can write the radial component of Eq. (\ref{eq1}) in the functional form

\begin{equation}
\label{eq3}
\frac{-\hbar^{2}}{2m}[\frac{1}{r^{2}}\frac{d}{dr}(r^{2}\frac{d}{dr})-\frac{l(l+1)}{r^{2}}]R(r)+V(r)R(r)=\mu R(r),
\end{equation}
where $m$ is the reduced mass, $r$ is the relative position, $l$ is the orbital angular momentum quantum number, $V(r)$ is the Gaussian interaction, and $\mu$ is the eigenvalue.  Using the substitution $u(r)=rR(r)$, we can recast Eq. (\ref{eq3}) as  

\begin{equation}
\label{eq4}
-\frac{\hbar^{2}}{2m}u''(r)+V_{\textrm{eff}}(r)u(r)=\mu u(r) 
\end{equation}
where $V_{\textrm{eff}}(r)$ is the effective potential defined as

\begin{equation}
  \label{eq5}
      V_{\textrm{eff}}(r)=-V_{0}e^{-\kappa^{2} r^{2}}+\frac{l(l+1)\hbar^{2}}{2m r^{2}},
  \end{equation}
   with our chosen parameters $V_{0}=400 $ MeV, $l=1$, $\kappa=0.15528440003849432114$  $\textrm{fm}^{-1}$, and $\frac{\hbar^{2}}{2m}=41.470984280273533723$ MeV $\textrm{fm}^{2} $. 
   
 Our motivation for using this potential is that it has been studied in another approach in Ref. \cite{source16}.  It is a purely central potential which is deep enough to support a number of bound states in our $l = 1$ application.  However, it cannot be considered a realistic NN interaction which has no bound states in the $l=1$ channel.  In addition, our interaction lacks the tensor, spin-orbit, spin-spin, and isospin components which are known to be influential in NN potentials (see Ref. \cite{source17} for a more detailed discussion on the structure of realistic NN interactions).  Instead, our aim is to provide guidance for applications to bound state problems in many-body systems which leads us to an attractive Gaussian interaction with multiple bound states that entail multiple distance scales as we will elaborate.

In a basis function approach, we can decompose the functional form of the solution $R_{\alpha}(r)$ (or $u_{\alpha}(r)$) corresponding to state $\alpha$ into a linear combination of generic basis functions $\Phi_{n}(r)$: 

\begin{equation}
\label{eq6}
R_{\alpha}(r)=\sum_{n=n_{i}}^{N} a_{\alpha,n} \Phi_{n}(r),
\end{equation}
where $N$ is the basis truncation.  Since all basis function representations must be truncated in order to have a finite and numerically tractable matrix eigenvalue problem, we anticipate and investigate the resulting inaccuracies tied to this truncation.  For a given basis function, we aim to relate the truncation $N$ as well as any other basis-specific parameter to identifiable deficiencies in the solutions for the wave equation.

We can quantify those deficiencies by introducing long-range infrared (IR) and short-range ultraviolet (UV) characteristic scales given by $\lambda_{\textrm{basis}}$ and $\Lambda_{\textrm{basis}}$, respectively (where the subscript shall denote either the sinc or HO basis).  These limits provide characteristic boundaries beyond which the basis-space approximation breaks down.  While the relation between the characteristic scales and the basis parameters depend on the basis function, we require that the range of accurate approximation of the basis increases with truncation.

 In addition to the IR and UV limits, there are basis-independent scales that depend solely on the physical problem.  We refer to scales that explicitly depend on the parameters of $V(r)$ as \textit{intrinsic scales}, and characteristic scales that emerge from the solution to the physical problem and are not apparent in the initial equation as \textit{emergent scales}.  In the Gaussian problem, the intrinsic scale is
 \begin{equation}
\label{eq7}
\lambda_{\textrm{potential}}=\kappa
\end{equation}
whereas the emergent scales are

\begin{equation}
\label{eq8}
\lambda_{\textrm{WBS}}=\frac{\sqrt{2m|E_{\textrm{WBS}}|}}{\hbar}
\end{equation}
where $E_{\textrm{WBS}}$ is the energy of the WBS and

\begin{equation}
\label{eq9}
\lambda_{\textrm{DBS}}=\frac{\sqrt{2m|E_{\textrm{DBS}}|}}{\hbar}
\end{equation}
where $E_{\textrm{DBS}}$ is the energy of the DBS.  As we will see in Section \ref{tail corrections}, the emergent scales shown here characterize the tail behavior of the wave functions whereas the intrinsic scales characterize the parameters of the potential.

 In addition to defining the basis-dependent and basis-independent scales, we want to establish a relationship between them.  We conjecture that one must satisfy the following inequalities in order to achieve high eigenvalue accuracy across the spectrum of solutions:

\begin{equation}
\label{eq10}
\lambda_{\textrm{basis}} < \textrm{min}({\lambda_{\textrm{WBS}},\lambda_{\textrm{potential}},\lambda_{\textrm{DBS}}}) < \textrm{max}({\lambda_{\textrm{WBS}},\lambda_{\textrm{potential}},\lambda_{\textrm{DBS}}}) < \Lambda_{\textrm{basis}}.
\end{equation}
This informs our choice of parameters; we must choose the basis parameters in such a way that the wave function approximation is inaccurate only where the wave function is small enough to have inconsequential effects on the observables of interest.

\subsection{The HO Basis Function Approach}

\subsubsection{The Eigenvalue Problem in the HO Basis}

We expand the radial part of the wave function $R(r)$ corresponding to a particular state $\alpha$ as a linear combination of radial HO basis states $\Phi_{nl}(r)$ following Eq. (\ref{eq6}).  In our case, a particular $\alpha$ corresponds to a definite energy $\mu$ and angular momentum $l$.  As the cutoff $N$ of the radial quantum number goes to infinity, our approximation of $R_{\alpha}(r)$ approaches the exact solution due to the completeness of the HO basis.  Our goal is to substitute Eq. (\ref{eq6}) into Eq. (\ref{eq3}) and solve for the coefficients $a_{\alpha,nl}$ via the eigenvalue problem given by Eq. (\ref{eq1}).  

The full 3-dimensional HO basis functions are orthonormal, with 

\begin{equation}
\label{eq11}
\int_{0}^{2\pi}\int_{0}^{\pi}\int_{0}^{\infty}\Phi_{n'l'}^{*}(r)Y_{l'm'}^{*}(\theta,\phi)\Phi_{nl}(r)Y_{lm}(\theta,\phi)r^{2}\sin{\theta} drd\theta d\phi=\delta_{n,n'}\delta_{l,l'}\delta_{m,m'},
\end{equation}
where $Y_{lm}(\theta,\phi)$ are spherical harmonics.  The explicit functional form of $\Phi_{nl}(r)$ is

\begin{equation}
\label{eq12}
\Phi_{nl}(r)=\sqrt{\frac{2n!}{b^{3} \Gamma(n+l+\frac{3}{2})}}(\frac{r}{b})^{l}e^{-\frac{r^{2}}{2b^{2}}}L_{n}^{l+\frac{1}{2}}(\frac{r^{2}}{b^{2}}),
\end{equation}
where $L_{n}^{l+\frac{1}{2}}(\frac{r^{2}}{b^{2}})$ is the generalized Laguerre polynomial, $\Gamma(n+l+\frac{3}{2})$ is a gamma function, and 
\begin{equation}
\label{eq13}
b=\sqrt{\frac{\hbar}{m\Omega}}
\end{equation}
is the oscillator length.  The HO energy eigenvalue associated with this basis state is

\begin{equation}
\label{eq14}
\epsilon_{nl} = (2n + l + 3/2) \hbar \Omega.
\end{equation}


Using the properties of the HO basis functions and following the methods outlined in Ref. \cite{source3}, we can derive the HO matrix elements of $H$.  In particular,
\begin{equation}
\label{eq15}
H_{nl,n'l'}=\frac{\hbar \Omega}{2}[(2n+l+\frac{3}{2})\delta_{n',n}+\sqrt{n(n+l+\frac{1}{2})}\delta_{n'+1,n}+\sqrt{(n+1)(n+l+\frac{3}{2})}\delta_{n'-1,n}]\delta_{l',l}+\int_{0}^{\infty}\delta_{l',l}\Phi_{n'l'}^{*}(r)V(r)\Phi_{nl}(r)r^2 dr
\end{equation}
(note here that the centrifugal term in Eq. (\ref{eq3}) is included in the kinetic term).  We therefore have a closed form for the kinetic energy matrix elements.  In general however, we do not have a closed form for the interaction matrix elements, so we must integrate numerically.   

Since both the kinetic energy and the interaction components of $H$ are real and symmetric in the radial quantum number, $H$ is hermitian and is therefore an operator representation of a physically observable quantity, the system's energy.  Because $H$ is symmetric and we choose to normalize our eigenvectors to unity, the resulting eigenvectors ($\textbf{v}_{\alpha}=\textrm{N-dimensional vector of the coefficients in Eq. (\ref{eq6})}$) are orthonormal.  In particular,

\begin{equation}
\label{eq16}
\textbf{v}_{\alpha'}^{*}\cdot \textbf{v}_{\alpha}=\delta_{\alpha',\alpha}.
\end{equation}
In practice, numerically diagonalizing the Hamiltonian introduces some errors so that the orthonormality condition given by Eq. (\ref{eq16}) is not exact even though Eq. (\ref{eq11}) is an exact analytic property.  To remedy this, we renormalize our calculated eigenvectors.

\subsubsection{The HO Basis Characteristic Scales}
Since the HO basis we employ is finite, there are both IR and UV characteristic scales beyond which our approximation of the wave functions are expected to break down.  Those characteristic scales are expressed in terms of $N_{\textrm{max}}=2N+l$ and $\hbar \Omega$ \cite{source3}.  Here, $N$ now represents the limit of the radial quantum number $n$ since we work with a fixed orbital momentum $l=1$.  As noted in Ref. \cite{source3} and \cite{source18}, with earlier analysis introduced in Ref. \cite{source19}, we can approximate these IR and UV limits as


\begin{equation}
\label{eq17}
\lambda_{\textrm{HO}}=\sqrt{\frac{m\Omega}{2(N_{\textrm{max}}+\frac{7}{2})\hbar}}
\end{equation}
and

\begin{equation}
\label{eq18}
\Lambda_{\textrm{HO}}=\sqrt{\frac{2m\Omega(N_{\textrm{max}}+\frac{7}{2})}{\hbar}},
\end{equation}
respectively.

\subsection{the Sinc Collocation Approach}\label{section2.2}

\subsubsection{The Eigenvalue Problem in the Sinc Basis}

We next solve the same problem using sinc basis functions.  We will primarily follow the procedure outlined in Ref. \cite{source13} and \cite{source20}.  As in the HO basis, we decompose $u_{\alpha}(r)$ into a linear combination of sinc basis functions:  

\begin{equation}
\label{eq19}
u_{\alpha}(r)=\sum_{k=-m_{\textrm{Val}}}^{m_{\textrm{Val}}}u_{\alpha k }(kh_{\gamma})sinc(\frac{r-kh_{\gamma}}{h_{\gamma}}),
\end{equation}
where $sinc(\zeta) \equiv \frac{sin(\pi \zeta)}{\pi \zeta}$, $m_{\textrm{Val}}$ is the truncation parameter (the sinc analog to $N_\textrm{max}$) and $h_{\gamma}$ is a step size (in fm) that is a measure of the spacing between collocation points.  In particular, $h_{\gamma}$ is   

\begin{equation}
\label{eq20}
h_{\gamma}=\frac{\pi}{\gamma\sqrt{m_{\textrm{Val}}}},
\end{equation}
where $\gamma$ is an adjustable parameter given in $\textrm{fm}^{-1}$.  In a way, it is analogous to $b$ in the HO basis since it sets an elementary length scale.  However, unlike in $b$, $h_{\gamma}$ is proportional to the inverse of the square root of $m_{\textrm{Val}}$ which controls the longest distances accessed in the basis.

The next step is to choose a conformal map based on the boundary conditions of the problem.  We know from basic quantum mechanics that the bound state wave functions span a semi-infinite space, behave algebraically near $r=0$, and exponentially decay for large $r$.  Following Ref. \cite{source5}, we choose the conformal map to be 

\begin{equation}
\label{eq21}
\theta_{\beta,\sigma,\xi}(r)=\beta  \ arcsinh(e^{\frac{r}{r_{0}}+\sigma}+\xi),
\end{equation}
where $\beta$, $\sigma$, $r_{0}$, and $\xi$ are adjustable parameters.  $\beta$ and $r_{0}$ are in fm whereas $\sigma$ and $\xi$ are dimensionless.  Throughout this work, we will set ($r_{0}$,$\beta$,$\sigma$,$\xi$,$\gamma$)=(1 fm, 1 fm ,0,0,1 $\textrm{fm}^{-1}$) and we will abbreviate $h_{\gamma}$ as $\theta_{\beta,\sigma,\xi}(r)$ to $h$ and $\theta(r)$, respectively.  Investigating alternative choices for these parameters is a potential topic for future study.

Next, we use the symmetrization transform

\begin{equation}
\label{eq22}
v(r)=(\sqrt{\eta'(r)}u(r)) \hspace {1mm} o \hspace {1mm} \theta(r)
\end{equation}
following Ref. \cite{source13} and \cite{source20}, where $\eta(r)=ln(sinh(r))$, the inverse of $\theta(r)$, $\eta'(r)=\frac{d{\eta}}{dr}$, and $F(r) \hspace {1mm} o \hspace {1mm} \theta(r)= F(\theta(r))$ ($F(r)$ in this case is $\sqrt{\eta'(r)}u(r)$).  As a result, we recast Eq. (\ref{eq4}) as 

\begin{equation}
\label{eq23}
-v''(r)+\tilde{V}(r)v(r)=\frac{2m\mu}{\hbar^{2}}\theta'(r)^{2}v(r),
\end{equation}
where

\begin{equation}
\label{eq24}
\tilde{V}(r)=-\sqrt{\theta'(r)}\frac{d}{dr}(\frac{1}{\theta'(r)}\frac{d}{dr}(\sqrt{\theta'(r)} ))+ \frac{2m}{\hbar^{2}}(\theta'(r))^2V_{\textrm{eff}}(\theta(r)).
\end{equation}

We multiply $\frac{2m}{\hbar^{2}}$ on both sides of Eq. (\ref{eq4}) so that the form of the Schrödinger equation is identical to that in Ref. \cite{source13} and \cite{source20}.  If we collocate the equation at points $r=$-$h$ $\cdot$ $m_{\textrm{Val}}$,...,$+h$ $\cdot$ $m_{\textrm{Val}}$, we can use the following identity for the sinc derivatives

\begin{equation}
\label{eq25}
\delta_{j,k}^{(i)}=h^{i}(\frac{d}{dr})^{i}sinc(\frac{r-jh}{h})|_{r=k\cdot h}
\end{equation}
as used in Ref. \cite{source13} and \cite{source20}.  We will need the cases $i=0$ (zeroth derivative) and $i=2$ (second derivative).  In the $i=0$ case,

\begin{equation}
\label{eq26}
   \delta_{j,k}^{(0)}=\ \begin{cases} 
      1 & j = k \\ 
      0 & j \neq k 
   \end{cases}  
\end{equation} 
whereas in the $i=2$ case, Eq. (\ref{eq25}) is

\begin{equation}
\label{eq27}
   \delta_{j,k}^{(2)}=\ \begin{cases} 
      \frac{-\pi^2}{3} & j = k \\ 
      \frac{(-2)(-1)^{j-k}}{(j-k)^{2}} & j \neq k 
   \end{cases}  
\end{equation}
\cite{source13}.  The result is the generalized eigenvalue problem 

\begin{equation}
\label{eq28}
(A-\frac{2m\mu}{\hbar^{2}} D^{2})v=0
\end{equation}
where the matrix elements of $A$ and $D^{2}$ are given by

\begin{equation}
\label{eq29}
    A_{j,k}=\frac{-1}{h^2}\delta_{j,k}^{(2)}+\tilde{V}(kh) \delta_{j,k}^{(0)}
\end{equation}
and

\begin{equation}
\label{eq30}
D^{2}_{j,k}=(\theta'(kh))^2\delta_{j,k}^{(0)}.
\end{equation}
 Note that $A$ and $D^{2}$ are square matrices with dimension $2m_{\textrm{Val}}+1$.  Because $D^{2}$ is diagonal and our choice of conformal map ensures that it has non-zero entries, we can easily recast Eq. (\ref{eq28}) as an eigenvalue problem.  In particular,
 
 \begin{equation}
 \label{eq31}
 D^{-2}Av=\frac{2m\mu}{\hbar^{2}} v.
 \end{equation}
 
 There are issues with numerically evaluating the matrix $D^{-2}A$.  First, even though it gives the expected eigenvalues (as we will see in the numerical results), it is not diagonal.  In our case, we find that all the eigenvalues (including those for unbound states) are real but the eigenvectors are not orthonormal.  Second, there are many orders of magnitude difference between the largest and the smallest eigenvalues of $D^{-2}A$.  In addition, as we discuss in \ref{appendix:a}, we encounter large off-diagonal matrix elements which lead to a badly-conditioned matrix and can cause convergence issues for large $m_{\textrm{Val}}$.  To get around this particular problem, we diagonalize numerically at 35-digit precision (see \ref{appendix:b}).
 
 We then obtain the functional form of the bound state eigenvectors employed to compute other observables besides energy.  The inverse of the transformation rule (it is $u(r)$, not $v(r)$ in the original eigenvalue problem that we seek) given by Eq. (\ref{eq21}) is
 
 \begin{equation}
 \label{eq32}
 u(r)=(\frac{1}{\sqrt{\eta'(r)}})(v \hspace{1mm} o \hspace{1mm} \eta(r)).
 \end{equation}
 Our non-normalized wave function for a particular eigenstate $\alpha$ is therefore,
 
 \begin{equation}
 \label{eq33}
 u_{\alpha}(r)=\sum_{k=-m_{\textrm{Val}}}^{m_{\textrm{Val}}}v_{\alpha k}\sqrt{\frac{1}{\eta'(r)h}}sinc(\frac{\eta(r)-kh}{h})=\sum_{k=-m_{\textrm{Val}}}^{m_{\textrm{Val}}}v_{\alpha k}S(k,h,r),
 \end{equation}
  where the $v_{\alpha k}$ values are obtained by solving Eq. (\ref{eq31}). It is straightforward to normalize by dividing Eq. (\ref{eq33}) by  $\sqrt{\int_{0}^{\infty} |u_{\alpha}(r)|^{2} dr}$ to obtain the normalized functional form of the wave function
  
  \begin{equation}
  \label{eq34}
      u_{\alpha }^{\textrm{norm}}(r)=\sum_{k=-m_{\textrm{Val}}}^{m_{\textrm{Val}}}\tilde{v}_{\alpha k}\sqrt{\frac{1}{\eta'(r)h}}sinc(\frac{\eta(r)-kh}{h})=u_{\alpha }^{\textrm{norm}}(r)=\sum_{k=-m_{\textrm{Val}}}^{m_{\textrm{Val}}}\tilde{v}_{\alpha k}S(k,h,r).
  \end{equation}
  
  To obtain the mean square radius of state $\alpha$, we simply take the integral $\int_{0}^{\infty} r^{2}|u_{\alpha}^{\textrm{norm}}(r)|^{2}dr$.  Assuming large enough $m_{\textrm{Val}}$, the bound state integrals are numerically stable at the machine precision level (12-digit precision, see \ref{appendix:b}).
  
\subsubsection{The Sinc Basis Characteristic Scales}  

Like in the HO or in any other basis, there is at least one point where the sinc basis approximation breaks down.  Here, we aim to develop information about the sinc characteristic scales that depend on $m_{\textrm{Val}}$ and $h$ inspired by how the HO scales depend on  $N_{\textrm{max}}$ and $\hbar \Omega$.  That is, we aim to provide an estimate of the UV and IR scales in the sinc basis defined with our choice of parameters.  We do this by first rewriting Eq. (\ref{eq21}) in a more suggestive form:

\begin{equation}
\label{eq35}
\theta(r)=ln(e^{r}+\sqrt{1+e^{2r}}).
\end{equation}
Since the UV and the IR characteristic scales correspond to small and large values of $r$ respectively, we are interested in what happens at the asymptotics.  Note that $\theta(r)$ behaves linearly for both small and large values of $r$.  Next, we plug in the appropriate collocation points ranging from $-h\cdot$ $m_{\textrm{Val}}$ to $+h$ $\cdot$ $m_{\textrm{Val}}$.  In the UV (small $r$) case, we set $r=h$, since it is the smallest nonzero value.  In the IR case, we set $r=h$ $\cdot $ $m_{\textrm{Val}}$ since it is the largest.  As a consequence, the IR and UV characteristic scales in the sinc basis are

\begin{equation}
\label{eq36}
 \lambda_{\textrm{sinc}}=\frac{C_{1}\gamma }{\pi \sqrt{m_{\textrm{Val}}}}   
\end{equation}
and

\begin{equation}
\label{eq37}
\Lambda_{\textrm{sinc}}=\frac{C_{2}\gamma \sqrt{m_{\textrm{Val}}}}{\pi},    
\end{equation}
respectively, where $C_{1}$ and $C_{2}$ are constants to be determined.  For the sinc basis in the present application, we choose $C_{1}$ and $C_{2}$ by setting $m_{\textrm{ValIR}}$ and $m_{\textrm{ValUV}}$ to be the values of $m_{\textrm{Val}}$ such that five-digit accuracy is satisfied for $E_{\textrm{WBS}}$ and $E_{\textrm{DBS}}$, respectively.  To obtain $C_{1}$, we equate $\lambda_{\textrm{sinc}}|_{m_{\textrm{Val}}=m_{\textrm{ValIR}}}$ with $\textrm{min}({\lambda_{\textrm{WBS}},\lambda_{\textrm{potential}},\lambda_{\textrm{DBS}}})$.  We obtain $C_{2}$ in a similar way, equating it with $\textrm{max}({\lambda_{\textrm{WBS}},\lambda_{\textrm{potential}},\lambda_{\textrm{DBS}}})$.  Eigenvalue calculations in the sinc basis discussed in Section \ref{results and comparisons} show that five-digit precision for the DBS is achieved at $m_{\textrm{ValIR}}=15$ and $m_{\textrm{ValUV}}=65$ for the WBS.  They also show that $\textrm{min}({\lambda_{\textrm{WBS}},\lambda_{\textrm{potential}},\lambda_{\textrm{DBS}}})=\kappa$ and $\textrm{max}({\lambda_{\textrm{WBS}},\lambda_{\textrm{potential}},\lambda_{\textrm{DBS}}})=\lambda_{{\textrm{DBS}}}$.  Hence,

\begin{equation}
\label{eq38}
C_{1}=\frac{\pi\kappa}{\gamma}\sqrt{m_{Val_{\textrm{IR}}}}    
\end{equation}
and

\begin{equation}
\label{eq39}
C_{2}=\frac{\pi}{\gamma\hbar}\sqrt{\frac{2m|E_{\textrm{DBS}}|}{m_{Val_{\textrm{UV}}}}}.
\end{equation}
 As a result, we obtain $C_{1}=3.93309$ and $C_{2}=2.19786$.  This implies that the WBS will be more challenging since it requires a larger $m_{\textrm{Val}}$.  Correspondingly, a value of $m_{\textrm{Val}}=65$ or greater will be needed for at least 5 digit accuracy of both the WBS and DBS states.

\section{Results and Comparisons}\label{results and comparisons}

We now compute the bound state eigenvalues and mean square radii using the sinc and HO bases.  Our first step is choosing the parameters and computing the corresponding cutoffs.

\subsection{Choice of Parameters}
As indicated in Section \ref{theory}, in order to achieve wave functions of suitable precision for all the bound states, we select our basis parameters constrained by Ineq. (\ref{eq10}).  For the HO basis, we have found, after some exploration, that we can choose the parameters $\hbar\Omega=20$ MeV and $N_{\textrm{max}}=60$ to obtain good overall convergence among all but the WBS.  If $\hbar\Omega$ is too small, the short-range (UV) behavior of the wave functions will not be sufficiently accurate.  If it is too large, the long-range (IR) behavior will not be sufficiently accurate.  The accuracy of the DBS (WBS) is more closely linked with the UV (IR) characteristic scale.  The choice of $\hbar\Omega=20$ MeV appears to be a reasonable compromise for our chosen application.  The margin of error quoted in Table \ref{tab1} is based on comparisons with results obtained with $N_{\textrm{max}}=50$ and with the same $\hbar\Omega$.

For the sinc basis, we choose $m_{\textrm{Val}}=200$ in order that the DBS eigenvalue precision matches the eigenvalue precision obtained via the HO basis at $N_{\textrm{max}}=60$ and $\hbar\Omega=20$ MeV (i.e. approximately 18 significant figures, as seen in Table \ref{tab1} and in \ref{appendix:c}).  Moreover, when we compare all the bound state energies we find that the HO basis and sinc basis results have about the same accuracy with our choices of basis parameters (with the exception of the WBS).  As an internal check, we compared the results to those obtained with $m_{\textrm{Val}}=150$.

Our parameter choices provide the following characteristic scales for the sinc and HO bases:

\begin{itemize}
 \item $\lambda_{\textrm{sinc}}=0.0885$ $\textrm{fm}^{-1}$, $\Lambda_{\textrm{sinc}}=9.894$ $\textrm{fm}^{-1}$ for $m_{\textrm{Val}}=200.$
  \item $\lambda_{\textrm{sinc}}=0.102$ $\textrm{fm}^{-1}$, $\Lambda_{\textrm{sinc}}=8.568$ $\textrm{fm}^{-1}$ for $m_{\textrm{Val}}=150.$
 
  \item $\lambda_{\textrm{HO}}=0.0436$ $\textrm{fm}^{-1}$, $\Lambda_{\textrm{HO}}=5.534$ $\textrm{fm}^{-1}$ for $N_{\textrm{max}}=60.$
  \item $\lambda_{\textrm{HO}}=0.0475$ $\textrm{fm}^{-1}$, $\Lambda_{\textrm{HO}}=5.079$ $\textrm{fm}^{-1}$ for $N_{\textrm{max}}=50.$
\end{itemize}

From the energies of the bound states we find that the emergent scales are

\begin{itemize}
  \item $\lambda_{\textrm{WBS}}=0.441$ $\textrm{fm}^{-1}$
  \end{itemize}
   \begin{itemize}
  \item $\lambda_{\textrm{DBS}}=2.710$ $\textrm{fm}^{-1}.$
\end{itemize}

and the intrinsic scale is

\begin{itemize}
\item $\lambda_{\textrm{potential}}=0.155$ $\textrm{fm}^{-1}.$
\end{itemize}

The parameters satisfy Ineq. (\ref{eq10}).

\subsection{Numerical Results}
In this section we present the main results for the $l=1$ bound states.  These results include both the quantum observable and quantifications of its calculated uncertainties.  Table \ref{tab1} presents the energies of the seven bound states and Table \ref{tab2} presents the square radii of each of these states.

We introduce two metrics of numerical uncertainty that are shown in Tables \ref{tab1} and \ref{tab2}.  The first is the uncertainty identifiable within a selected basis, which we call the \textit{internal error}.  We quantify the internal error as the difference between the eigenvalues or mean square radii obtained by using the same basis with parameters $m_{\textrm{Val}}=200$ and $m_{\textrm{Val}}=150$ for the sinc basis and $N_{\textrm{max}}=60$ and $N_{\textrm{max}}=50$ for the HO basis.  We define a second form of uncertainty through a comparison of results in the two separate bases, which we refer to as the \textit{inter-basis error}.  We quantify the inter-basis error as the difference between the eigenvalues or mean square radii obtained using the sinc basis for $m_{\textrm{Val}}=200$ and the HO basis with $N_{\textrm{max}}=60$.  As an additional cross-check, we compare the data for energy to the Gaussian energies obtained by Crandall \cite{source16}, who applied a Prüfer transform \cite{source21} on Eq. (\ref{eq4})  (with the same Gaussian potential but in natural units) and implemented a discretized RK4 method \cite{source22} to solve for the eigenvalues.  We found no mean square radius values reported in the literature.

Table \ref{tab1} shows that the HO and sinc bases eigenvalues agree to well-beyond five significant figures except for the WBS.  Indeed, the agreement between the HO and sinc energies indicates the DBS results are accurate to 17 figures as displayed in the column labelled "Inter-Basis Error".  With increasing excitation energy, the results become less precise, dropping to 9 figures for the next-to-weakest bound state.  For the WBS, the inter-basis error shows agreement through only 4 figures.

The drop reflects the decreasing accuracy of the HO basis in calculating states approaching the continuum as indicated by its internal uncertainty (whereas the sinc basis maintains consistency regardless of state except for the WBS: see Table \ref{tabC6} in \ref{appendix:c}).  With the exception of the WBS state, both the sinc and HO bases agree with Crandall's results to at least nine significant figures (and possibly more than that had he provided more significant figures).  

Similar trends follow for the mean square radii except that the differences are larger, as seen in Table \ref{tab2}.  The inter-basis error consistently increases with the energy level.  As in Table \ref{tab1}, five-digit precision is achieved except for the WBS.  Moreover, as one approaches the continuum, internal error increases for the HO basis, reflecting the decreasing accuracy of the HO basis with increasing energy.  In addition to numerical diagonalization error, numerical integration error plays a role in the mean square radius error.  We further discuss integration techniques and numerical precision in appendices A and B. 

\begin{table}[H]
\renewcommand{\arraystretch}{2}
 \caption{Comparison between the sinc and HO bases bound state eigenvalues at $m_{\textrm{Val}}=200$, $\gamma=1$ $\textrm{fm}^{-1}$ and $N_{\textrm{max}}=60$, $\hbar\Omega=20$ MeV.  They are compared with the results of Crandall \cite{source16} who used separate methods.  Note that the inter-basis error (the difference in energy values obtained between the sinc and HO values at the aforementioned parameters) increases with energy levels. Blue digits indicate agreement with either of the Crandall or HO basis values.  Red digits denote  possible or certain disagreement with both the Crandall and HO values. The numbers in parenthesis are the internal uncertainties between $m_{\textrm{Val}}=200$ and $m_{\textrm{Val}}=150$ and $N_\textrm{max}=60$ and $N_\textrm{max}=50$ for the two bases.  The inter-basis error is the absolute difference between the HO results at $N_\textrm{max}=60$ and the sinc results at $m_{\textrm{Val}}=200$.   }
 
 \vspace{3mm}
 
 \centering
   \begin{adjustbox}{max width=30cm}
   
    \begin{tabular} {|c|c|c|c|}
    \hline
    HO Energy (MeV) & Sinc Energy (MeV)  & Inter-Basis Error (MeV) & Crandall's Energies (MeV) \\ \hline
      $-304.462838518739310(1)$& $-304.4628385\textcolor{blue}{1873931(2)}$& $4.9\cdot10^{-17}$ & -304.46283852 \\ \hline
      -235.4500423784240(1)&-235.4500423\textcolor{blue}{784240(2)} & $5.2\cdot10^{-16}$& -235.45004238 \\ \hline
      -173.244320477591(5)&-173.2443204\textcolor{blue}{77591}\textcolor{red}{0(5)} &$8.5\cdot10^{-15}$ & -173.24432048 \\ \hline
     -118.3839812228(1)& -118.3839812\textcolor{blue}{228}\textcolor{red}{132(10)}& $1.3\cdot10^{-13}$& -118.3839812 \\ \hline
      -71.6235513471(7)&-71.623551\textcolor{blue}{347}\textcolor{red}{070(2)} & $1.2\cdot10^{-12}$& -71.6235514\\ \hline
       -34.12993490(11)&-34.129934\textcolor{red}{899529(2)}& $2.6\cdot10^{-9}$& -34.1299349 \\ \hline
      -8.0826(31) & -8.08\textcolor{blue}{33}\textcolor{red}{299755(3)}& $6.9\cdot10^{-4}$& -8.08333 \\ \hline
    \end{tabular}
   \end{adjustbox}
   \label{tab1}
    \end{table}

\begin{table}[H]
\renewcommand{\arraystretch}{2}
\caption{Comparison between the sinc and HO bases bound state mean square radii for the same states presented in Table \ref{tab1}.  Uncertainties are larger and like in Table \ref{tab1}, increase with energy level.  Red digits denote possible or certain disagreement of the sinc result with the corresponding HO result.  Unlike in Table \ref{tab1}, we found no published mean square radii results with which to compare. }

 \vspace{3mm}
 
     \centering
    \begin{tabular} {|c|c|c|}
    \hline
    HO Square Radii ($\textrm{fm}^{2}$) & Sinc Square Radii ($\textrm{fm}^{2}$) & Inter-Basis Error ($\textrm{fm}^{2}$) \\ \hline
     5.701845132386(3)& 5.70184513238\textcolor{red}{5(1)}& $2.0\cdot10^{-13}$ \\ \hline
     11.30629956490(2)& 11.306299564\textcolor{red}{89(2)}&$1.6\cdot10^{-12}$  \\ \hline
    18.24968732879(12)& 18.249687328\textcolor{red}{90(5)}& $1.1\cdot10^{-10}$ \\ \hline
    27.2767730589(4)&27.276773058\textcolor{red}{3(2)} & $6.5\cdot10^{-10}$\\ \hline
   39.93321853(8)& 39.933218\textcolor{red}{45(1)}& $8.4\cdot10^{-8}$\\ \hline
   60.27815(5)& 60.2781\textcolor{red}{10103(3)}& $4.3\cdot10^{-5}$\\ \hline
    106.44(25)& 106\textcolor{red}{.5271867(2)}& $8.7\cdot10^{-2}$ \\ \hline
    \end{tabular}
  \label{tab2}
    \end{table}

\subsection{Structure of the Sinc Wave Functions as Compared to the HO Wave Functions}

We next investigate the properties of the two deepest-bound wave functions.  By plotting these wave functions, we expect to learn about the source(s) of the inter-basis error presented in Tables \ref{tab1} and \ref{tab2}.  Graphical representations of the wave functions should also illustrate the scales where the finite sinc and HO basis expansions break down.  We also anticipate that analyzing the wave functions will indicate a means through which to improve the accuracy of the results apart from simply increasing the size of the basis truncation.  We note that, for a simple problem like this, increasing $m_{\textrm{Val}}$ or $N_{\textrm{max}}$ to a large value could improve precision but, for more complicated problems, diagonalizing a matrix with a larger $m_{\textrm{Val}}$ or $N_{\textrm{max}}$ is computationally expensive.

\begin{figure}[H]
 
\begin{subfigure}{0.5\textwidth}
\includegraphics[width=1\linewidth, height=5cm]{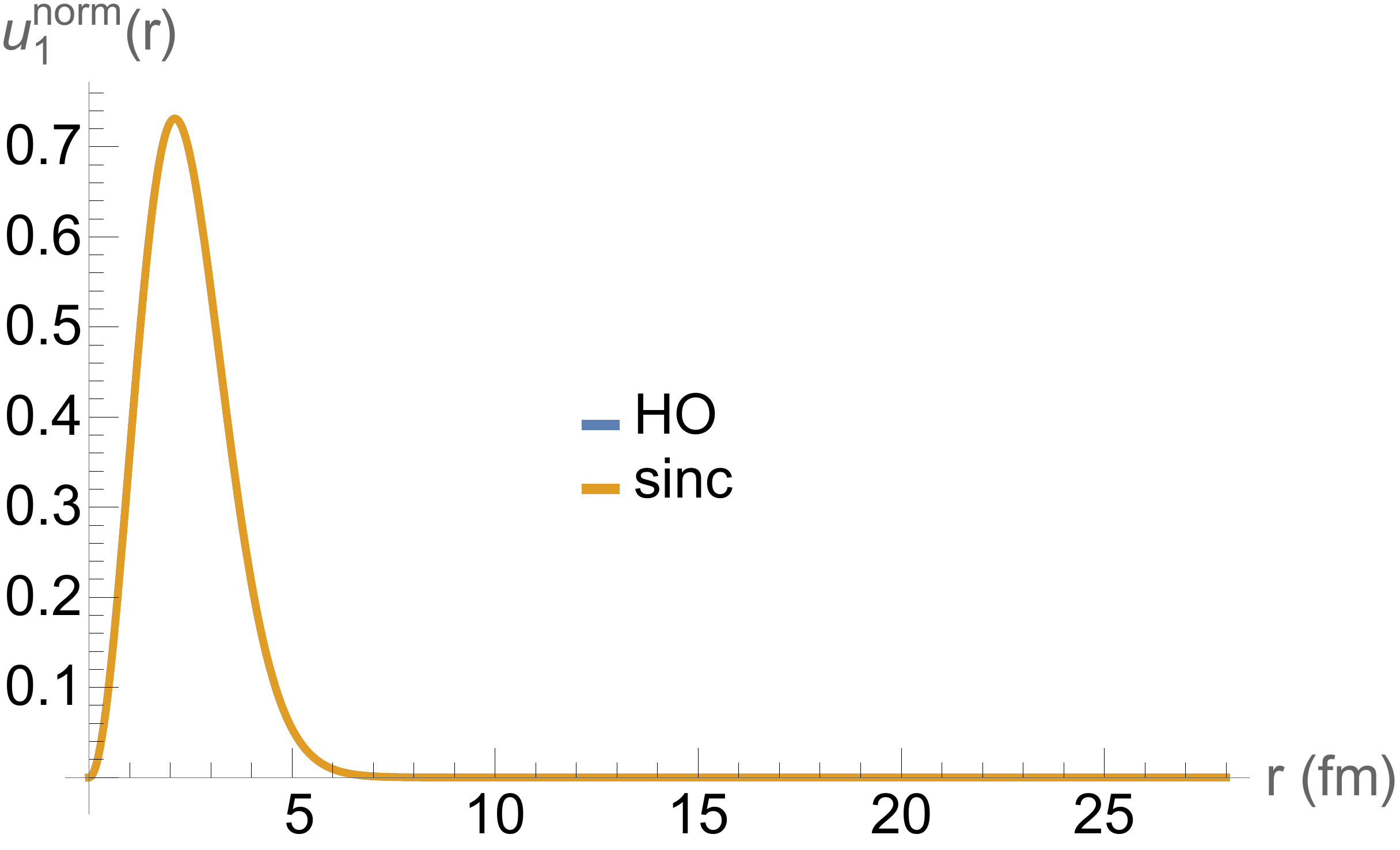} 
\caption{}
\label{fig1a}
\end{subfigure}
\begin{subfigure}{0.5\textwidth}
\includegraphics[width=1\linewidth, height=5cm]{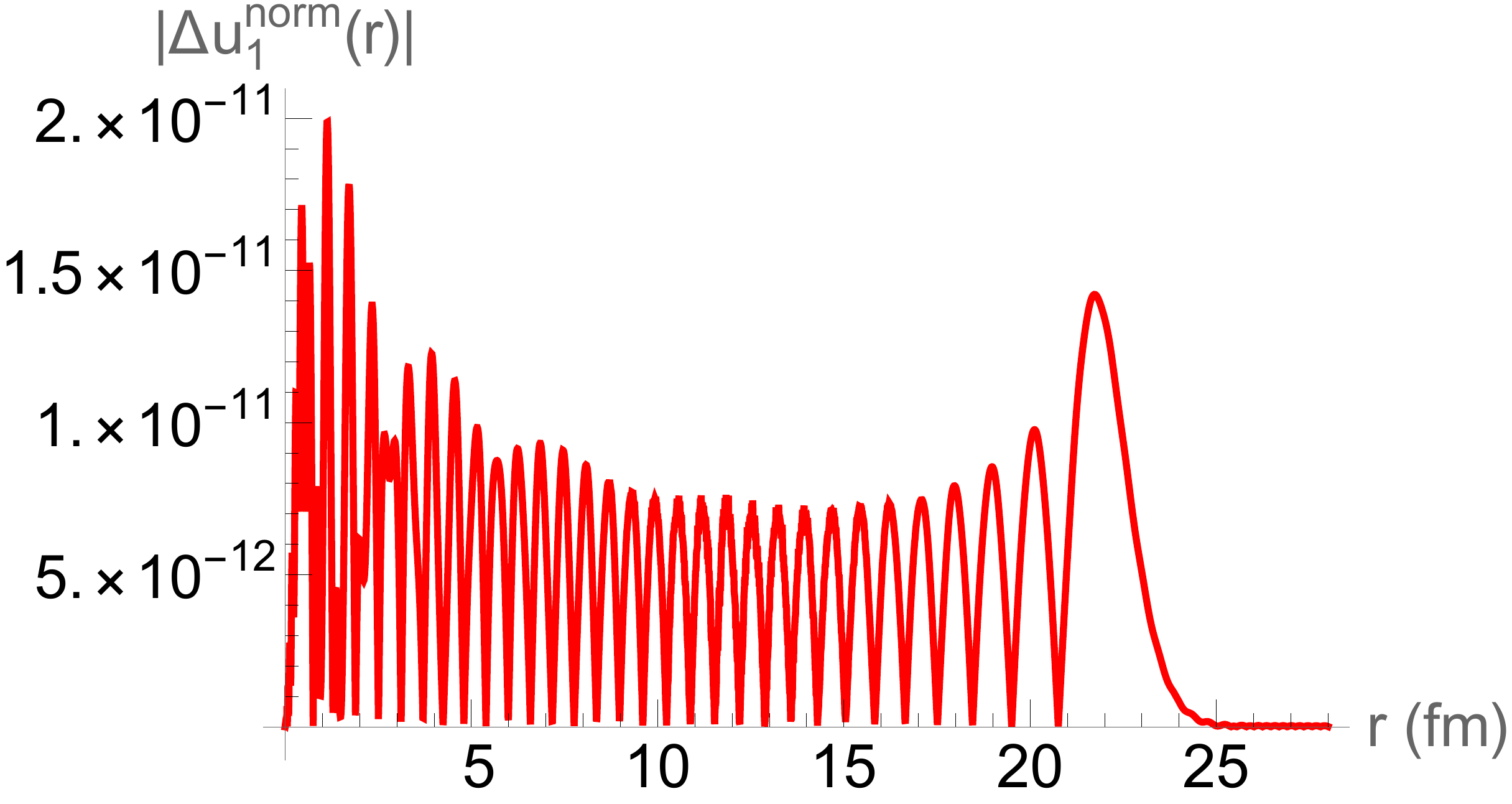}
\caption{}
\label{fig1b}
\end{subfigure}
\begin{subfigure}{0.5\textwidth}
\includegraphics[width=1\linewidth, height=5cm]{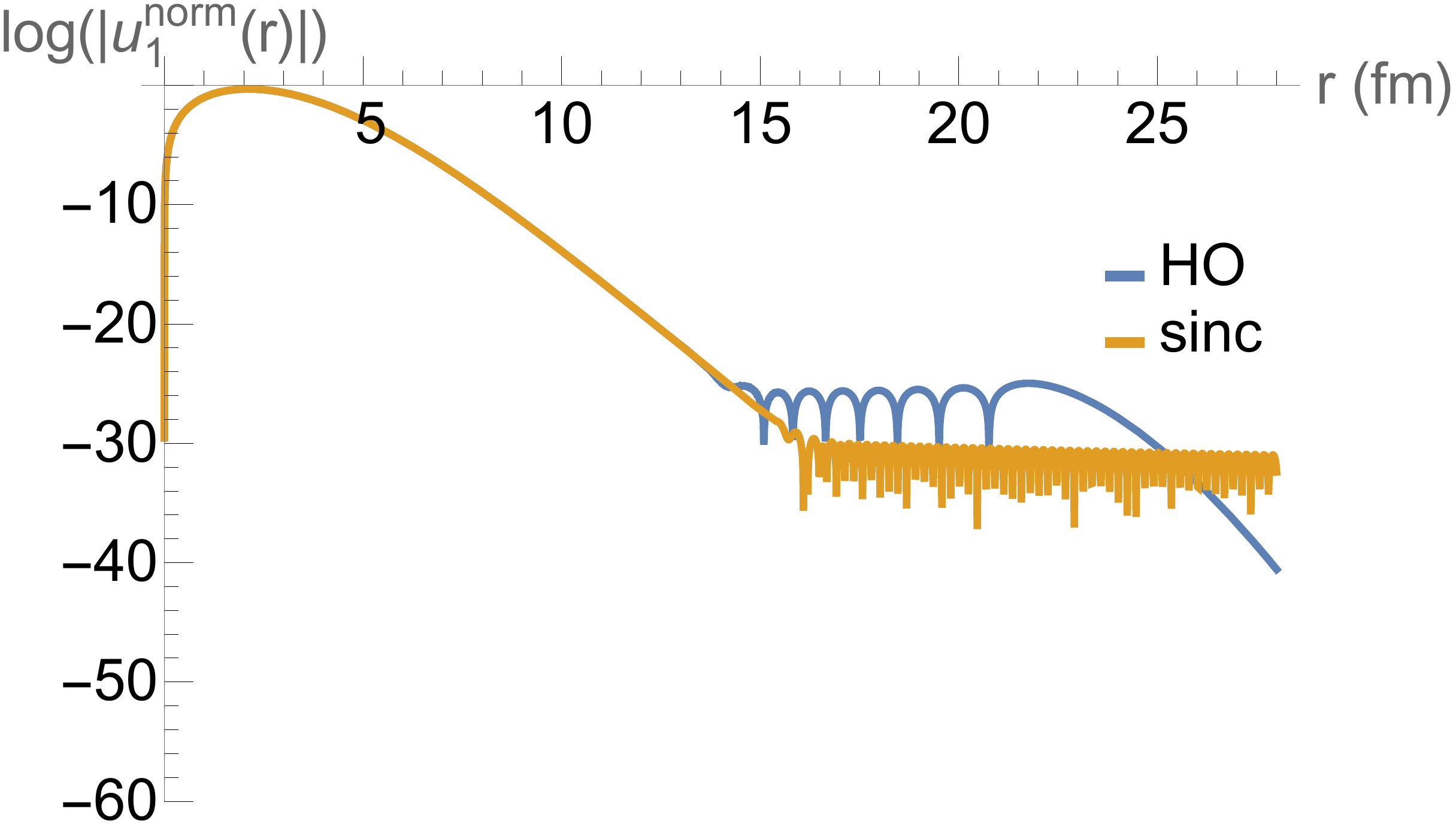}
\caption{}
\label{fig1c}
\end{subfigure}
\begin{subfigure}{0.5\textwidth}
\includegraphics[width=1\linewidth, height=5cm]{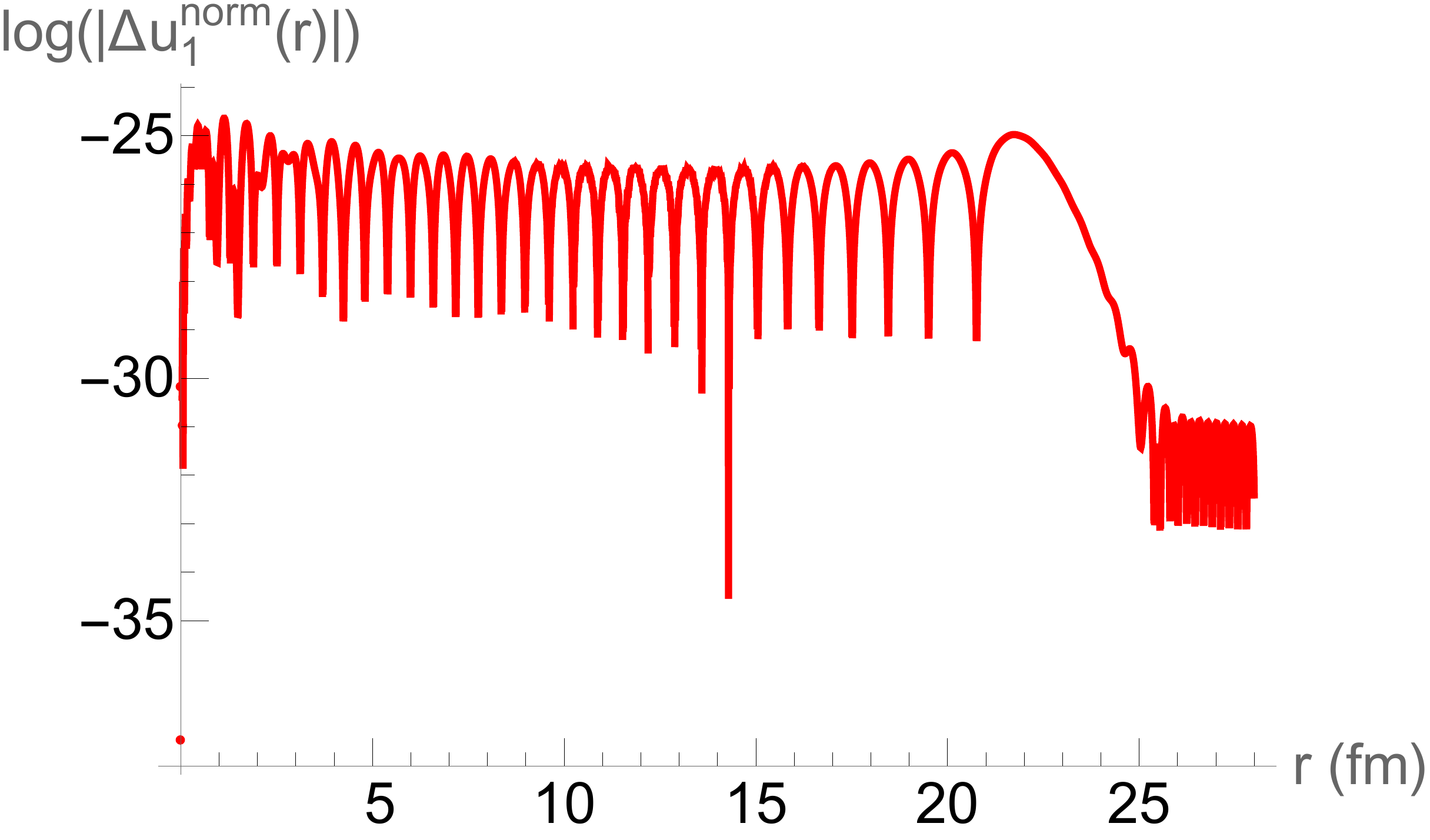}
\caption{}
\label{fig1d}
\end{subfigure}
 
\caption{
(a)  DBS $u_{1}^{\textrm{norm}}(r)$ for the sinc and HO bases.  The functions are virtually indistinguishable and appear visually as a single curve.
(b)  Absolute difference between the DBS approximations in the two bases.  Note the oscillatory nature of the error.  The bump and subsequent drop in error at large $r$ reflect the asymptotic behavior of the wave functions and the breakdown scale of the wave function approximations.  
(c)  Semi-log plot of the absolute wave functions of the DBS in the sinc and HO bases.  Note the oscillations in the respective tails of the two approximations.  They illustrate the breakdown scale beyond which the basis approximations deviate substantially from the expected analytic behavior (see \ref{appendix:d}).  Because there are no physical nodes for the DBS, the nodes in the tail end of both the sinc and HO bases wave functions arise from the small contributions of long-range basis states.
(d)  Semi-log plot of Fig. \ref{fig1b}.}

\label{fig1}
\end{figure}

We plot the functional forms of the sinc and HO wave functions for the DBS in Fig. \ref{fig1a} and observe that they are practically indistinguishable at the scale shown.  As expected, the DBS behaves algebraically near $r=0$ and exponentially decays as $r$ goes to infinity.  Fig. \ref{fig1b} shows the absolute difference between the sinc and HO wave functions.  The oscillatory nature of the error for $r$ less than about 20 fm reflects numerical noise whereas the bump and subsequent decay reflects the tails of the respective basis representations beyond their breakdown points.  

\begin{figure}[H]
 
\begin{subfigure}{0.5\textwidth}
\includegraphics[width=1\linewidth, height=5cm]{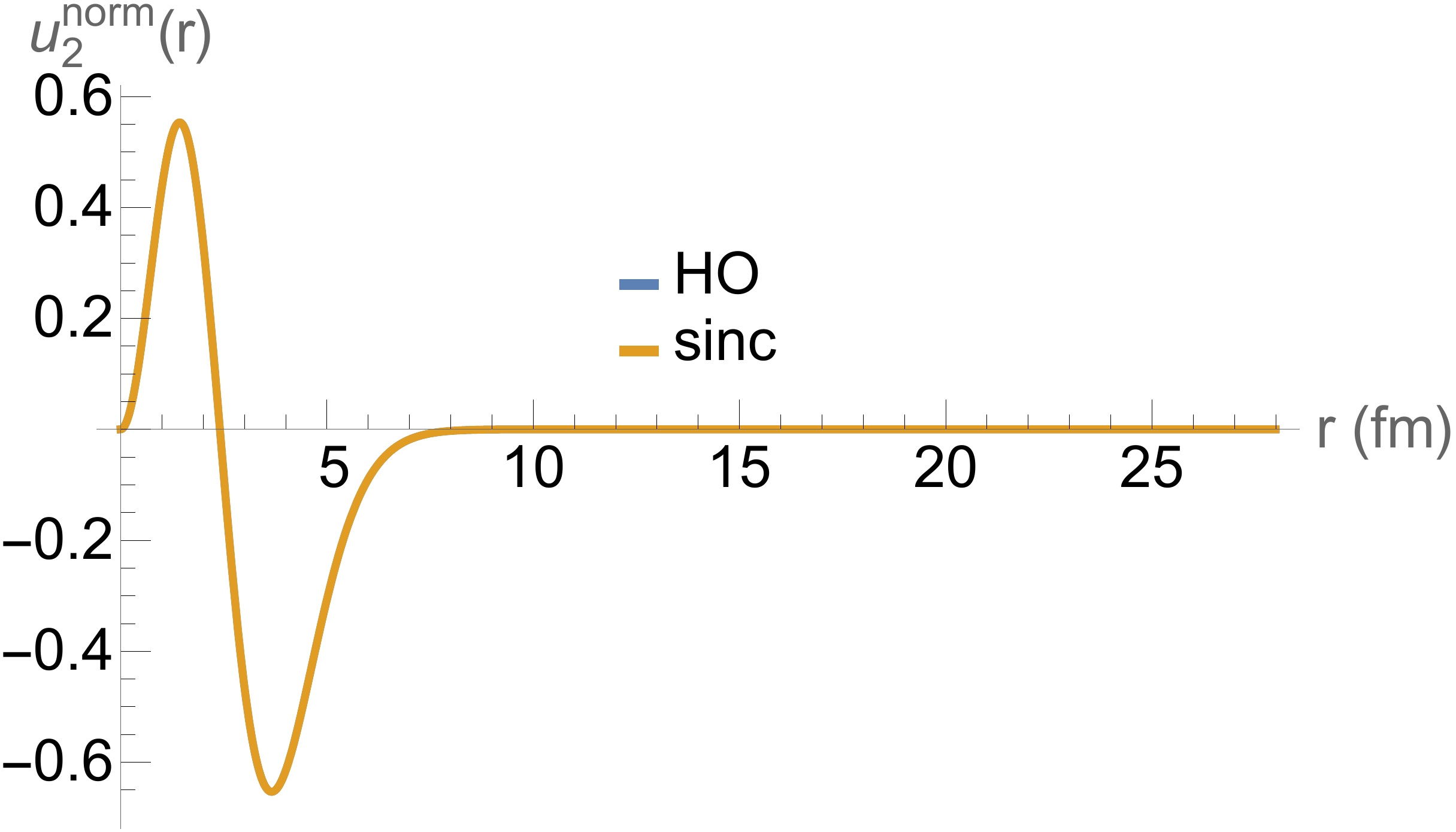} 
\caption{}
\label{fig2a}
\end{subfigure}
\begin{subfigure}{0.5\textwidth}
\includegraphics[width=1\linewidth, height=5cm]{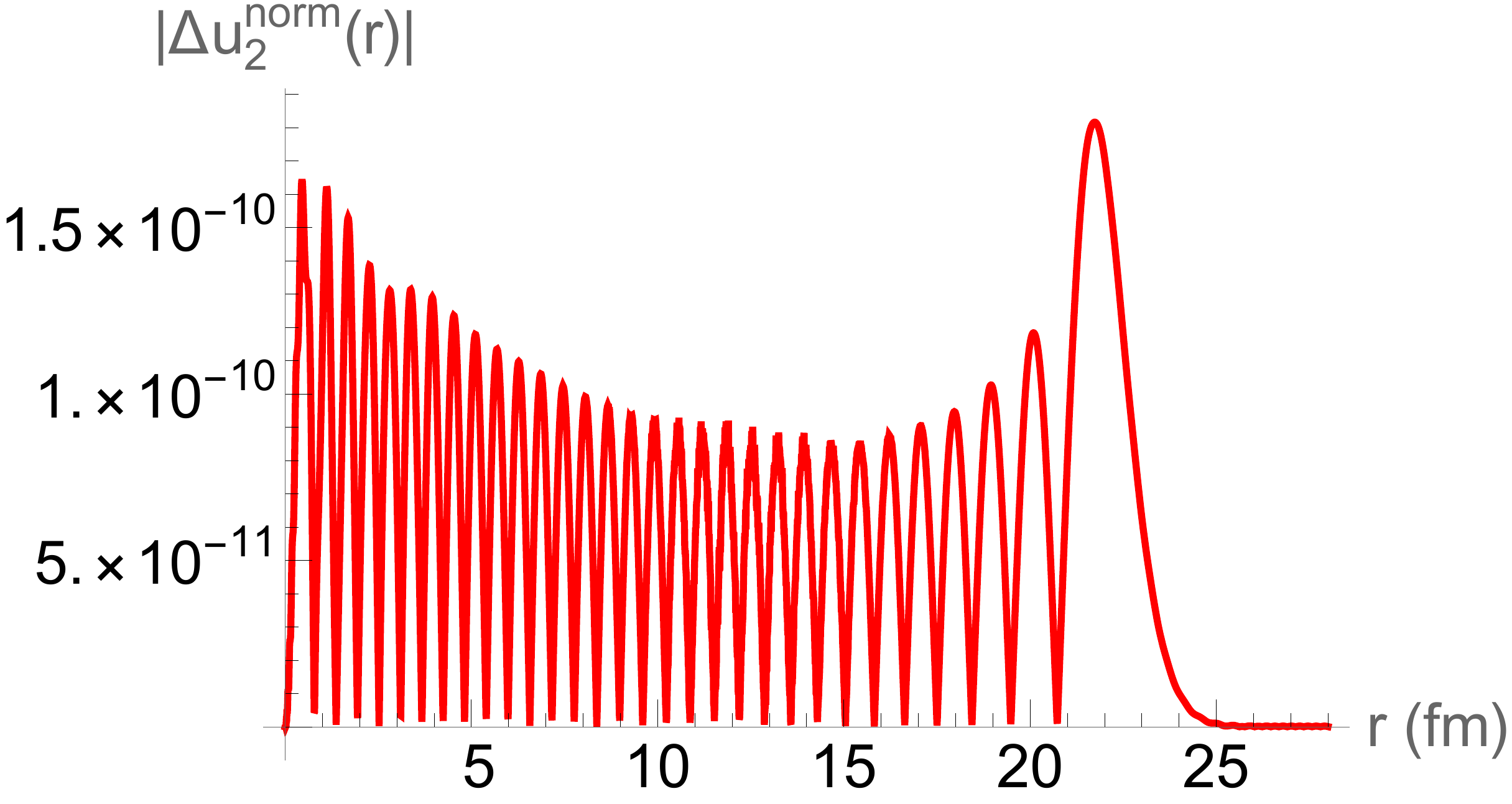}
\caption{}
\label{fig2b}
\end{subfigure}
\begin{subfigure}{0.5\textwidth}
\includegraphics[width=1\linewidth, height=5cm]{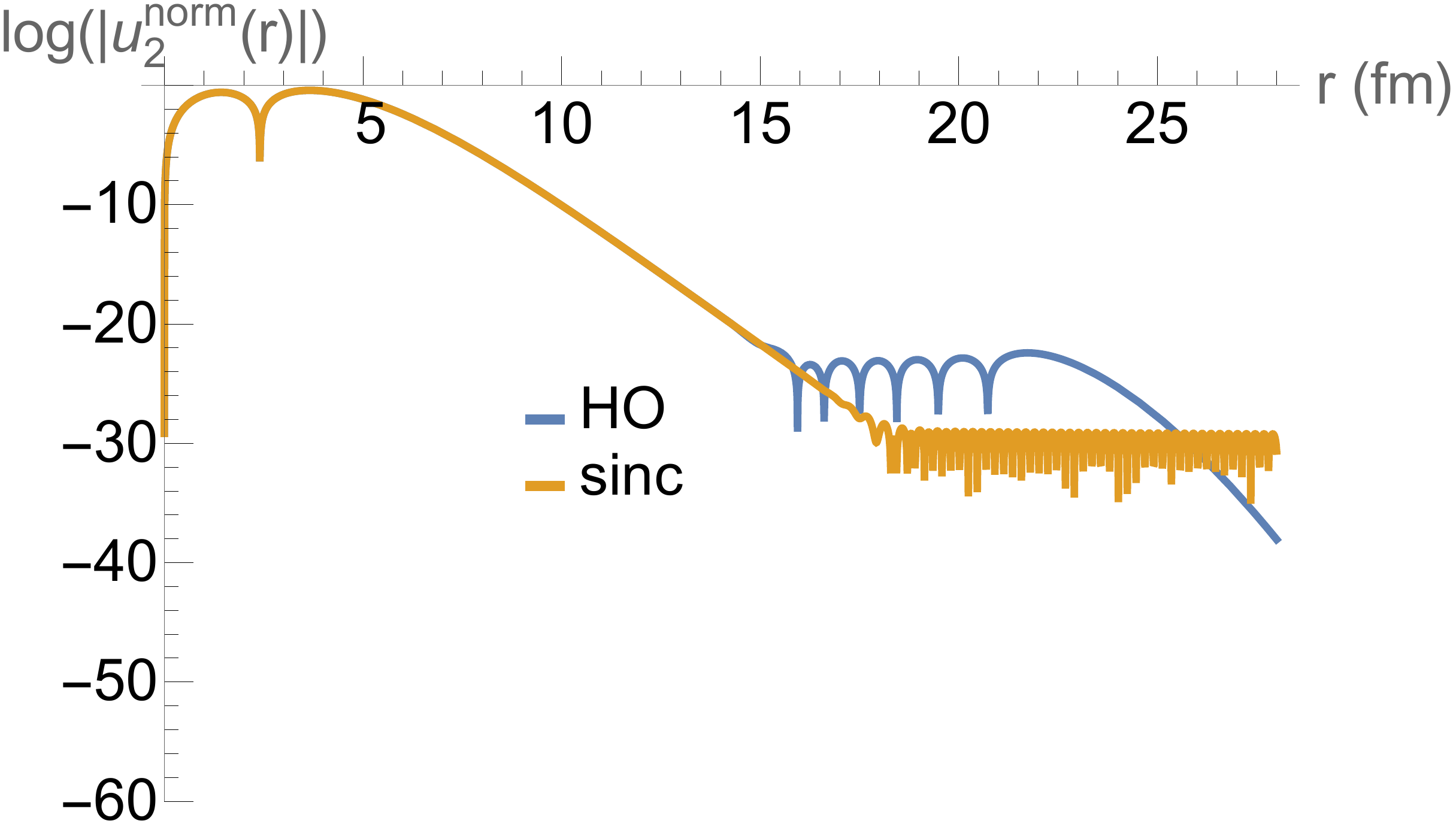}
\caption{}
\label{fig2c}
\end{subfigure}
\begin{subfigure}{0.5\textwidth}
\includegraphics[width=1\linewidth, height=5cm]{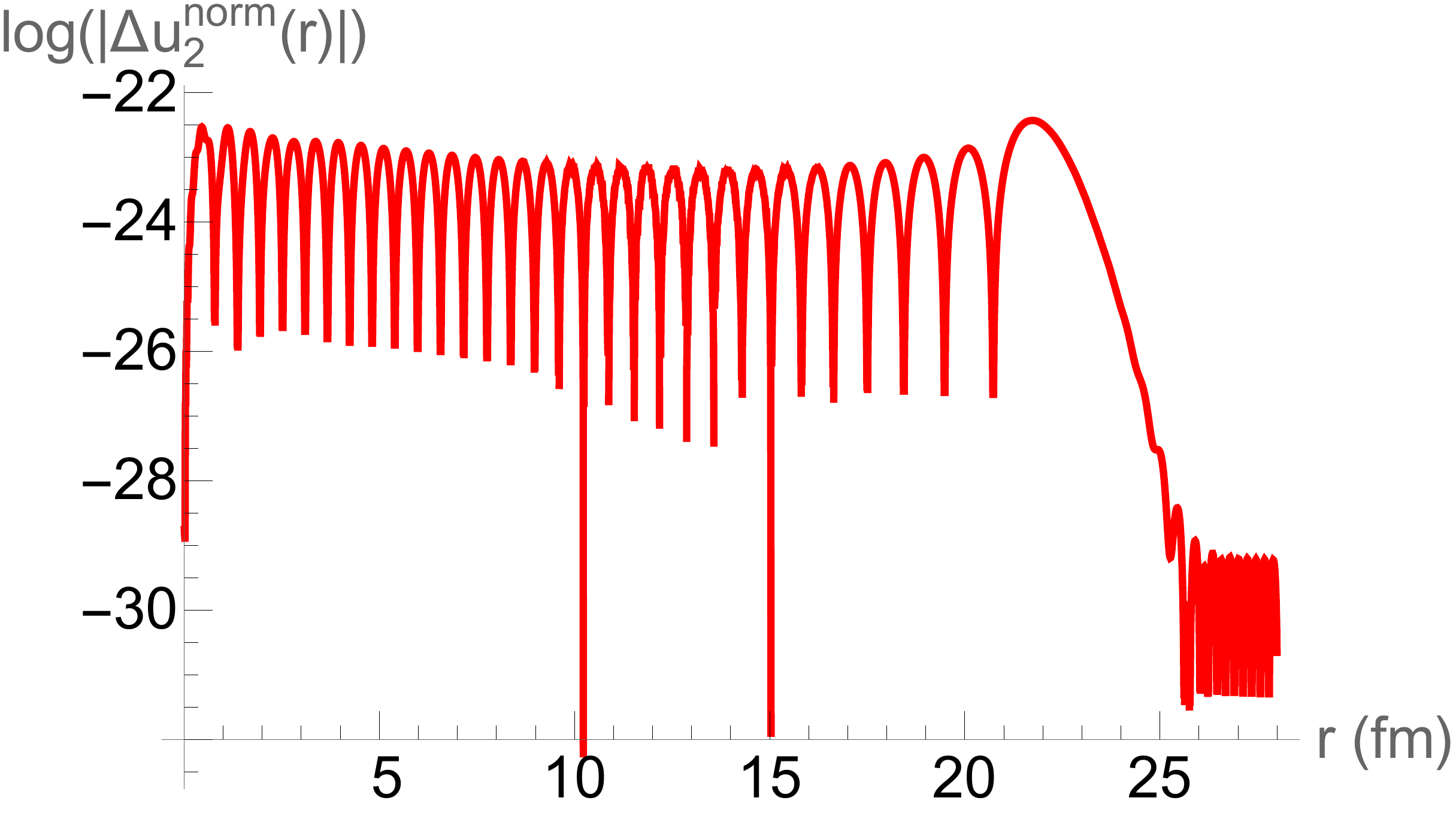}
\caption{}
\label{fig2d}
\end{subfigure}
 
\caption{(a)  First excited state wave function $u_{2}^{\textrm{norm}}(r)$ for the sinc and HO bases.  The wave function crosses zero once between the asymptotes at around $r=2.5$ fm.  The nodes in the tail occur beyond the numerical breakdown scale.  Compare with Fig. \ref{fig1a}.  (b)  Absolute difference between the first excited state wave function approximations.  Compare with Fig. \ref{fig1b}.  (c)  Semi-log plot of the absolute wave functions of the first excited state in the sinc and HO bases.  Note the oscillations in the respective tails of the two approximations.  As in Fig. \ref{fig1c}, they reflect breakdown scale beyond which the wave function approximations deviate substantially from the expected analytic behavior.  (d)  Semi-log plot of Fig. \ref{fig2b}.  Compare with Fig. \ref{fig1d}.}
\label{fig2}
\end{figure}

To gain more insight, we plot the log of the absolute wave function in Fig. \ref{fig1c}.  In the figure, sinc and HO representations of the DBS are practically indistinguishable and the error below about 15 fm appears to be negligible.  Beyond that however, oscillations occur at around 15 fm and 18 fm in the HO and sinc basis representations, respectively.  Beyond about 20 fm, the Gaussian nature of the HO basis becomes apparent, and the HO DBS representation decays in that fashion.  Since there are no physical nodes at nonzero $r$ in the DBS, the nodes at the tail reflect both numerical noise for both the bases and the respective breakdown scales of the sinc and HO DBS representations (in \ref{appendix:d}), we test the accuracy of the IR cutoff of the sinc basis given by Eq. (\ref{eq36}) by comparing it with the point in which the DBS sinc wave function breaks down and starts oscillating).  Fig. \ref{fig1d} shows the logarithm of the absolute error. We can draw similar conclusions on the sinc and HO basis representations of the first excited state shown in a corresponding set of panels in Fig. \ref{fig2}.  The main difference from the results in Fig. \ref{fig1} is that the error is one order of magnitude larger and that there is one more physical node.  A similar trend occurs up to the fifth excited state, where the nonphysical nodes of the HO basis disappear (see \ref{appendix:d}).

From our observations in Fig. \ref{fig1} and Fig. \ref{fig2}, both the sinc and HO wave function tails break down, but they do so differently; the sinc wave function dies exponentially, then oscillates around a small constant whereas the HO wave function tail oscillates (much less frequently than its sinc counterpart), then falls off like a Gaussian (a quadratic function in the semi-log figure).  This is due to the influence of the highest HO basis states at large $r$.  Because those underlying states are highly oscillatory before the Gaussian behavior takes over, the HO wavefunction tails break down in that manner.  Similarly, the sinc tail breakdown occurs due to the nature of the underlying basis function.  As we will see in the following section, the Gaussian behavior of the HO basis at large $r$ becomes more consequential in the WBS state than in the ground and first excited state.  This is the cause of the increasing inter-basis error with energy levels in Tables \ref{tab1} and \ref{tab2}. 

\section{Tail Corrections}\label{tail corrections}

As seen in the previous section, the main numerical issue with the DBS and first excited states shown in Fig. \ref{fig1} and \ref{fig2}, is the behavior of the sinc and HO wave function tails.  We associate the tail region with the IR properties of the solution.  Since we are interested in long-range observables such as the mean square radius, discussed in Section \ref{results and comparisons}, we want to explore possible improvements to the IR properties of our wave functions that are more efficient than further increasing the basis dimension.  Hence, in this section, we attempt to address this by replacing the tails of one or both of the wave functions with exponential functions and observe if numerical results improve by calculating the mean square radii of the corrected wave functions.  As in the previous sections, we use the parameters $m_{\textrm{Val}}=200$, $\gamma=1$ $\textrm{fm}^{-1}$ for the sinc basis and $N_{\textrm{max}}=60$, $\hbar \Omega=20$ MeV for the HO basis.

\subsection{The DBS Tail Correction}\label{section4.1}

If we solve Eq. (\ref{eq1}) for a given bound state $n$ with energy $E_{n}$ at $r$ large enough so that the potential vanishes, we expect a solution of the form $e^{-\frac{\sqrt{2mE_{n}}}{\hbar}}$.  Our first task is to identify a threshold point $r_{\textrm{threshold}}$ beyond which the exponential behavior dominates.  We linearly fit points of the log of the absolute sinc DBS in the vicinity of $r_{\textrm{threshold}}$.  The reason we use a linear fit in the semi-log and not the linear scale is that the DBS is small in the region where it behaves exponentially (starting at around $r=8$ fm).  We then use the linear function obtained from the fit to attach to the log of the absolute sinc or HO DBS.  For details on fitting, see \ref{appendix:e}.  Fig. \ref{fig3} shows Fig. \ref{fig1c} in the region where both the log of the absolute sinc and the log of the absolute HO DBS deviate from the expected linear form for the DBS.  From the figure, it is clear that the log of the absolute sinc DBS stays linear until around $r=16$ fm whereas the log of the absolute HO DBS deviates from linearity at around $r=14$ fm.  We therefore choose $r_{\textrm{threshold}}=13.8$ fm, as it is the approximate point in which the log of the absolute HO DBS starts oscillating whereas its sinc counterpart is linear.  At around $r=23$ fm (which corresponds to $\frac{1}{\lambda_{\textrm{HO}}}$), the log of the absolute HO DBS starts behaving quadratically, corresponding to the Gaussian tail of the underlying HO basis function.  The contrast between the DBS without the exponential tail and that with the exponential tail is illustrated in Fig. \ref{fig3b}.

\begin{figure}[H]
 
\begin{subfigure}{0.5\textwidth}
\includegraphics[width=1\linewidth, height=5cm]{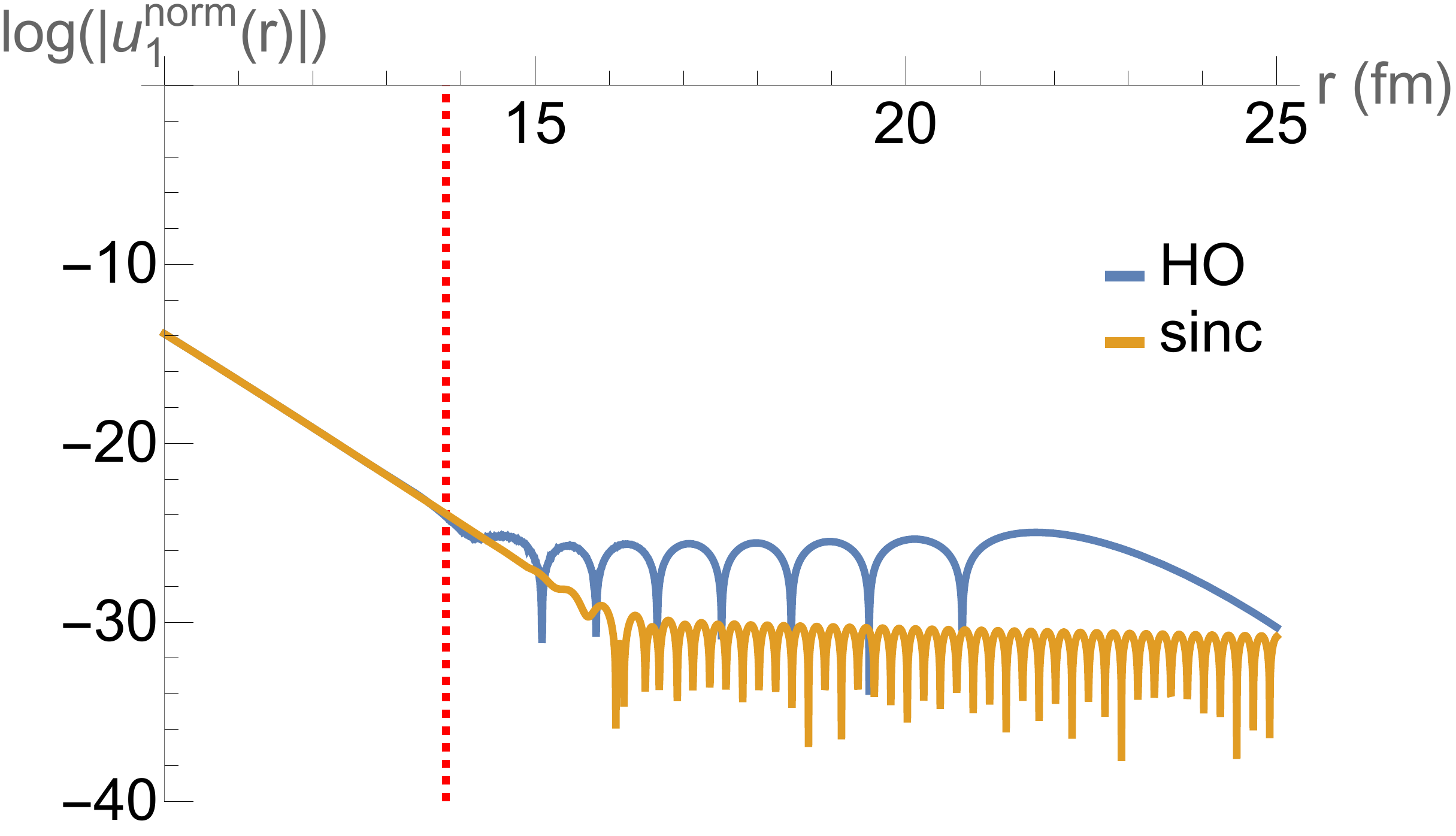} 
\caption{}
\label{fig3a}
\end{subfigure}
\begin{subfigure}{0.5\textwidth}
\includegraphics[width=1\linewidth, height=5cm]{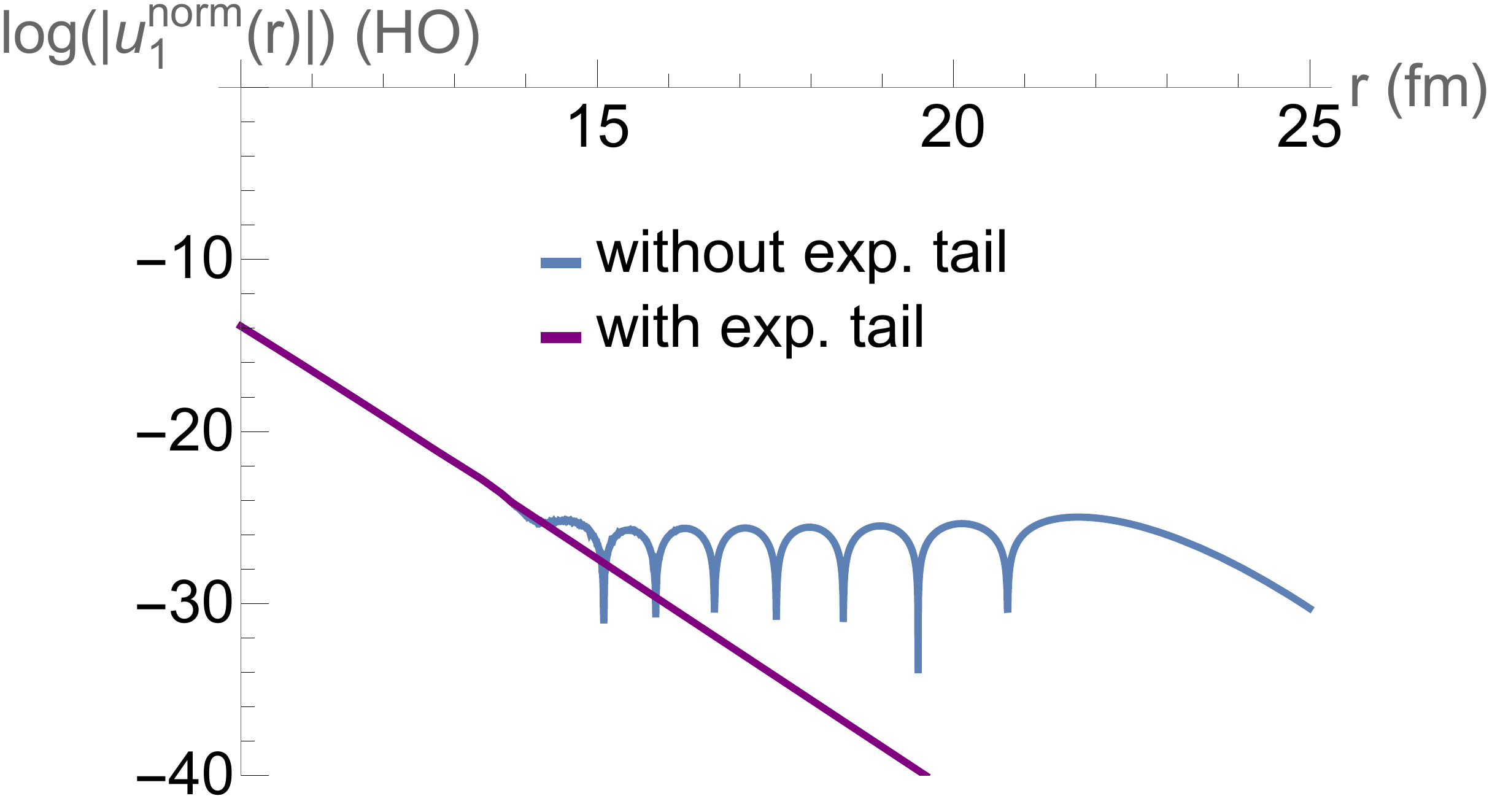}
\caption{}
\label{fig3b}
\end{subfigure}

\caption{(a)  Semi-log plot of the absolute sinc and HO DBS as in Fig. \ref{fig1c}.  The vertical line coincides with the chosen $r_{\textrm{threshold}}=13.8$ fm, which is the approximate point at which the HO DBS starts oscillating whereas its sinc counterpart is linear.   (b)  Semi-log plot of the corrected absolute HO DBS as compared to that without the linear tail.}
\label{fig3}
\end{figure}

In order to attach the chosen DBS tail to the wave function (sinc or HO), we redefine the sinc and HO DBS as piecewise (pw) functions:

\begin{equation}
\label{eq40}
     u_{\textrm{pw DBS}}^{\textrm{HO/sinc}}(r)=\ K\begin{cases} 
      u_{\textrm{HO/sinc}}(r) & r<r_{\textrm{threshold}} +\delta \\
     Ae^{-\lambda'r} & r>r_{\textrm{threshold}} +\delta
   \end{cases} 
\end{equation}
where $u_{\textrm{pw DBS}}^{\textrm{HO/sinc}}(r)$ is the function that is equal to the sinc or HO DBS below $r_{\textrm{threshold}}+\delta$ (up to a normalization factor K) and A is chosen such that the corrected wave function is continuous and has a continuous first derivative at the point $r_{\textrm{threshold}}+\delta$.  $\lambda'$ is a parameter to be fitted (following the procedure outlined in \ref{appendix:e}).  

\begin{figure}[H]
\begin{subfigure}{0.5\textwidth}
\includegraphics[width=1\linewidth, height=5cm]{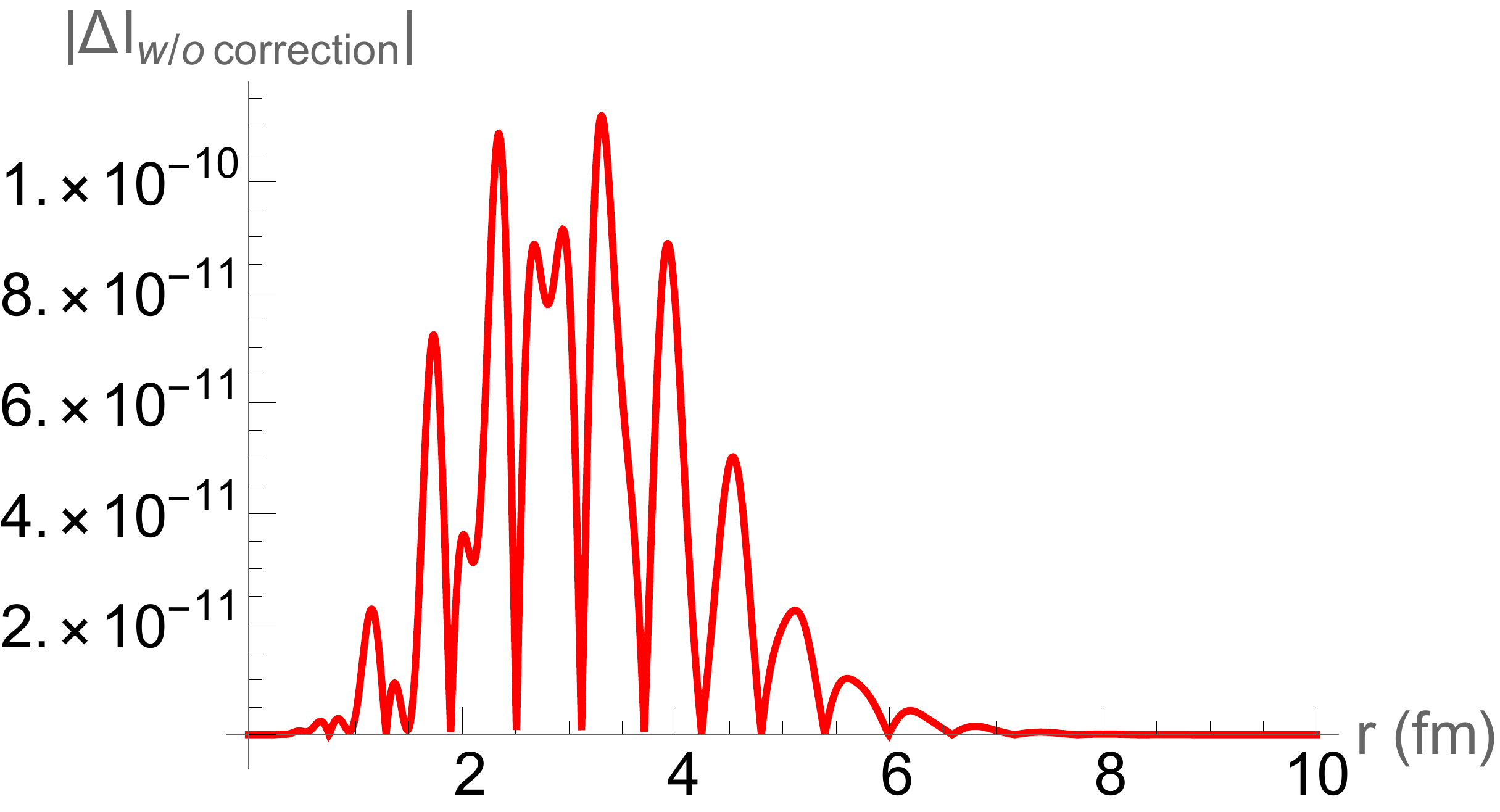} 
\caption{}
\label{fig4a}
\end{subfigure}
\begin{subfigure}{0.5\textwidth}
\includegraphics[width=1\linewidth, height=5cm]{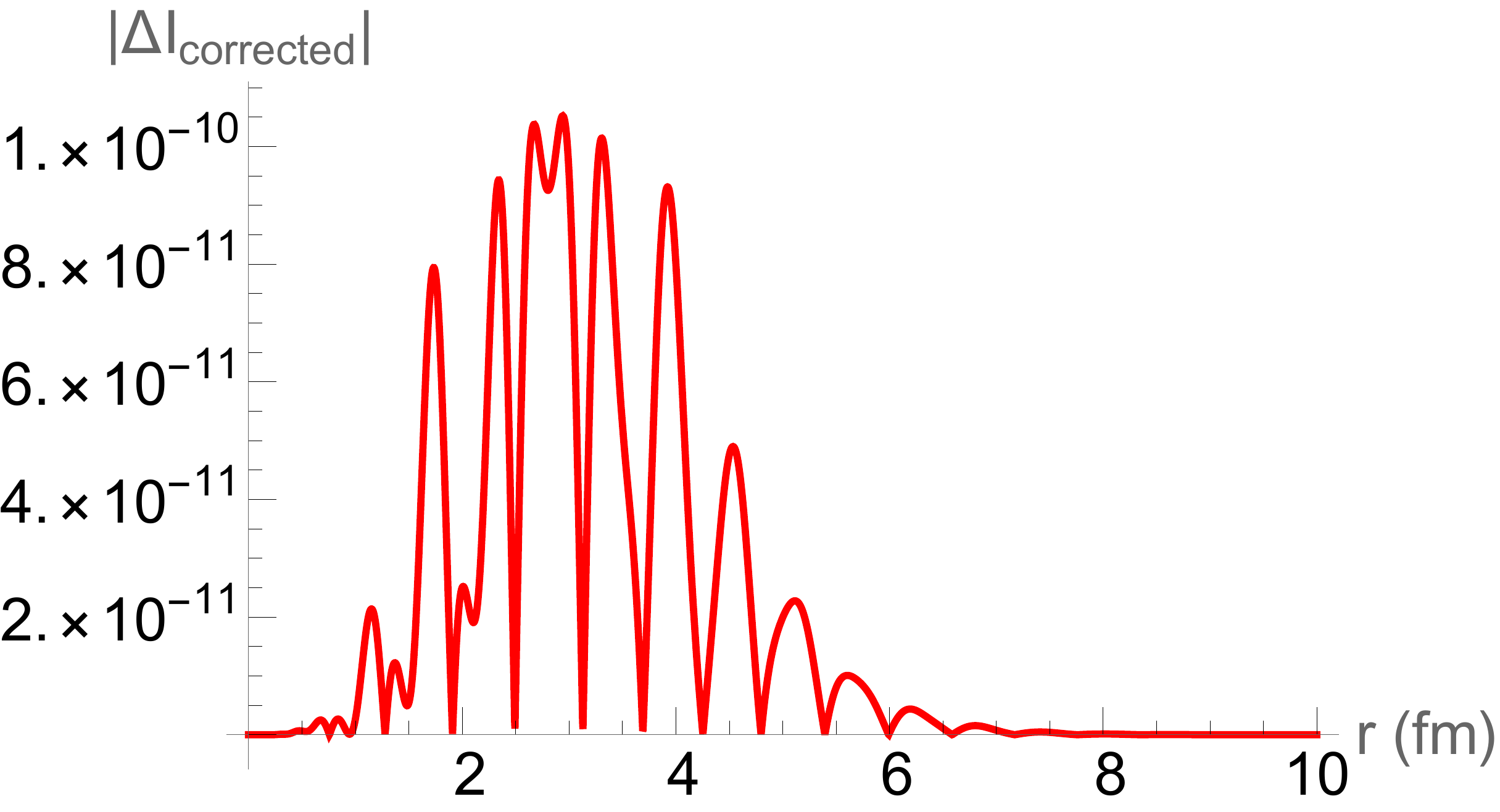}
\caption{}
\label{fig4b}
\end{subfigure}
 
\caption{(a)  Absolute difference between the integrand of the DBS mean square radii in the sinc and the HO basis.  The error is negligible in the vicinity of $r_{\textrm{threshold}}$ and much of it is concentrated in the 2-3 fm range, where the wave function does not exhibit exponential behavior.  (b)  Absolute difference between the integrands of the DBS mean square radii in the sinc and the HO bases after fitting the exponential tail.  The error beyond $r_{\textrm{threshold}}$ is eliminated, but at the cost of increasing the error elsewhere due to renormalization.}
\label{fig4}
\end{figure}

For the HO basis function, if we set $\delta=0$, fit an exponential tail to the DBS in the HO basis, and multiply by the normalization factor $K$, we obtain a new piecewise HO DBS whose semi-log plot is shown in Fig. \ref{fig4}.  We do a similar procedure with $\delta=0$ for the sinc DBS.

For the piecewise HO DBS, the corrected mean square radius yields $5.70184513221$ $\textrm{fm}^{2}$, which is identical to that of the HO DBS without the attached tail up to 11 significant figures ($5.701845132386(3)$ $\textrm{fm}^{2}$).  For the sinc DBS, the corrected mean square radius is practically the same as that in Table \ref{tab2} ($5.701845132385(1)$ $\textrm{fm}^{2}$), with the sinc DBS and the sinc piecewise DBS square radii being identical to 22 significant figures. As a result, the error increased from $2\cdot10^{-13}$ $\textrm{fm}^{2}$ (as quoted in Table \ref{tab2}) to $2.0\cdot10^{-10}$ $\textrm{fm}^{2}$. In order to know why the error \footnote{Throughout Sections \ref{section4.1} and \ref{section4.2}, we will refer to "inter-basis error" as "error".} increased despite the tail being fixed, it is instructive to look at Fig. \ref{fig4a} and Fig. \ref{fig4b}.

On the one hand, the exponential tails completely eliminate the errors beyond $r_{\textrm{threshold}}$.  However, when we plot the absolute difference between the sinc and HO DBS square radius integrands (which we denote by $\Delta I_{\textrm{w/o   correction}}=r^{2}|u_{\textrm{DBS}}^{\textrm{sinc}}(r)|^{2}-r^{2}|u_{\textrm{DBS}}^{\textrm{HO}}(r)|^{2}$), Fig. \ref{fig4a} shows us that the error in the vicinity of $r_{\textrm{threshold}}$ is virtually minuscule compared to that in the vicinity of $r=3$ fm.  Moreover, the wave function is sizable at the region where no exponential tail is attached.  Therefore, correcting the error where the integrand is virtually zero gives no numerical improvement, as seen in Fig. \ref{fig4b} (here, we plot the corrected difference between the integrands which we denote by $\Delta I_{\textrm{corrected}}=r^{2}|u_{\textrm{pw DBS}}^{\textrm{sinc}}(r)|^{2}-r^{2}|u_{\textrm{pw DBS}}^{\textrm{HO}}(r)|^{2}$).  In attaching the DBS tail, we must multiply the overall wave function by a normalization factor to keep it normalized to unity.  The value of this factor is virtually unity (up to 24 and 12 decimal places for the sinc and HO DBS, respectively). In addition, we need to multiply the wave function tail by another factor in order to keep the wave function continuous at $r_{\textrm{threshold}}$.  The error introduced by this normalization is significant enough to outweigh any error eliminated beyond $r_{\textrm{threshold}}$.  As a result, the net error increases.  Thus, as one might expect, we find that attaching an exponential tail to the DBS (and presumably other well-converged, strongly bound states) can potentially prove counterproductive.  This is because the tail of the DBS is sufficiently suppressed and (in this case) well-approximated by the basis expansion that the impact on normalization from introducing a piecewise tail outweighs any improvement provided by that tail. We now turn our attention to the same implementation of a piecewise tail but for the WBS, where we expect more significant improvements to appear.

\subsection{The WBS Tail Correction}\label{section4.2}

\begin{figure}[h]
 
\begin{subfigure}{0.5\textwidth}
\includegraphics[width=1\linewidth, height=5cm]{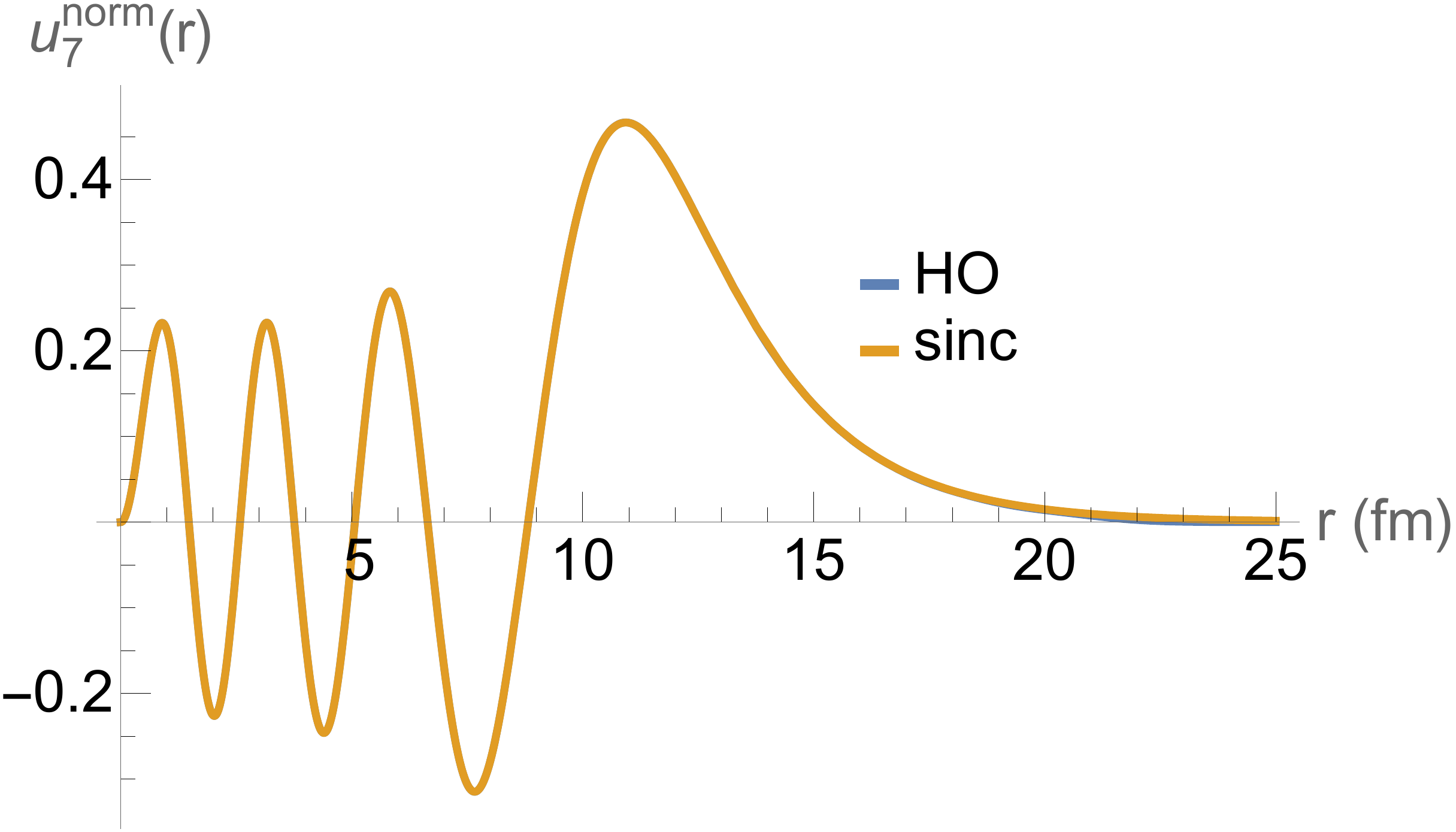} 
\caption{}
\label{fig5a}
\end{subfigure}
\begin{subfigure}{0.5\textwidth}
\includegraphics[width=1\linewidth, height=5cm]{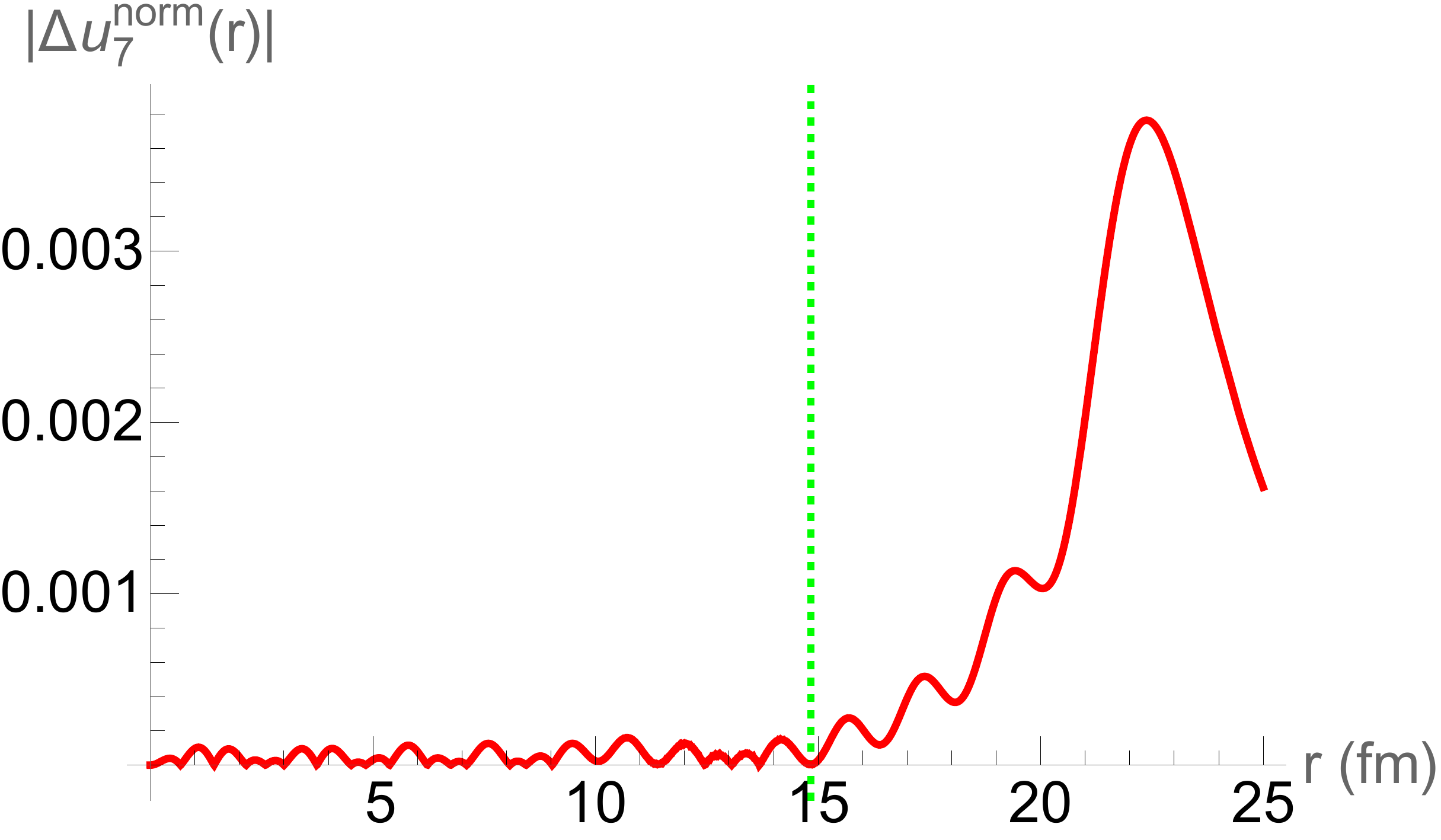}
\caption{}
\label{fig5b}
\end{subfigure}
\begin{subfigure}{0.5\textwidth}
\includegraphics[width=1\linewidth, height=5cm]{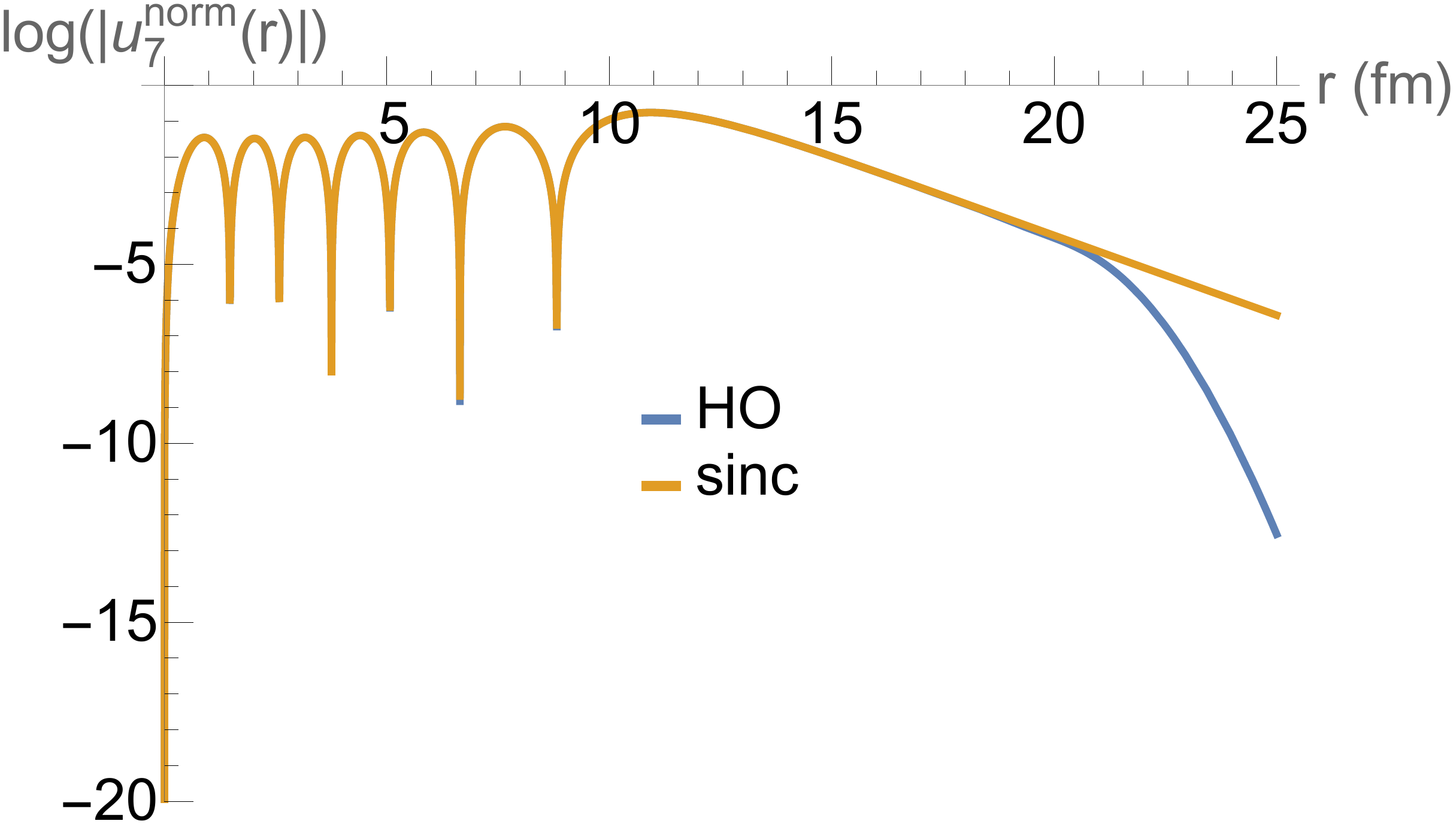}
\caption{}
\label{fig5c}
\end{subfigure}
\begin{subfigure}{0.5\textwidth}
\includegraphics[width=1\linewidth, height=5cm]{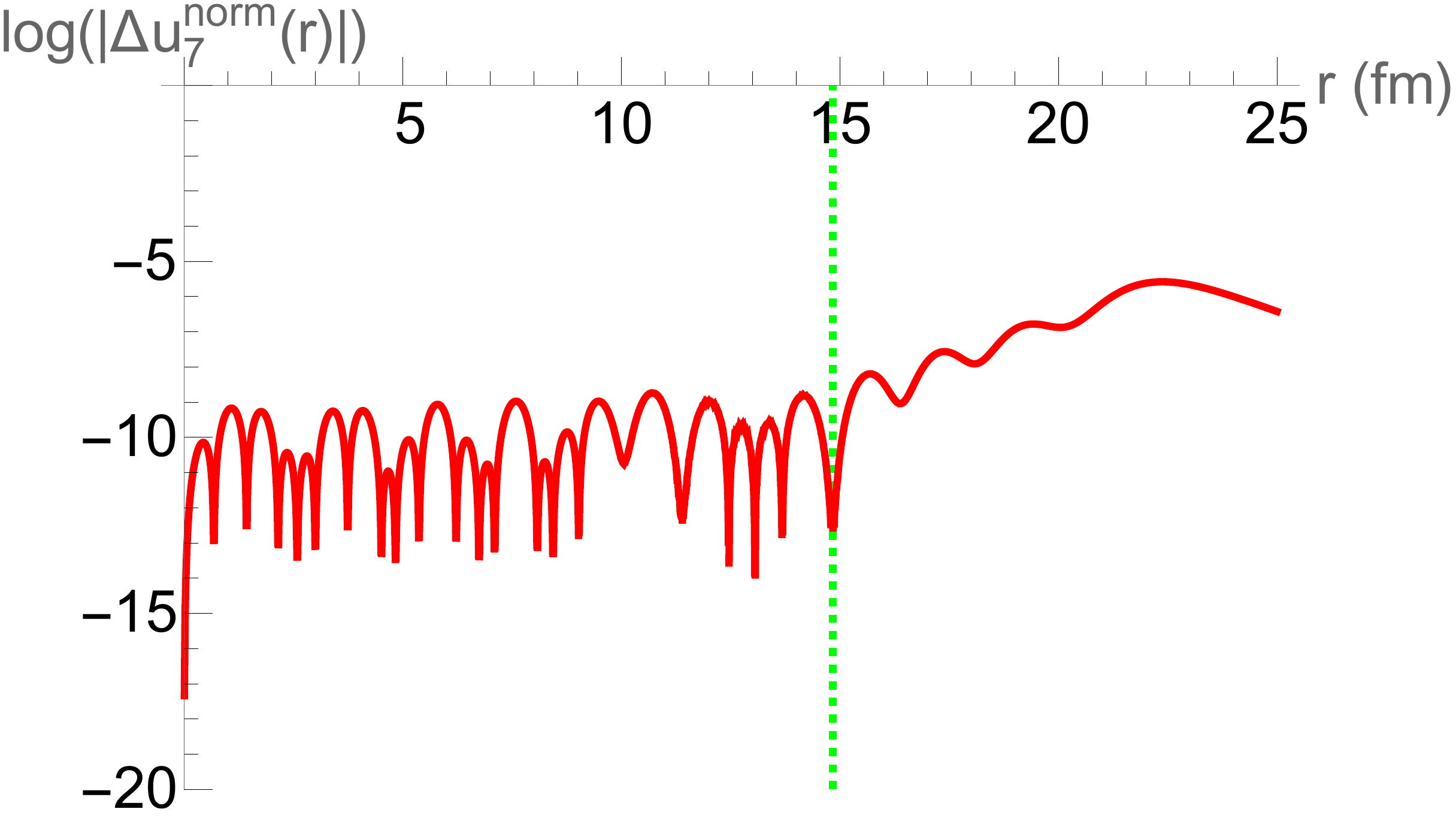}
\caption{}
\label{fig5d}
\end{subfigure}
 
\caption{(a)  The WBS in functional form as expanded in both the sinc and the HO basis (which are practically on top of each other in this scale).  The wave function appears to behave as expected in terms of asymptotic behavior and the number of axis crossings.  (b)  Absolute difference between the WBS approximations in the two bases.  The vertical line indicates the threshold from which the error abruptly increases by one to two orders of magnitude.  Note that the value is close to the length corresponding to $\lambda_{\textrm{WBS}}$.  (c)  Semi-log plot of the absolute sinc basis WBS and the HO basis WBS.  Note the quadratic behavior of the HO basis as $r$ goes to infinity in contrast to the linear behavior of the sinc basis out to at least 25 fm.  (d)  Semi-log plot of Fig. \ref{fig5b}.}
\label{fig5}
\end{figure}

Our main motivation for fixing the WBS is to address the large error in Tables \ref{tab1} and \ref{tab2}.  Fig. \ref{fig5} shows this error visually.  As in Fig. \ref{fig1} and \ref{fig2}, Fig. \ref{fig5} compares the sinc and HO basis WBS.  In Fig. \ref{fig5a}, the two approximations are practically indistinguishable.  The difference becomes apparent beyond $r_{\textrm{threshold}}$ as seen in Fig. \ref{fig5b}.  The semi-log plots in Fig. \ref{fig5c} and Fig. \ref{fig5d} reflect those properties.  This illustrates the HO basis's vulnerability to IR effects in the tail region of nuclear WBS.  

From this, we observe that Ineq. (\ref{eq10}) is to be understood as a rough guide.  As in the breakdown scales discussed in \cite{source23}, there is no guarantee that convergent solutions will be obtained merely satisfying Ineq. (\ref{eq10}).  In other words, while our particular parameter selection seems to satisfy this condition, it does not indicate that we can naively translate the IR parameter of our basis to a condition on the accuracy of our wave functions in coordinate space.  From our eigenvalue results in Table \ref{tab1} we can see how the Ineq. (\ref{eq10}) translates into quantified precision in the eigenvalues.  In Fig. \ref{fig5}, we observe that the wave function of the WBS appears accurate out to $r=r_{\textrm{threshold}}$, as noted.  This is a practical result and should serve as a cautionary note for the interpretation of our Ineq. (\ref{eq10}).  That is, to achieve accuracy in long-range observables for a WBS, comparable to the accuracy for a DBS, puts additional burden on the accuracy of the tail of the WBS wave function, well beyond the scale indicated by $\textrm{min}({\lambda_{\textrm{WBS}},\lambda_{\textrm{potential}},\lambda_{\textrm{DBS}}})$.   

As in the DBS, our first task is to find an appropriate threshold value $r_{\textrm{threshold}}$ where we can pick a point to fit an exponential function and possibly relate this point to one of the physical cutoffs in Ineq. (\ref{eq10}).  Fig. \ref{fig5a} shows the sinc and HO WBS.  The behavior reflects what we expect: six crossings of zero, algebraic behavior near $r=0$, and exponential behavior for large $r$.

From Fig. \ref{fig5b} and Fig. \ref{fig5d}, it is clear that we can identify the threshold point $r_{\textrm{threshold}}=14.8$ fm to be the point in which the error is a minimum at the vicinity of $r=15$ fm.  Note that this is close to $\frac{2\pi}{\lambda_{\textrm{WBS}}}=14.2$ fm.  Whether this proximity is coincidental requires further investigation, for example, by studying the WBS in other channels which is beyond the scope of the present work.  Looking at Fig. \ref{fig5c}, it is important to note that unlike in the DBS case, neither the sinc nor the HO WBS suffer from the severe oscillations in the tail region (at least not until further out in $r$, where the wave function practically vanishes).  However, the log of the absolute sinc WBS behaves linearly at large $r$ whereas its HO counterpart initially behaves linearly, then behaves quadratically after a certain breakdown point.

It is apparent that the main issue with the HO basis WBS is its long-range rather than its short-range behavior.  To get a more quantitative indicator of the error, we can integrate the HO and sinc basis WBS to obtain the mean square radius.  In the sinc basis, $\langle r^{2}\rangle =106.5271867(2)$ $\textrm{fm}^{2}$ compared to $106.44(25)$ $\textrm{fm}^{2}$ in the HO basis.  This yields an error of $8.7\cdot10^{-2}$ $\textrm{fm}^{2}$ which is large compared to that of the other bound state mean square radii as seen in Table \ref{tab2}.

As in the DBS, we will try to address the issue by fitting the tail of the HO WBS.  Unlike in the previous section however, we will try and fit only the HO WBS tail as the sinc wave function itself is reasonably exponential in the vicinity of $r_{\textrm{threshold}}$.  We outline the details in fitting the WBS and verifying that the sinc basis WBS tail is in fact exponential in \ref{appendix:e}.  The results there justify the exponential behavior of the sinc WBS tail (until the basis itself breaks down).  We therefore define the corrected HO WBS as a piecewise function: from the intervals 0 to $r_{\textrm{threshold}}+\delta$, the wave function is built as the standard HO wave function and from $r_{\textrm{threshold}}+\delta$ to infinity, we append the sinc wave function tail.  To ensure continuity and differentiability, we multiply the sinc WBS by a correction factor $K_{1}$ that is equal to the ratio of the value of the HO WBS at $r=r_{\textrm{threshold}}+\delta$ to that of the sinc WBS at the same value.  For $\delta=0$, that ratio is 0.99998.  To ensure that the wave function normalizes to 1, we multiply by a factor $K_{2}$ that is equal to 1 over the square root of the integral of the new wave function squared.  For $\delta=0$, this is equal to 0.99986.  Our new wave function is therefore

\begin{equation}
\label{eq41}
 u_{\textrm{pw}}(r)=\ K_{2}\begin{cases} 
      u_{\textrm{HO}}(r) & r<r_{\textrm{threshold}}+\delta \\
     K_{1}u_{\textrm{sinc}}(r) & r>r_{\textrm{threshold}} +\delta,
   \end{cases} 
\end{equation}
where $K_{1}$ is adjusted for continuity and differentiability of the wave function at $r=r_{\textrm{threshold}} +\delta$, $K_{2}$ is adjusted for the normalization of the piecewise wave function, and $u_{\textrm{HO(sinc)}}(r)$ is the normalized HO (sinc) WBS.  Using this new wave function to compute the WBS mean square radius, we obtain $\langle r^{2} \rangle=106.51703$ $\textrm{fm}^{2}$ and thus, the error compared to the sinc value is $1.0\cdot10^{-2}$ $\textrm{fm}^{2}$, which is nearly an order of magnitude smaller than the error quoted in Table \ref{tab2}.  

\begin{figure}[h]
 
\begin{subfigure}{0.5\textwidth}
\includegraphics[width=1\linewidth, height=5cm]{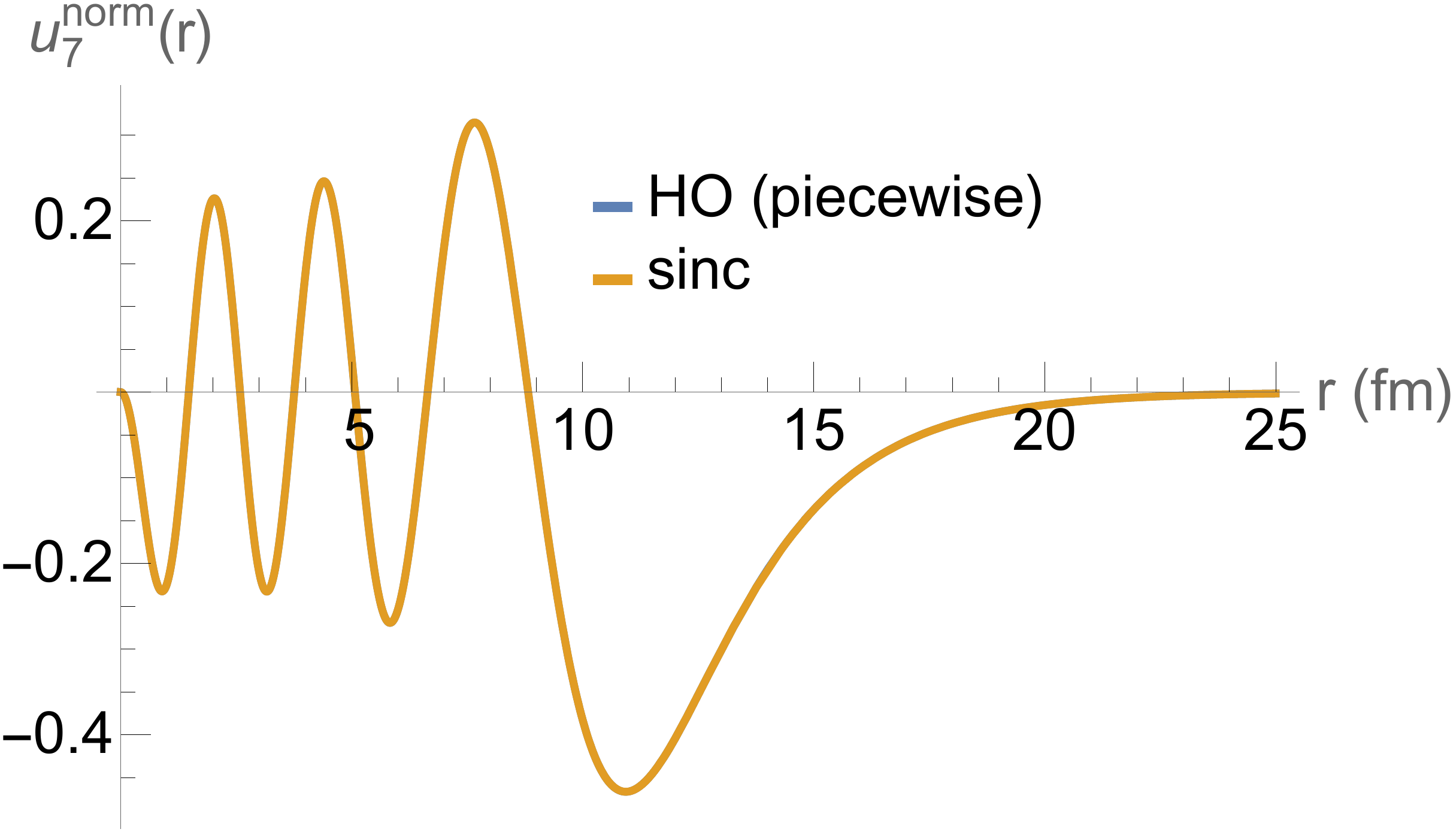} 
\caption{}
\label{fig6a}
\end{subfigure}
\begin{subfigure}{0.5\textwidth}
\includegraphics[width=1\linewidth, height=5cm]{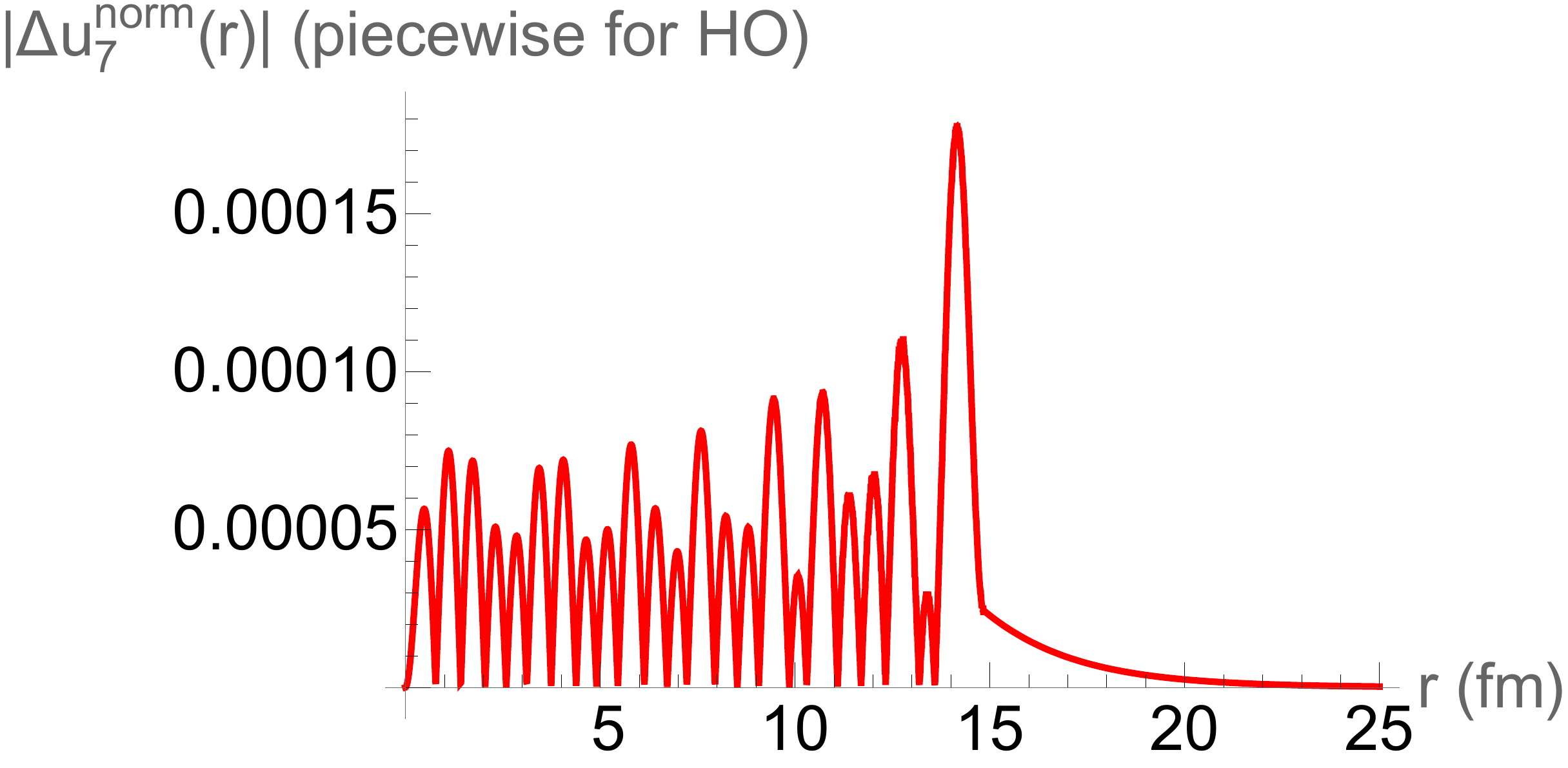}
\caption{}
\label{fig6b}
\end{subfigure}
\begin{subfigure}{0.5\textwidth}
\includegraphics[width=1\linewidth, height=5cm]{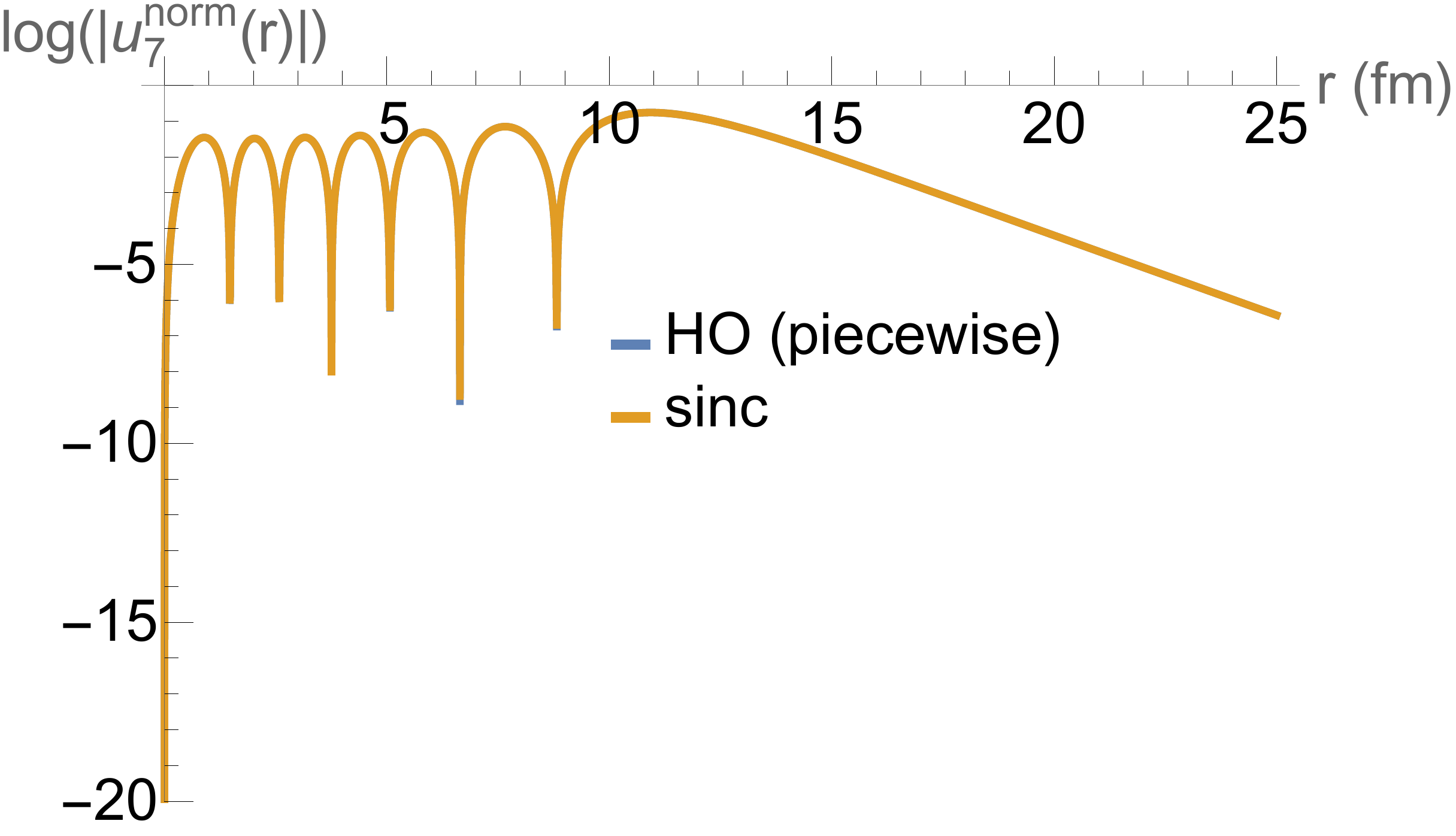}
\caption{}
\label{fig6c}
\end{subfigure}
\begin{subfigure}{0.5\textwidth}
\includegraphics[width=1\linewidth, height=5cm]{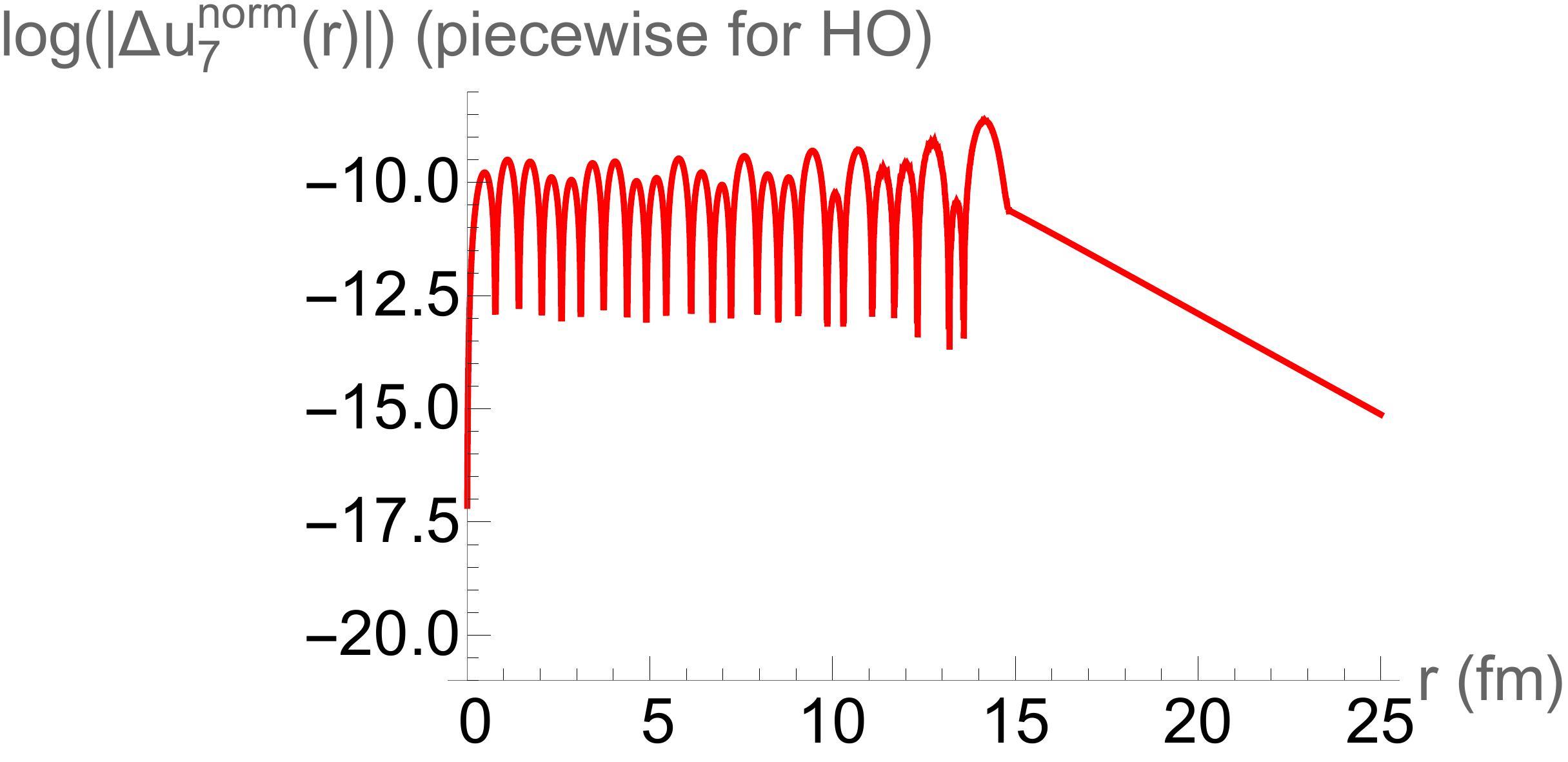}
\caption{}
\label{fig6d}
\end{subfigure}
 
\caption{  (a)  Corrected piecewise HO WBS according to Eq. \eqref{eq41} plotted along with the sinc WBS.  Compare with Fig. \ref{fig5a}.  (b)  Absolute difference between the wave functions plotted in Fig. \ref{fig6a}.  Compare with Fig. \ref{fig5b} and note the scale of the y-axis.  (c)  Semi-log plot of the absolute sinc WBS and the absolute piecewise HO WBS.  Compare with Fig. \ref{fig5c}.  (d)  Semi-log plot of Fig. \ref{fig6b}.  Compare with Fig. \ref{fig5d}.}
\label{fig6}
\end{figure}

Fig. \ref{fig6} compares the sinc basis WBS with that of the HO WBS with the corrected tail indicated in Eq. \eqref{eq41}.  The difference between the wave function is not apparent in Fig. \ref{fig6a}, but it is clear that in Fig. \ref{fig6b}, the error drastically decreases for larger $r$ relative to the error in Fig. \ref{fig5b}, especially for $r>r_{\textrm{threshold}}$.  This is also reflected in the semi-log plots in Fig. \ref{fig6c} and \ref{fig6d}.  Note the scales of the y-axes of Fig. \ref{fig6b} and \ref{fig6d}  as compared to those of Fig. \ref{fig5b} and Fig. \ref{fig5d}, respectively.

\begin{figure}[H]
 
\begin{subfigure}{0.5\textwidth}
\includegraphics[width=1\linewidth, height=5cm]{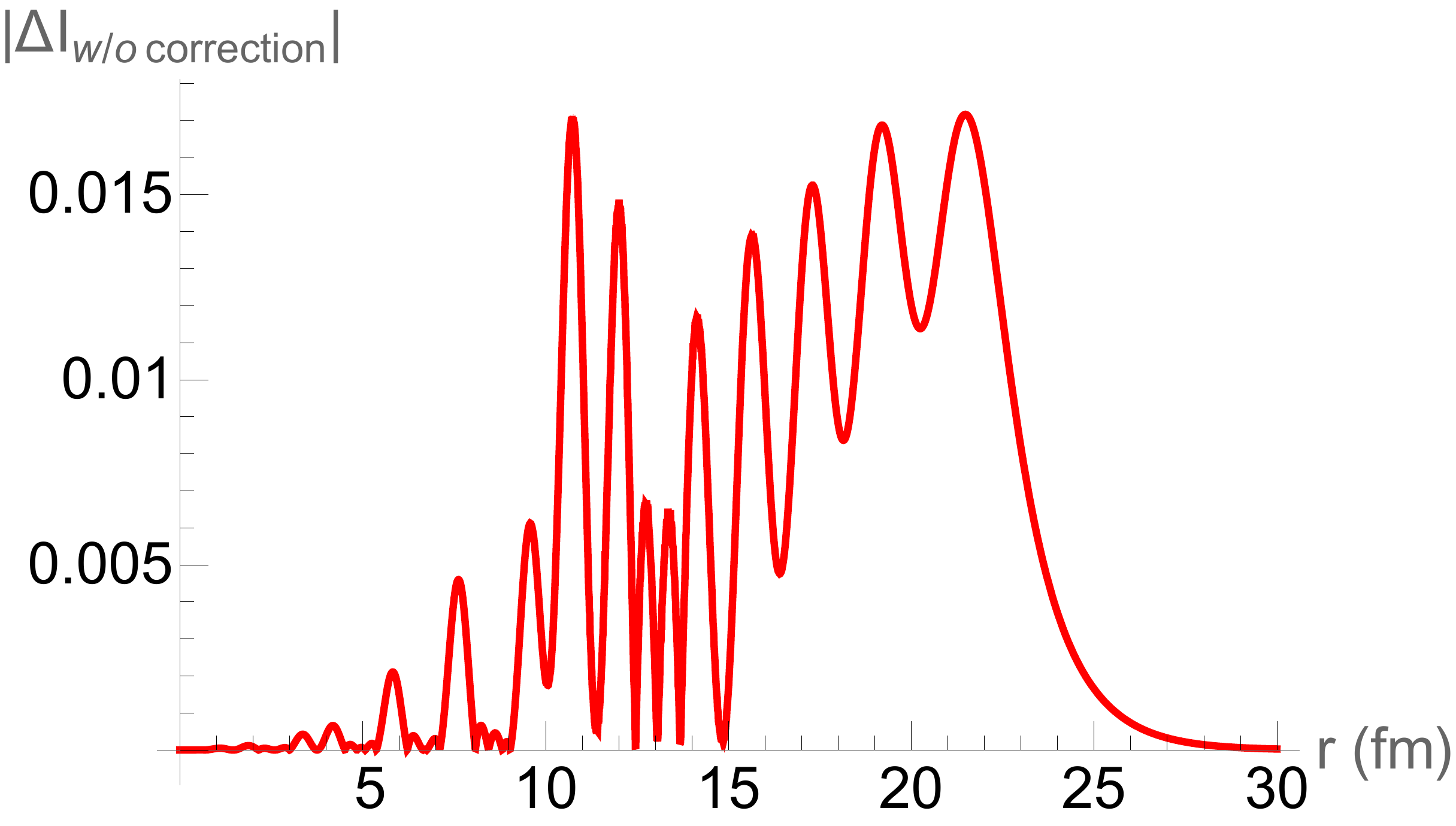}
\caption{}
\label{fig7a}
\end{subfigure}
\begin{subfigure}{0.5\textwidth}
\includegraphics[width=1\linewidth, height=5cm]{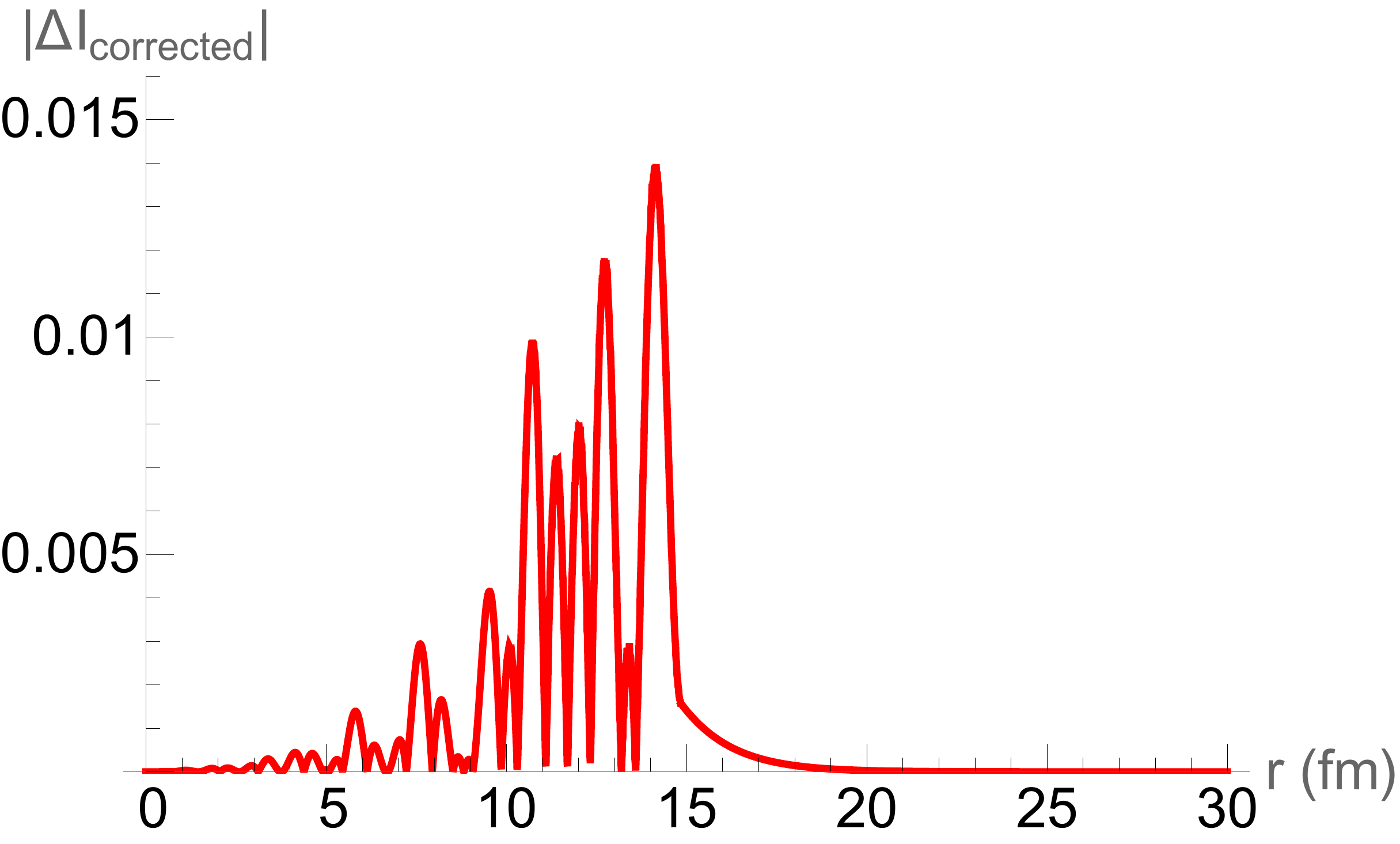}
\caption{}
\label{fig7b}
\end{subfigure}
\caption{  (a)  Absolute difference between the integrand of the WBS mean square radii in the sinc and the HO basis.  In contrast to Fig. \ref{fig4a}, the bulk of the error is concentrated in the interval beyond $r_{\textrm{threshold}}$ and hence, the sinc WBS fit to the tail of the HO WBS is more suitable for reducing the error in mean square radius.  (b)  Absolute difference between the integrands of the WBS mean square radii of the piecewise HO WBS and the sinc WBS.}
\label{fig7}
\end{figure}

As in the DBS, it is instructive to look at the differences in the integrand for the WBS shown in Fig. \ref{fig7}.  Here, $\Delta I_{\textrm{w/o correction}}=r^{2}|u_{\textrm{WBS}}^{\textrm{sinc}}(r)|^{2}-r^{2}|u_{\textrm{WBS}}^{\textrm{HO}}(r)|^{2}$ in  Fig. \ref{fig7a} whereas $\Delta I_{\textrm{corrected}}=r^{2}|u_{\textrm{WBS}}^{\textrm{sinc}}(r)|^{2}-r^{2}|u_{\textrm{pw}}(r)|^{2}$ in Fig. \ref{fig7b}.  Unlike what we have seen in Fig. \ref{fig4a} and Fig. \ref{fig4b}, the error is weighed towards the region near $r_{\textrm{threshold}}$.  Moreover, the WBS tail is larger at long distances compared with the DBS.  Consequently, while the inclusion of the piecewise tail in the DBS appeared counterproductive after normalization, we find the benefit of the piecewise tail to the WBS outweighs the cost of normalization in this case. 

Thus, we have observed that DBS can become less accurate by the introduction of a piecewise exponential tail, while WBS can become more accurate. This matches our expectation, as a wave function's tail becomes more significant with weaker binding. We therefore surmise that the use of a piecewise tail may be very beneficial when considering weakly bound nuclei, where the conventional HO basis faces the demanding challenge of producing an approximate exponential falloff.

\section{Conclusion}\label{conclusion}

The results in Tables \ref{tab1}, \ref{tab2}, and in \ref{appendix:c}, \ref{appendix:d}, and \ref{appendix:e} suggest that the main advantage of the sinc basis is higher precision at relatively small $m_{\textrm{Val}}$ for more weakly-bound states in which long-range features of the wave function provide an emergent IR scale ($\lambda_{\textrm{WBS}}$) significantly below the intrinsic scale of the potential.  At $m_{\textrm{Val}}=200$, the WBS eigenvalue is precise to 12 significant figures.  The ability to choose the asymptotic behavior of the wave function through choice of conformal map up to a breakdown scale beyond which no important physics takes place allows us to achieve this precision.

 While the sinc basis is advantageous in accurately solving the radial Schrödinger equation (and more generally, the Sturm-Liouville problem), it is not without its disadvantages.  First, the magnitude of the difference among matrix elements and eigenvalues of $D^{-2}A$ increases rapidly as $m_{\textrm{Val}}$ increases.  This leads eventually to a badly-conditioned matrix and the potential for numerical obstacles (see \ref{appendix:a}).  Second, a non-orthonormal set of discrete eigenvectors can complicate the process of transforming operators into effective operators through renormalization.  Third, transforming between position and momentum space generally must be done numerically.  This is often an important step in more complex problems since most realistic potentials are often given in momentum space.  Because position and momentum are not treated on an equal footing as they are in the HO basis, transformation between position and momentum space may introduce additional numerical error.  Fourth, for applications to quantum many-body systems, there may be additional complications in transitioning between single-particle and relative coordinates which can be done in closed form in the HO basis.  

At its current stage, the present cutoff scheme introduced in the theory is incomplete.  In the HO basis, $\lambda_{\textrm{HO}}$ is smaller than $\textrm{min}({\lambda_{\textrm{WBS}},\lambda_{\textrm{potential}},\lambda_{\textrm{DBS}}})$ (see Ineq. (\ref{eq10})).  This seems to be at odds with the low precision obtained by the HO WBS at $N_{\textrm{max}}=60$ as seen in the HO WBS's energy and mean square radius.  This suggests that care must be exercised in adopting basis spaces that nominally satisfy the Ineq. (\ref{eq10}).  Our computational results suggest that $\lambda_{\textrm{basis}}$ may need to be much smaller than $\textrm{min}({\lambda_{\textrm{WBS}},\lambda_{\textrm{potential}},\lambda_{\textrm{DBS}}})$ (for an exploration of other breakdown scales and how the HO breakdown scale relates to convergence, see \ref{appendix:d} and \cite{source23}, respectively).  In addition, the breakdown scale approximations in the sinc basis have limited utility (see Fig. \ref{figD8b} in \ref{appendix:d}).   

In light of this, the methods introduced in this work could be further refined.  For instance, one could devise a more sophisticated approach to quantifying the sinc UV and IR characteristic scales that is not based on a first approximation of the conformal map and is less dependent on fitting high-precision energy values (which might not be easily accessible in more complex problems).  An improved cutoff scheme must also account for different bound state approximations that break down at different points.  Moreover, one could look for additional breakdown scales that have dependence on both the problem and/or the basis space used (i.e. improving Ineq. (\ref{eq10}) to insure that all relevant emergent scales are encompassed by the basis selected).  In addition, one could further improve the method of tail attachment by choosing a more accurate $\delta$ defined in the piecewise function based on knowledge of characteristic scales.  One could also introduce a more sophisticated function for the region leading into the asymptotic tail region such as those used in the WKB approach \cite{source15}.

With regards to physics applications, future work may focus on the evaluation and application of discrete representations of scattering states.  Other potential topics for future study include more complex potentials arising from strong interactions and on other observables such as quadrupole transitions.  The challenges of  transforming from relative coordinates (or momenta) to single-particle coordinates need to be investigated for applications to many-body systems.  High-performance parallel computing will play a significant role in these calculations since the calculation of two-body matrix elements, for example, should be highly parallelizable.

\appendix
\label{appendix}


\section{Sizes of Matrix Elements and Eigenvalues in the Sinc and HO Bases} \label{appendix:a}

For the sinc and HO bases, it is worth examining the scope of the matrix elements for different basis truncations ($N_{\textrm{max}}$ or $m_{\textrm{Val}}$) for the $l=1$ channel.  This gives us a feel for the numerical tractability of the matrices we are diagonalizing.  Tables \ref{tabA3} and \ref{tabA4} give the spread of the diagonal matrix elements (the difference between the largest and smallest diagonal elements) and the largest off-diagonal matrix element (in magnitude) for the HO and sinc basis, respectively.  

For both the HO and sinc bases, both the spread and the largest off-diagonal matrix element increase with truncation.  However, they do so more slowly for the HO basis whereas they rapidly increases by orders of magnitude for the sinc basis.  For the sinc basis, this rapid divergence of matrix elements could lead to an upper limit of $m_{\textrm{Val}}$ for which accurate solutions can be obtained due to the increasingly ill-conditioned nature of the matrix with growing $m_{\textrm{Val}}$. 

\begin{table}[H]
 \renewcommand{\arraystretch}{2}
 \caption{Scope of the HO basis matrix elements and eigenvalues (in absolute value) for the $l=1$ channel.  The spread of the diagonal matrix elements is the difference between the largest and smallest diagonal elements.  In the HO basis, both the spread and the largest off-diagonal matrix element increase with truncation.   }
 
\vspace{3mm}

     \centering
    \begin{tabular} {|c|c|c|}
    \hline
    Spread & Largest Off-Diagonal Matrix Element &$N_{\textrm{max}}$ \\ \hline
     $2.15\cdot10^{2}$ &$3.77\cdot10^{1}$ &10 \\ \hline
     $3.74\cdot10^{2}$& $3.77\cdot10^{1}$&20 \\ \hline
   $4.99\cdot10^{2}$&$6.84\cdot10^{1}$ &30  \\ \hline
    $8.29\cdot10^{2}$& $2.36\cdot10^{2}$ &60 \\ \hline

    \end{tabular}
    \label{tabA3}
    \end{table}

\begin{table}[H]
 \renewcommand{\arraystretch}{2}
 \caption{Scope of the sinc basis matrix elements and eigenvalues (in absolute value) for the $l=1$ channel.  As in Table \ref{tabA3}, both the spread and the largest off-diagonal matrix element increase with truncation. However, this increase is much more rapid in the sinc basis, which could lead to numerical intractability for large $m_{\textrm{Val}}$.  }
 
\vspace{3mm}

   \centering
    \begin{tabular} {|c|c|c|}
    \hline
    Spread & Largest Off-Diagonal Matrix Element &$m_{\textrm{Val}}$ \\ \hline
     $4.66\cdot10^{14}$ &$2.23\cdot10^{14}$ &25 \\ \hline
     $3.73\cdot10^{20}$&$2.00\cdot10^{20}$ &50 \\ \hline
   $1.17\cdot10^{25}$&$6.51\cdot10^{24}$ &75  \\ \hline
    $1.38\cdot10^{35}$&$8.00\cdot10^{34}$ &150 \\ \hline

    \end{tabular}
    \label{tabA4}
    \end{table}

\section{Details of Numerical Methods}\label{appendix:b}
For all calculations, we use Wolfram Mathematica\textsuperscript{\textregistered} version 10.3.1.0 on a Mac OS X x86\textsuperscript{\textregistered} (32-bit, 64-bit Kernel).  We use the built-in Mathematica\textsuperscript{\textregistered} functions, Eigensystem, Eigenvalues, and Eigenvectors to diagonalize the matrices of the HO and sinc collocation  bases.  For cross-checking eigenvalue and eigenvector results, we use the C language version 1.31.1 (using the gcc 6.4.0 compiler).  The eigensolver was from the gsl package.  In this work, we present no result from the C program.  For the fits, we use NonlinearModelFit.

In computing the eigenvalues of the sinc basis, we multiply the factor $\frac{2m}{\hbar^{2}}$ on both sides of Eq. (\ref{eq4}).  After calculation, we divide the eigenvalues by $\frac{2m}{\hbar^{2}}$ to re-scale the eigenvalues.  We perform the calculation to 35-digit precision, using SetPrecision throughout and checking precision by the Mathematica\textsuperscript{\textregistered} function Precision.  For the sinc results in Table \ref{tab1}, we quote the non-hermitian $D^{-2}A$ bound state eigenvalues.  For both the sinc and HO values of the mean square radii, we use NIntegrate with integrand $r^{2}|u_{\alpha}^{\textrm{norm}}(r)|^{2}$ to compute the mean square radii quoted in Table \ref{tab2}.  However, we use Integrate to compute the matrix elements of the potential matrix component $V$ of the HO-basis Hamiltonian $H$.  This is due to numerical issues in using NIntegrate.  For NIntegrate, we use a 15-digit working precision.  The intervals of integration for the sinc and HO bases mean square radii are from zero to infinity with the exception of the HO WBS, where the interval is from 0 to 7000 fm due to numerical precision issues.  For calculations shown here, $\gamma=1$ $\textrm{fm}^{-1}$ and $\hbar \Omega=20$ MeV.

\section{Trends in Eigenvalue and Mean Square Radius Convergence with Increasing $m_{\textrm{Val}}$}\label{appendix:c}

It is instructive to investigate convergence of the bound state eigenvalues and mean square radii in both the sinc and HO bases with respect to their truncation parameters.  Throughout the main text of this paper we use $m_{\textrm{Val}}=200$ and $N_{\textrm{max}}=60$ for the eigenvalues and wave functions in the sinc and HO basis, respectively.  We use $m_{\textrm{Val}}=150$ and $N_{\textrm{max}}=50$ only to define an internal error.  Here, we use different $N_{\textrm{max}}$ and $m_{\textrm{Val}}$ to observe convergence trends in the two bases.  The tables below show the values of the eigenvalues and mean square radii for different HO and sinc truncations in intervals of 25 and 4, respectively. Red denotes digits that disagree with those of the subsequent increases in the truncation parameter (except for the last entry at the highest value of the truncation parameter, where red denotes disagreement with the previous truncation quoted).  

As expected, precision increases with increasing basis truncation.  In the HO basis however, convergence in the WBS occurs slowly.  In general, more weakly-bound states converge more slowly than the more deeply-bound states.  For the sinc basis, note the near-uniformity in precision of the eigenvalues (with the exception of the WBS).  Although convergence is arrested at some point (unless we further increase precision), we obtain results with precision well-beyond the digits displayed here.  Similar trends occur in the mean square radii.  We quote fewer figures for the mean square radii because we calculate the integral using machine precision (which is 12 digits in the Mathematica\textsuperscript{\textregistered} version we use).  This is due to numerical issues for integrals with small $m_{\textrm{Val}}$.  Because there are numerical issues with integrating to infinity for small $m_{\textrm{Val}}$, we truncate the integration interval in Tables \ref{tabC7} and \ref{tabC8} to be from 0 to 490 fm.

\begin{table}[H]
 \renewcommand{\arraystretch}{2}
   \caption{ Convergence of bound state eigenvalues (in MeV) for different values of $N_{\textrm{max}}$.  Red indicates changing digits with respect to the next increment of $N_{\textrm{max}}$ (with the exception of the last column $N_{\textrm{max}}=58$, where red denotes disagreement with the previous increment).  Note the increasing precision with increasing $N_{\textrm{max}}$.  The $N_{\textrm{max}} = 60$ results have been presented in Table \ref{tab1}. }
   
   \vspace{3mm}
   
 \centering
    \begin{tabular} {|c|c|c|c|}
    
    \hline
    $N_{\textrm{max}}=14$ & $N_{\textrm{max}}=18$ & $N_{\textrm{max}}=22$ & $N_{\textrm{max}}=26$ \\ \hline
    -304.46\textcolor{red}{180034953630449}&-304.4628\textcolor{red}{0042505374853}&-304.46283\textcolor{red}{704690714718} & -304.462838\textcolor{red}{45820012126}\\ \hline
     -235.4\textcolor{red}{1063228245754779}&-235.44\textcolor{red}{809989097428283}&-235.4\textcolor{red}{4995013609140492} &-235.4500\textcolor{red}{3794712042736} \\ \hline
   -17\textcolor{red}{2.81460477214573011}&-173.2\textcolor{red}{1428948792143886} &-173.24\textcolor{red}{252645149418605} &-173.244\textcolor{red}{21881674776662} \\ \hline
     -11\textcolor{red}{6.42423600634452243}&-118\textcolor{red}{.18194563710865897} &-118.3\textcolor{red}{6843131236470376} &-118.38\textcolor{red}{293334573129006} \\ \hline
  \textcolor{red}{-67.308239150793282425}&-7\textcolor{red}{0.949374681196123512} &-71\textcolor{red}{.555120534232507812} &-71.6\textcolor{red}{18031280214355913}\\ \hline
     \textcolor{red}{-23.997933794262901965}&-3\textcolor{red}{2.396986069649200081} &-3\textcolor{red}{3.907140302618468931} &-34.1\textcolor{red}{07265370271591936}  \\ \hline
      - & \textcolor{red}{-4.3996482018309850446} &\textcolor{red}{-6.9757914705624342779} &\textcolor{red}{-7.6562503972891170260}\\ \hline
     $N_{\textrm{max}}=30$ & $N_{\textrm{max}}=34$  & $N_{\textrm{max}}=38$& $N_{\textrm{max}}=42$ \\ \hline
     -304.46283851\textcolor{red}{608349585}&-304.462838518\textcolor{red}{61524289}&-304.46283851873\textcolor{red}{315629} &-304.46283851873\textcolor{red}{898740} \\ \hline
     -235.450042\textcolor{red}{15866757726}&-235.4500423\textcolor{red}{6707695796}&-235.45004237\textcolor{red}{781197158} & -235.450042378\textcolor{red}{38951404}\\ \hline
      -173.2443\textcolor{red}{1475445900111}&-173.244320\textcolor{red}{15047982249}&-173.2443204\textcolor{red}{5840335407} &-173.24432047\textcolor{red}{642968508}\\ \hline
  -118.3839\textcolor{red}{1414096431504} &-118.3839\textcolor{red}{7698230382463}&-118.38398\textcolor{red}{095283743236} &-118.3839812\textcolor{red}{0531609992} \\ \hline
    -71.623\textcolor{red}{150220687023334}&-71.6235\textcolor{red}{23443298726937}&-71.6235\textcolor{red}{49431506946331} &-71.623551\textcolor{red}{215152176841}\\ \hline
     -34.12\textcolor{red}{7810519539795499}&-34.129\textcolor{red}{728511193162741}&-34.1299\textcolor{red}{12046504850535} &-34.12993\textcolor{red}{1820077297275}  \\ \hline
  \textcolor{red}{-7.9017348605359610791}&-8.0\textcolor{red}{039386178403708573}&-8.0\textcolor{red}{478118740945418950} &-8.0\textcolor{red}{669997721125902561} \\ \hline
       $N_{\textrm{max}}=46$ & $N_{\textrm{max}}=50$  & $N_{\textrm{max}}=54$ & $N_{\textrm{max}}=58$ \\ \hline  
       -304.462838518739\textcolor{red}{29259}&-304.4628385187393\textcolor{red}{0944}&-304.4628385187393104\textcolor{red}{2}&-304.4628385187393104\textcolor{red}{8} \\ \hline
    -235.45004237842\textcolor{red}{199361}&-235.45004237842\textcolor{red}{390195}&-235.4500423784240\textcolor{red}{1896} & -235.4500423784240\textcolor{red}{2643}\\ \hline
      -173.2443204775\textcolor{red}{1830848}&-173.2443204775\textcolor{red}{8630576}&-173.244320477590\textcolor{red}{70306} &-173.244320477590\textcolor{red}{99750} \\ \hline
    -118.38398122\textcolor{red}{165201949}&-118.383981222\textcolor{red}{73406681}&-118.3839812228\textcolor{red}{0768464} & -118.3839812228\textcolor{red}{1283810}\\ \hline
      -71.6235513\textcolor{red}{37872276239}&-71.62355134\textcolor{red}{6417215961}&-71.6235513470\textcolor{red}{22514503} &-71.6235513470\textcolor{red}{66250742} \\ \hline
       -34.129934\textcolor{red}{388292505247}&-34.129934\textcolor{red}{798580311377}&-34.1299348\textcolor{red}{76922277098} &-34.1299348\textcolor{red}{93995332465}  \\ \hline
      -8.0\textcolor{red}{756017733650966431}&-8.0\textcolor{red}{795686621945507431}&-8.08\textcolor{red}{14509081943926320} &-8.08\textcolor{red}{23684971203325319}\\ \hline
      
    \end{tabular}
\label{tabC5}
    \end{table}

 \begin{table}[H]
  \renewcommand{\arraystretch}{2}
   \caption{ Convergence of bound state eigenvalues (in MeV) for different values of $m_{\textrm{Val}}$.  Like in Table \ref{tabC5}, red indicates changing digits with respect to the next increment of $m_{\textrm{Val}}$ (with the exception of the last column $m_{\textrm{Val}}=300$, where red denotes disagreement with the previous increment).  Except for the WBS state, note the relative uniformity in eigenvalue precision.  The $m_{\textrm{Val}}=200$ results have been presented in Table \ref{tab1}.  }
   
   \vspace{3mm}
   
 \centering
    \begin{tabular} {|c|c|c|c|}
    
    \hline
    $m_{\textrm{Val}}=25$ & $m_{\textrm{Val}}=50$  & $m_{\textrm{Val}}=75$ & $m_{\textrm{Val}}=100$ \\ \hline
      -304.46\textcolor{red}{305726097202037}& -304.462838\textcolor{red}{84196667050}& -304.4628385\textcolor{red}{2079062518} &-304.4628385187\textcolor{red}{6761578} \\ \hline
      -235.450\textcolor{red}{92550539014729}&-235.45004\textcolor{red}{398509874567} &-235.4500423\textcolor{red}{8946716641}& -235.450042378\textcolor{red}{58465053}\\ \hline
   -173.24\textcolor{red}{598651327430706} & -173.24432\textcolor{red}{466024134392}& -173.244320\textcolor{red}{50879900417}& -173.24432047\textcolor{red}{806503731}\\ \hline
   -118.38\textcolor{red}{551942639803607}&-118.38398\textcolor{red}{866493311958} &-118.3839812\textcolor{red}{8406868778}& -118.38398122\textcolor{red}{378558364}\\ \hline
      -71.623\textcolor{red}{173405263499732} &-71.6235\textcolor{red}{61166555897802} &-71.623551\textcolor{red}{438234727767} &-71.62355134\textcolor{red}{8585763719}\\ \hline
       -34.12\textcolor{red}{6456643318970715}&-34.1299\textcolor{red}{44588711233056}& -34.12993\textcolor{red}{5003226797000}&-34.129934\textcolor{red}{901335373935}  \\ \hline
     \textcolor{red}{-7.9517854489912306088} &-8.08\textcolor{red}{28924208134428662} &-8.08332\textcolor{red}{44886140727857} & -8.083329\textcolor{red}{8392870321925} \\ \hline
     $m_{\textrm{Val}}=125$ & $m_{\textrm{Val}}=150$  & $m_{\textrm{Val}}=175$ & $m_{\textrm{Val}}=200$ \\ \hline
     -304.462838518739\textcolor{red}{95469}& -304.4628385187393\textcolor{red}{3142}&-304.46283851873931\textcolor{red}{137} &-304.462838518739310\textcolor{red}{53} \\ \hline
      -235.45004237842\textcolor{red}{784153}&-235.450042378424\textcolor{red}{15577} &-235.4500423784240\textcolor{red}{3266} & -235.450042378424027\textcolor{red}{29} \\ \hline
      -173.244320477\textcolor{red}{60261468}&-173.244320477591\textcolor{red}{42046} &-173.2443204775910\textcolor{red}{3791} &-173.2443204775910\textcolor{red}{2088}\\ \hline
    -118.3839812228\textcolor{red}{3767615}& -118.38398122281\textcolor{red}{410421}&-118.3839812228132\textcolor{red}{8311} & -118.38398122281324\textcolor{red}{602}\\ \hline
      -71.623551347\textcolor{red}{108864504}&-71.6235513470\textcolor{red}{71166468} &-71.623551347069\textcolor{red}{830226} &-71.62355134706976\textcolor{red}{9074}\\ \hline
      -34.1299348995\textcolor{red}{76375899}&-34.1299348995\textcolor{red}{30368170} &-34.129934899528\textcolor{red}{711879} &-34.12993489952863\textcolor{red}{5239}  \\ \hline
      -8.08332997\textcolor{red}{03031290389}&-8.083329975\textcolor{red}{2609370328} &-8.0833299755\textcolor{red}{167161754} & -8.08332997553\textcolor{red}{34166791}\\ \hline
       $m_{\textrm{Val}}=225$ & $m_{\textrm{Val}}=250$  & $m_{\textrm{Val}}=275$ & $m_{\textrm{Val}}=300$ \\ \hline  
        -304.46283851873931048& -304.46283851873931048&-304.46283851873931048 &-304.46283851873931048 \\ \hline
    -235.45004237842402\textcolor{red}{700}&-235.45004237842402698 &-235.45004237842402698 & -235.45004237842402698\\ \hline
      -173.244320477591019\textcolor{red}{94}& -173.24432047759101988&-173.24432047759101988 &-173.24432047759101988 \\ \hline
    -118.383981222813243\textcolor{red}{96}& -118.3839812228132438\textcolor{red}{2}& -118.38398122281324381& -118.38398122281324381\\ \hline
      -71.623551347069765\textcolor{red}{638}&-71.623551347069765\textcolor{red}{409} &-71.62355134706976539\textcolor{red}{1} &-71.62355134706976539\textcolor{red}{0} \\ \hline
       -34.129934899528630\textcolor{red}{896}& -34.129934899528630\textcolor{red}{604}&-34.12993489952863058\textcolor{red}{2} &-34.12993489952863058\textcolor{red}{0}  \\ \hline
      -8.083329975534\textcolor{red}{7312370}&-8.0833299755348\textcolor{red}{518296} &-8.08332997553486\textcolor{red}{44080} &-8.08332997553486\textcolor{red}{58722} \\ \hline
    \end{tabular}
\label{tabC6}
    \end{table}
    
    \begin{table}[H]
     \renewcommand{\arraystretch}{2}
 
   \caption{Convergence of the mean square radii (in $\textrm{fm}^{2}$) for different values of $N_{\textrm{max}}$.  Like in the eigenvalues tables, red indicates changing digits with respect to the next increment of $N_{\textrm{max}}$ (with the exception of $N_{\textrm{max}}=58$, where red denotes disagreement with the previous increment).  The trends here are similar to those of Table \ref{tabC5}.  The interval of integration used here is from 0 to 490 fm.  We quote fewer digits due to our using machine precision in the integration.  The $N_{\textrm{max}}=60$ results have been presented in Table \ref{tab2}. }
   
   \vspace{3mm}
   
 \centering
    \begin{tabular} {|c|c|c|c|}
    
    \hline
    $N_{\textrm{max}}=14$ & $N_{\textrm{max}}=18$ & $N_{\textrm{max}}=22$ & $N_{\textrm{max}}=26$ \\ \hline
     5.70\textcolor{red}{279694670} &5.7018\textcolor{red}{8522494}&5.70184\textcolor{red}{685013} &5.701845\textcolor{red}{20871} \\ \hline
    11.3\textcolor{red}{450112171} &11.30\textcolor{red}{85553906}&11.306\textcolor{red}{4200922} &11.306\textcolor{red}{3058711} \\ \hline
  18\textcolor{red}{.7003196259} &18.2\textcolor{red}{886310131}&18.2\textcolor{red}{523605832} &18.249\textcolor{red}{8596774} \\ \hline
   2\textcolor{red}{9.4808964537}&27\textcolor{red}{.5740749038}&27\textcolor{red}{.3038621685} &27.27\textcolor{red}{88135883} \\ \hline
      4\textcolor{red}{6.7731150414}&4\textcolor{red}{1.2427717469} &\textcolor{red}{40.0914079307} &39.9\textcolor{red}{476185828}\\ \hline
     6\textcolor{red}{3.0183320553} &6\textcolor{red}{1.6241802985} & 60\textcolor{red}{.4283245189}&60.2\textcolor{red}{680963991} \\ \hline
      - &\textcolor{red}{83.6802145901} &9\textcolor{red}{1.2910430193} &\textcolor{red}{97.1765608074}\\ \hline
     $N_{\textrm{max}}=30$ & $N_{\textrm{max}}=34$  & $N_{\textrm{max}}=38$& $N_{\textrm{max}}=42$ \\ \hline
    5.70184513\textcolor{red}{579}&5.701845132\textcolor{red}{38}&5.7018451322\textcolor{red}{1} &5.70184513220 \\ \hline
     11.306299\textcolor{red}{9018}&11.3062995\textcolor{red}{833}&11.30629956\textcolor{red}{59} &11.3062995649 \\ \hline
      18.2496\textcolor{red}{970769}&18.249687\textcolor{red}{9783}&18.2496873\textcolor{red}{678}&18.24968733\textcolor{red}{35}\\ \hline
   27.276\textcolor{red}{9149332}&27.2767\textcolor{red}{826676}&27.276773\textcolor{red}{6931}&27.276773\textcolor{red}{1013} \\ \hline
      39.93\textcolor{red}{43641106}&39.933\textcolor{red}{3044089}&39.9332\textcolor{red}{247719}&39.933218\textcolor{red}{9169}\\ \hline
      60.27\textcolor{red}{13843266}&60.27\textcolor{red}{63566833}&60.27\textcolor{red}{77141295}&60.2780\textcolor{red}{202218}  \\ \hline
  10\textcolor{red}{1.103781272}&10\textcolor{red}{3.443755882}&10\textcolor{red}{4.780766288}&105\textcolor{red}{.534534360} \\ \hline
       $N_{\textrm{max}}=46$ & $N_{\textrm{max}}=50$  & $N_{\textrm{max}}=54$ & $N_{\textrm{max}}=58$ \\ \hline  
       5.70184513220&5.70184513220&5.70184513220&5.70184513220 \\ \hline
    11.3062995649&11.3062995649&11.3062995649 &11.3062995649 \\ \hline
     18.249687331\textcolor{red}{7}&18.249687331\textcolor{red}{3}&18.2496873313&18.2496873313\\ \hline
    27.27677305\textcolor{red}{89}&27.276773056\textcolor{red}{6}&27.2767730564&27.2767730564 \\ \hline
     39.9332184\textcolor{red}{858}&39.93321845\textcolor{red}{37}&39.933218451\textcolor{red}{8}&39.933218451\textcolor{red}{5} \\ \hline
       60.278\textcolor{red}{0886006}&60.27810\textcolor{red}{46196}&60.27810\textcolor{red}{86217}&60.27810\textcolor{red}{96061}  \\ \hline
     10\textcolor{red}{5.958838041}&106\textcolor{red}{.198761927}&106\textcolor{red}{.335482385}&106\textcolor{red}{.414083454}\\ \hline
    
    \end{tabular}
\label{tabC7}
    \end{table}

     \begin{table}[H]
     
      \renewcommand{\arraystretch}{2}
     
     \caption{ Convergence of the mean square radii (in $\textrm{fm}^{2}$) for different values of $m_{\textrm{Val}}$.  The trends are similar to those of Table \ref{tabC6}.  The interval of integration used here is from 0 to 490 fm due to convergence issues with NIntegrate for small $m_{\textrm{Val}}$.  Here, we have fewer columns because subsequent values are identical to 12 significant figures.  The $m_{\textrm{Val}}=200$ results have been presented in Table \ref{tab2}.}
     
       \vspace{3mm}
       
  \centering
    \begin{tabular} {|c|c|c|}
    
    \hline
       $m_{\textrm{Val}}=25$ & $m_{\textrm{Val}}=50$  & $m_{\textrm{Val}}=75$ \\ \hline
       5.7018\textcolor{red}{0278514}& 5.701845\textcolor{red}{05944}&5.70184513\textcolor{red}{171}  \\ \hline
       11.306\textcolor{red}{1211257}&11.306299\textcolor{red}{1088} & 11.30629956\textcolor{red}{13} \\ \hline
     18.249\textcolor{red}{4746499}&18.24968\textcolor{red}{58559} &18.2496873\textcolor{red}{193}  \\ \hline
 27.27\textcolor{red}{74133849}&27.2767\textcolor{red}{697970} &27.2767730\textcolor{red}{236}  \\ \hline
       39.93\textcolor{red}{69583574}& 39.93321\textcolor{red}{26179}& 39.933218\textcolor{red}{3817} \\ \hline
       60.2\textcolor{red}{856517514}& 60.27810\textcolor{red}{01831}& 60.2781\textcolor{red}{099373} \\ \hline
 10\textcolor{red}{2.454212362} &106\textcolor{red}{.466746180} & 106.52\textcolor{red}{5590747} \\ \hline
  $m_{\textrm{Val}}=100$ & $m_{\textrm{Val}}=125$ & $m_{\textrm{Val}}=150$ \\ \hline
      5.7018451321\textcolor{red}{8}&5.7018451321\textcolor{red}{7} & 5.701845132\textcolor{red}{19}\\ \hline
        11.30629956\textcolor{red}{62}& 11.30629956\textcolor{red}{49}&11.30629956\textcolor{red}{50}  \\ \hline
       18.249687331\textcolor{red}{1}& 18.249687331\textcolor{red}{3}&18.2496873314  \\ \hline
       27.276773056\textcolor{red}{0}&27.2767730564 &27.2767730564  \\ \hline
  39.9332184\textcolor{red}{495}&39.9332184507 &39.9332184507  \\ \hline
           60.2781100\textcolor{red}{873} &60.278110093\textcolor{red}{5} & 60.27811009\textcolor{red}{37} \\ \hline
  106.5271\textcolor{red}{22340}&106.52718\textcolor{red}{3058} & 106.527186\textcolor{red}{429}  \\ \hline
  $m_{\textrm{Val}}=175$ & $m_{\textrm{Val}}=200$ & $m_{\textrm{Val}}=225$ \\ \hline
  5.701845132\textcolor{red}{21}& 5.701845132\textcolor{red}{19}&5.701845132\textcolor{red}{20}  \\ \hline
     11.306299564\textcolor{red}{8}&11.3062995649 &11.3062995649  \\ \hline
  18.2496873314& 18.2496873314& 18.2496873314 \\ \hline
  27.2767730564&27.2767730564 &27.2767730564  \\ \hline
     39.9332184507&39.9332184507 &39.9332184507  \\ \hline
  60.27811009\textcolor{red}{40} & 60.2781100938& 60.2781100938  \\ \hline
  106.5271866\textcolor{red}{56}& 106.52718667\textcolor{red}{5} & 106.52718667\textcolor{red}{7}  \\ \hline
    \end{tabular}
   \label{tabC8}
    \end{table}
    
    \section{Trends in the Wave Function Structure with Increasing Basis Size}\label{appendix:d}

We complement the figures in the main text and Tables \ref{tabC5} to \ref{tabC8} with semi-log plots of the absolute HO and sinc bound-states at representative values of $N_{\textrm{max}}$ and $m_{\textrm{Val}}$ for the DBS and the WBS.  We also investigate the properties of different bound states for fixed $N_{\textrm{max}}$ and $m_{\textrm{Val}}$. 

\begin{figure}[H]
 
\begin{subfigure}{0.5\textwidth}
\includegraphics[width=1\linewidth, height=5cm]{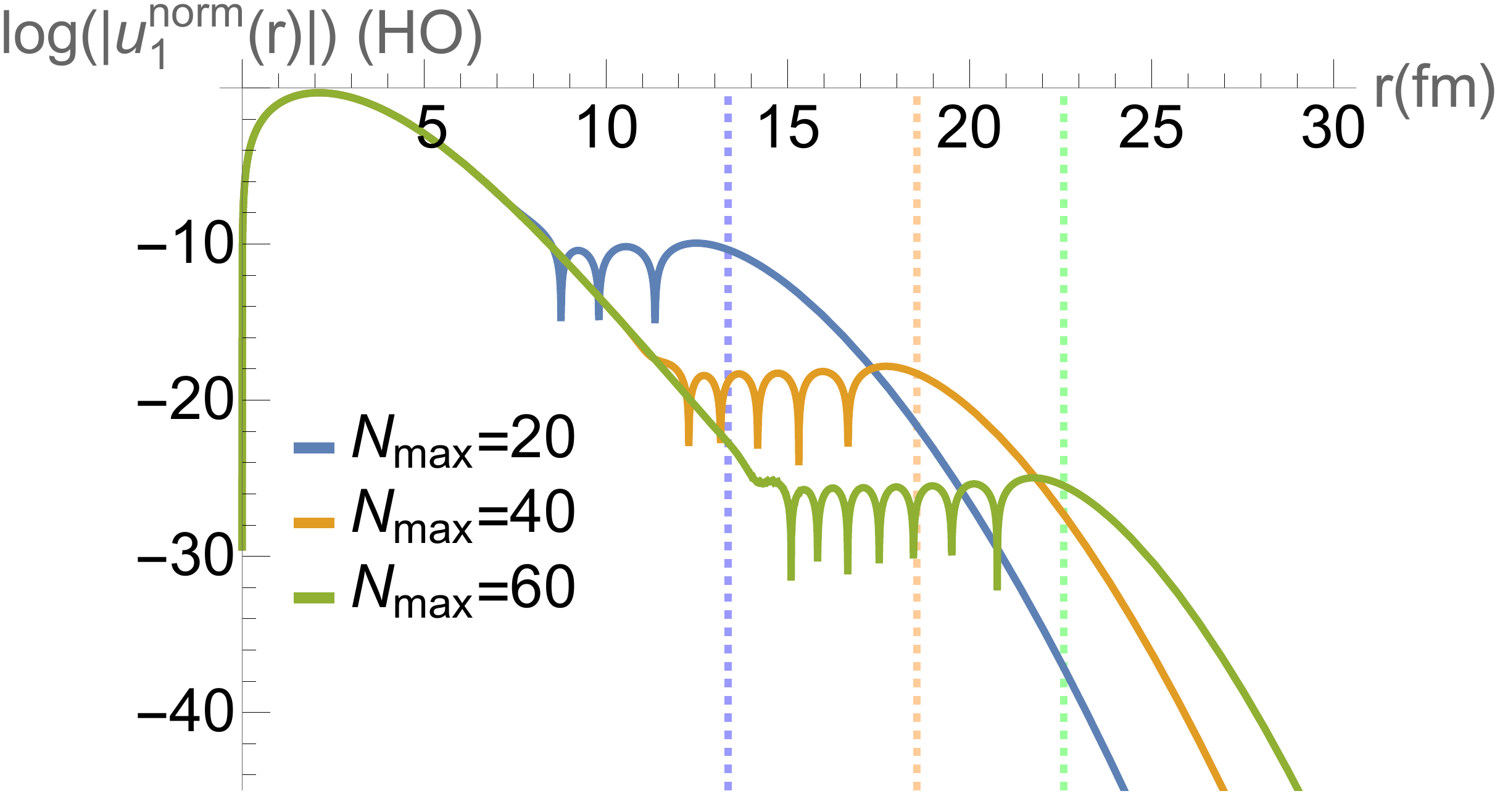} 
\caption{}
\label{figD8a}
\end{subfigure}
\begin{subfigure}{0.5\textwidth}
\includegraphics[width=1\linewidth, height=5cm]{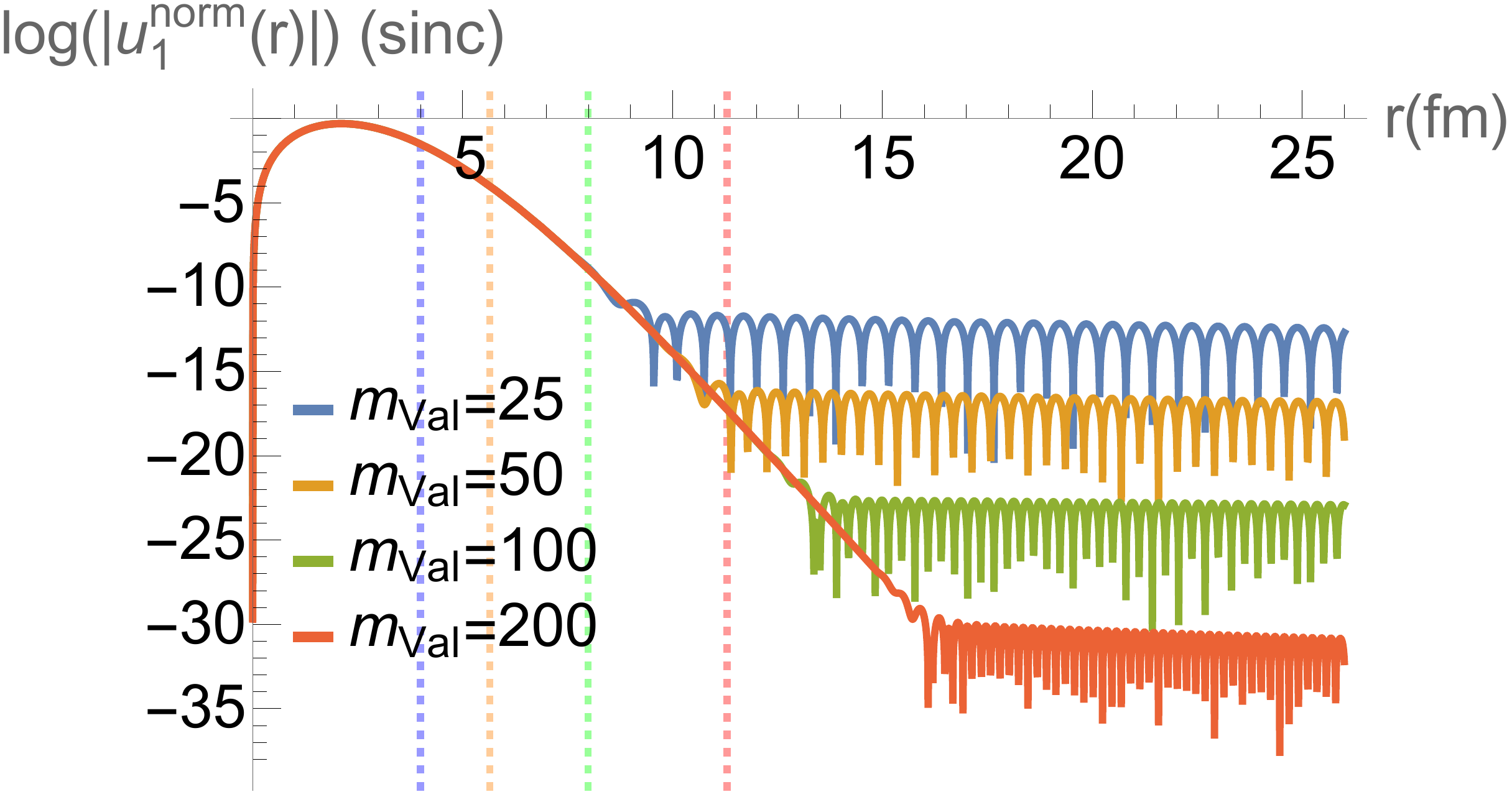}
\caption{}
\label{figD8b}
\end{subfigure}
 
\caption{(a)  Semi-log plot of the absolute HO DBS for representative choices of $N_{\textrm{max}}$.  Note that the breakdown and subsequent quadratic tail (on this semi-log plot) occurs at larger $r$ with increasing $N_{\textrm{max}}$.  Note also the increasing frequency of oscillation with increasing $N_{\textrm{max}}$.  The onset of the parabola extends further out with increasing $N_{\textrm{max}}$ and is represented by a vertical line corresponding to a given $N_{\textrm{max}}$.  They closely correspond to the inverse IR breakdown scale $\frac{1}{\lambda_{\textrm{HO}}}$.  (b)  Semi-log plot of the absolute sinc DBS for representative choices of $m_{\textrm{Val}}$ for the sinc DBS.  Unlike in the HO basis, there is no quadratic tail: the oscillations continue into large $r$.  Like in the HO basis, the asymptotic behavior (which is linear) extends further out with increasing $m_{\textrm{Val}}$ and the frequency of oscillation increases with increasing $m_{\textrm{Val}}$ whereas the oscillation amplitude decreases.  The vertical lines represent our coordinate space approximation of the sinc DBS IR breakdown scale given by $\frac{1}{\lambda_{\textrm{sinc}}}$.  }
\label{figD8}
\end{figure}

\begin{figure}[h]
 
\begin{subfigure}{0.5\textwidth}
\includegraphics[width=1\linewidth, height=5cm]{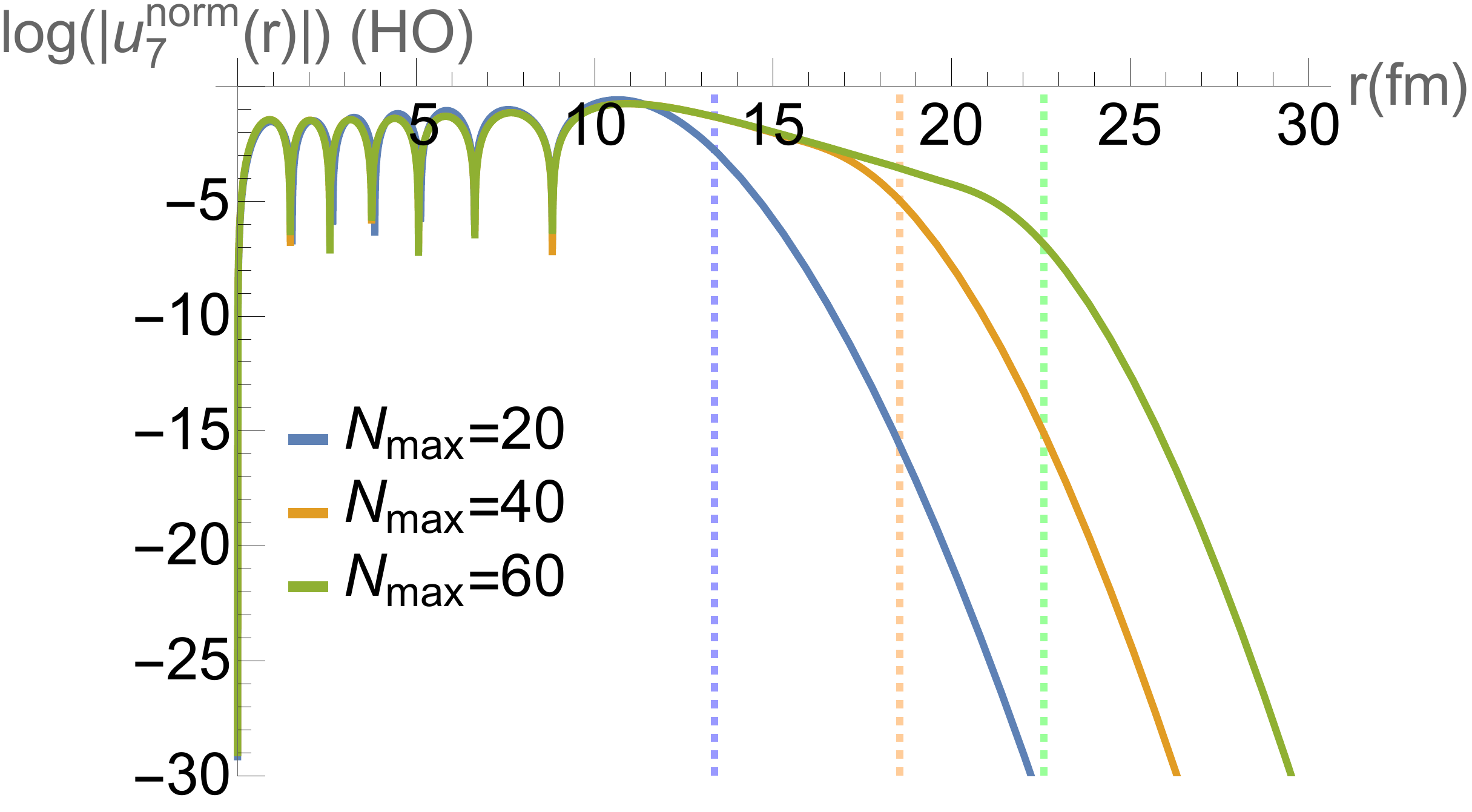} 
\caption{}
\label{figD9a}
\end{subfigure}
\begin{subfigure}{0.5\textwidth}
\includegraphics[width=1\linewidth, height=5cm]{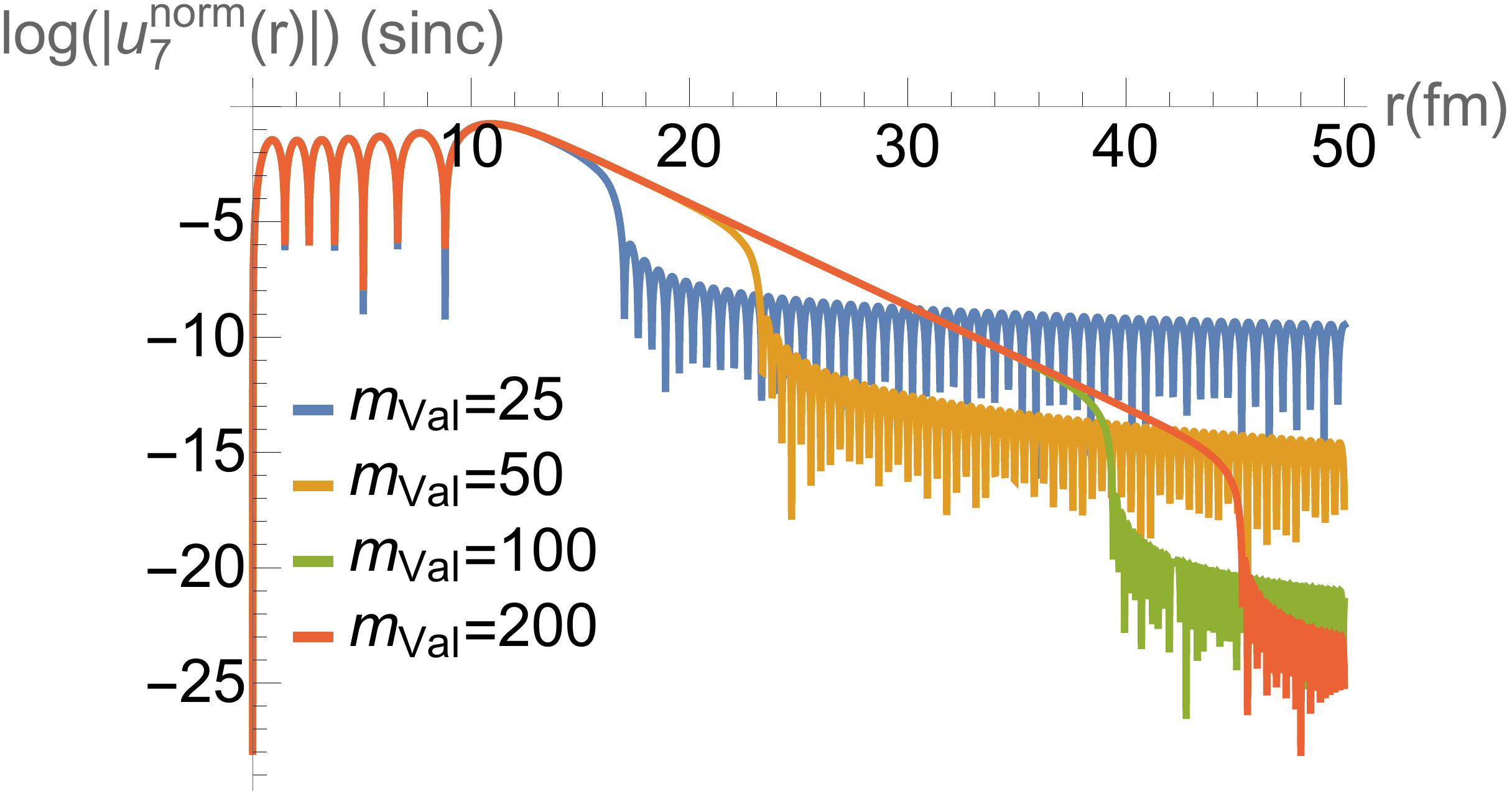}
\caption{}
\label{figD9b}
\end{subfigure}
 
\caption{  (a)  Semi-log plot of the absolute HO WBS for representative choices of $N_{\textrm{max}}$.  The main issue is the quadratic behavior occuring at large $r$.  Note that there are no oscillations and only one breakdown point $r_{0}$, namely the transition from a line to a parabola on the semi-log scale.  The onset of the parabola extends further out with increasing $N_{\textrm{max}}$ and is represented by a vertical line corresponding to a given $N_{\textrm{max}}$.  (b)  Semi-log plot of the absolute sinc WBS for representative choices of $m_{\textrm{Val}}$.  The trends of the oscillation frequency and the characteristic scales are similar to those of Fig. \ref{figD8b}, with the larger $m_{\textrm{Val}}$ providing an improved description of the WBS.  }
\label{figD9}
\end{figure}

In Fig. \ref{figD8a}, the log of the absolute HO DBS deviates from linearity at a certain point; it oscillates, then quadratically decays (reflecting the Gaussian behavior of the HO basis function tail).  Unsurprisingly, as seen in the succession of curves in Fig. \ref{figD9a}, this breakdown scale increases with increasing $N_{\textrm{max}}$.  It is notable that the number of nonphysical nodes occurring at the transition interval between the quadratic tail and the log of the absolute HO DBS before the first breakdown point increases with $N_{\textrm{max}}$ and the amplitude of oscillation decreases with $N_{\textrm{max}}$.  These non-physical nodes are from the remnant of the highest basis state at this $N_{\textrm{max}}$ appearing with non-zero amplitude in the wave function for this state. A good approximation of the bound state should therefore have those nonphysical oscillations entering only when the wave function magnitude has negligible impact on observables of interest.  For example, $N_{\textrm{max}}=20$ gives a significant error in approximating the DBS energy and mean square radius relative to the error at higher $N_{\textrm{max}}$ values since oscillatory behavior occurs where the wave function magnitude is significant.  By looking at Fig. \ref{figD8a}, we can identify two breakdown scales in addition to $\lambda_{\textrm{HO}}$: the scale beyond which nonphysical oscillations occur and the scale beyond which transition to a Gaussian (the form of the tail of the underlying basis) occurs.

A similar breakdown occurs in the sinc basis as seen in Fig. \ref{figD8b}.  Like in the HO DBS, the log of the absolute sinc DBS behaves smoothly until a certain breakdown point.  As expected, the tail behavior of this function is linear on the semi-log plot up to that point.  As in the HO case, this breakdown point increases with $m_{\textrm{Val}}$ along with the frequency of oscillation due to an increase in the number of collocation points.  Correspondingly, the oscillation amplitudes decrease with increasing $m_{\textrm{Val}}$.  Unlike in the HO basis, the oscillations continue indefinitely.  There is only one characteristic breakdown point: that in which rapid oscillations occur (like in the HO basis, this reflects the nature of the underlying basis).  As in the HO basis, a good approximation of the bound states should insure that the nonphysical oscillations (which are much more frequent in the sinc basis) occur only when the wave function magnitude has negligible impact on observables of interest.  The purpose of the vertical lines in Fig. \ref{figD8b} is to identify the breakdown scale with $\lambda_{\textrm{sinc}}$ beyond which the log of the absolute sinc DBS starts oscillating using Eq. (\ref{eq36}).  The vertical lines are obtained by taking $\frac{1}{\lambda_{\textrm{sinc}}}$.  If $\frac{1}{\lambda_{\textrm{sinc}}}$ gives an order of magnitude estimate of the transition point where exponential behavior gives way to oscillatory behavior, then our approximation of the sinc IR breakdown scale for the DBS is reasonable.  We see that our approximation of the IR breakdown scale is roughly half of the actual breakdown scale of the DBS.  This difference indicates that Eq. (\ref{eq36}) is only an order of magnitude estimate and whose value cannot be taken as a precise approximation of the breakdown scale.

\begin{figure}[H]
 
\begin{subfigure}{0.5\textwidth}
\includegraphics[width=1\linewidth, height=5cm]{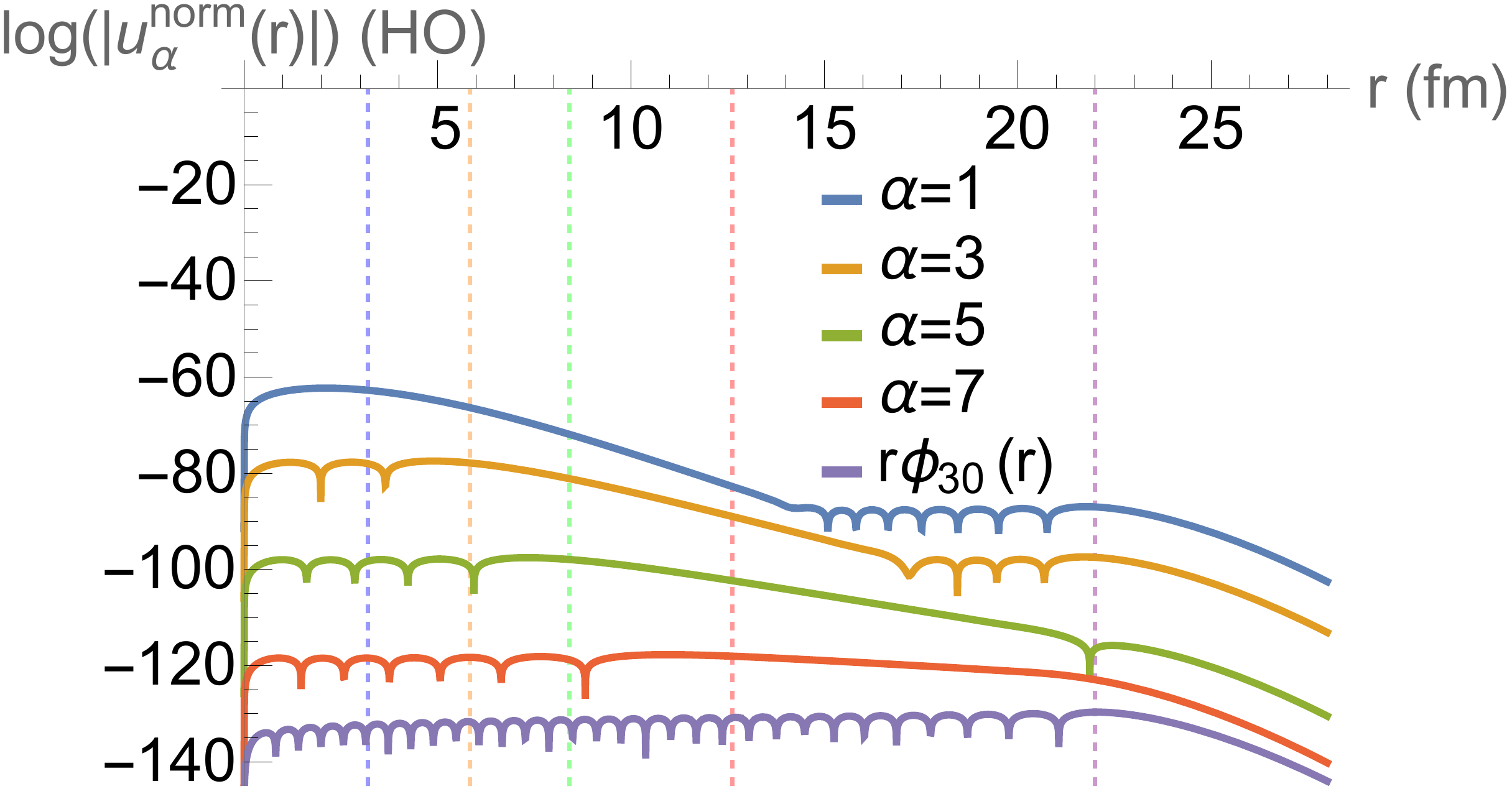} 
\caption{}
\label{figD10a}
\end{subfigure}
\begin{subfigure}{0.5\textwidth}
\includegraphics[width=1\linewidth, height=5cm]{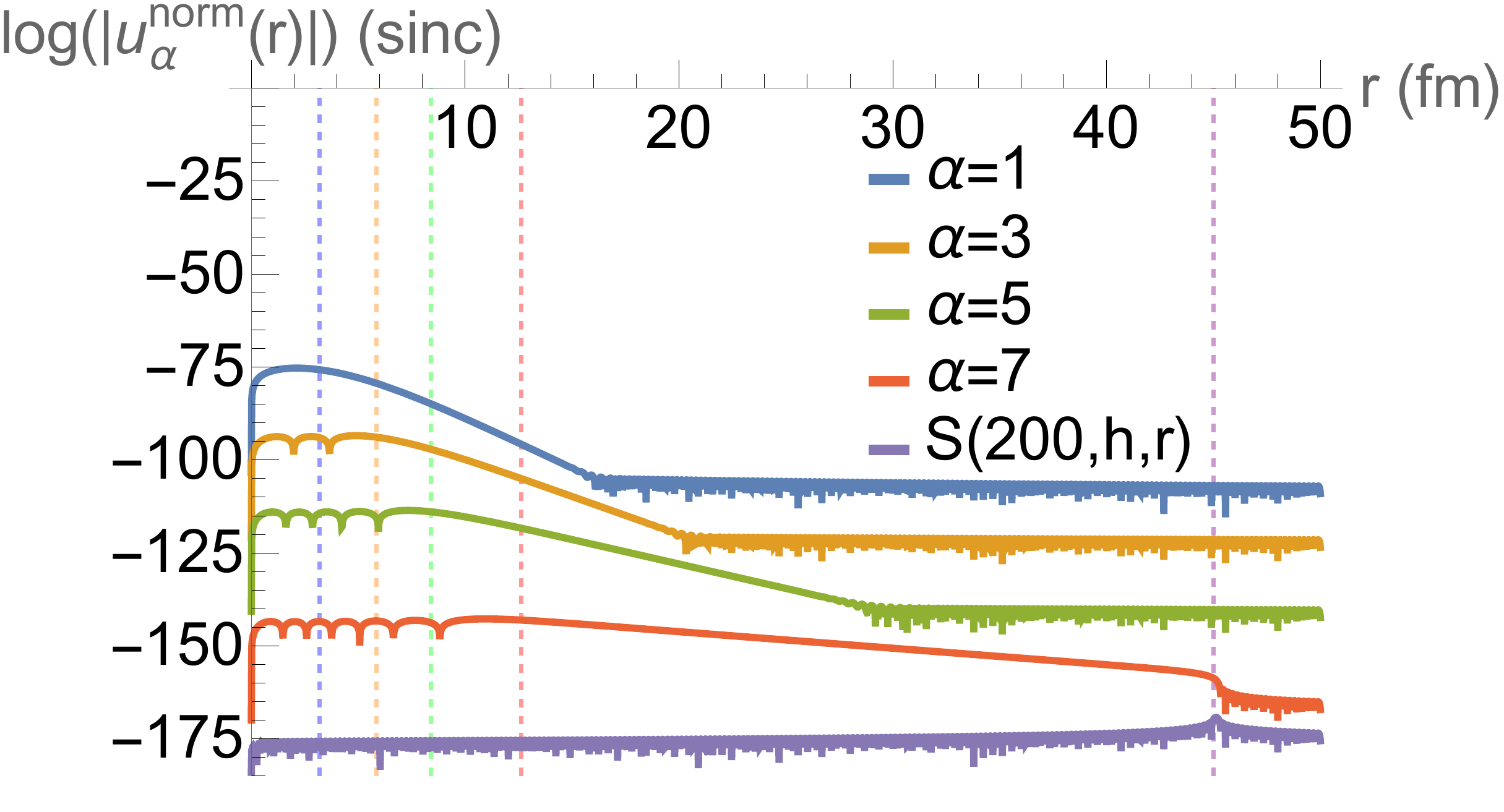}
\caption{}
\label{figD10b}
\end{subfigure}
 
\caption{  (a)  Semi-log plot of the absolute bound state HO wave functions (vertically shifted with different additive factors for clarity) for several bound states for $N_{\textrm{max}}=60$ and $\hbar\Omega=20$ MeV plotted against the HO basis function $r\Phi_{30}(r)$ with the longest range.  Note the merger between the first breakdown point (onset of oscillations) and second breakdown point (onset of quadratic behavior) at some energy $E_{\textrm{critical}}$ (here,$E_{\textrm{critical}}$ is close to $E_{5}$, or around $-71$ MeV.  Note also the approximate independence of the quadratic transition point (marked by the vertical line) from energy as well as the quadratic behavior of both the bound states and the basis beyond the second breakdown point.  For both this figure and Fig. \ref{figD10b}, the dotted vertical line corresponds to the local maximum of the basis function with the maximum radial extent (approximately the HO classical turning point for the HO basis) whereas the colored vertical lines matching the color of state $\alpha$ correspond to the Gaussian interaction turning point.  (b)  Semi-log plot of the absolute bound state sinc wave functions (also vertically shifted with different additive factors for clarity) for several bound states for $m_{\textrm{Val}}=200$ and $\gamma=1$ $\textrm{fm}^{-1}$ compared with the basis function $S(200,h,r)=\sqrt{\frac{1}{\eta'(r)h}}sinc(\frac{(\eta(r)-200h)}{h})$ (see Eq. (\ref{eq33}) in Section \ref{section2.2}).  There is only one breakdown point that depends on both the basis truncation $m_{\textrm{Val}}$ and the different energies.  As in the HO basis, the wave function tail beyond the breakdown point resembles that of the basis function at the highest collocation point. }
\label{figD10}
\end{figure}

For the log of the absolute HO WBS, the nonphysical oscillations do not appear.  To see why, it is instructive to look at Fig. \ref{figD10a}, which shows four different bound states with $N_{\textrm{max}}=60$ (the semi-log absolute wave functions are shifted vertically for clarity).  As we increase energy, the first breakdown point increases and moves towards the second breakdown point beyond which quadratic behavior occurs.  At some critical energy $E_{\textrm{critical}}$, the first breakdown point merges with the second breakdown point.  Hence, at energies $E$ beyond $E_{\textrm{critical}}$, the HO wave function breakdown is characterized only by the second breakdown point.  Note that the second breakdown point can be identified with the local maximum of the highest-energy basis state of the HO basis $rR_{30,1}(r)$ as seen in Fig. \ref{figD10a}.  This closely corresponds to the classical turning point $r_{0}$ (the point in which $\frac{1}{2}m\Omega^{2}r_{0}^{2}=\hbar\Omega(N_{\textrm{max}}+\frac{3}{2})$) of the HO oscillator potential for a particular HO excitation energy and can be contrasted with the turning point for the Gaussian interaction, which is represented by vertical lines with corresponding colors to the bound states in both Fig. \ref{figD10a} and Fig. \ref{figD10b}.  From the figure, the Gaussian turning point for a given bound state $\alpha$ occurs after the last crossing of zero (where linear behavior of the log of the absolute bound state takes over).  Moreover, the first breakdown point depends on both $N_{\textrm{max}}$ and the energy whereas the second breakdown point depends almost entirely on the basis parameters $N_{\textrm{max}}$ and $\hbar \Omega$.  Note that the second breakdown point is identified with $\frac{1}{\lambda_{\textrm{HO}}}$ up to an addition of 2 to $N_{\textrm{max}}$ in Eq. (\ref{eq17}).  

As in the HO DBS, the HO WBS approximation eventually breaks down, behaving quadratically on the semi-log plot for large $r$.  However, as we increase $N_{\textrm{max}}$, the expected linear behavior takes over in an intermediate interval between short- and long-range $r$ (in which $r$ is large enough so that the HO DBS is linear on the semi-log plot, but small enough to not exceed the second turning point).  Unsurprisingly, the larger $N_{\textrm{max}}$ is, the larger the breakdown point.  Because the WBS decays slowly with increasing $r$ (as compared to the DBS), we require a larger $N_{\textrm{max}}$ for computing WBS observables to our desired accuracy.  This explains the relatively slow convergence of the HO WBS energy and mean square radius in Tables \ref{tabC5} and \ref{tabC7}, respectively.  It also underscores the importance of the WBS asymptotic behavior (see the main text and Fig. \ref{fig5a} to Fig. \ref{fig5d}).  From this observation, we can conclude that the point where the linear tail becomes quadratic for the WBS is a good candidate for the threshold point for the piecewise function in Eq. (\ref{eq41}).

The situation is more straightforward for the sinc WBS approximation in Fig. \ref{figD9b}.  As in the DBS case in Fig. \ref{figD8b}, the larger the $m_{\textrm{Val}}$, the higher the oscillation frequency, the smaller the tail amplitude, and the larger the breakdown point (though a relatively high $m_{\textrm{Val}}$ is still required to obtain observables for the WBS at the desired accuracy).  Unsurprisingly, the asymptotic behavior of the log of the absolute sinc WBS is linear (as it should be since we prescribed the wave function to behave like that by virtue of the conformal map) until it breaks down.  Unlike in Fig. \ref{figD8b}, there does not appear to be a correspondence between the breakdown points and any of the characteristic scale equations introduced in Section \ref{theory}.  Moreover, there is only one breakdown point at the tail of the wave function beyond which its behavior corresponds to that of the basis state at the limit of the basis as seen in Fig. \ref{figD9b}.  We can draw similar conclusions in the intermediate sinc basis bound states shown in Fig. \ref{figD10b}.  Note that the onset of rapid oscillations occurs at a point that varies with both $m_{\textrm{Val}}$ and energy.  As in the HO basis, the wave function tail beyond the breakdown point reflects the furthest-reaching underlying basis function, namely $S(200,h,r)=\sqrt{\frac{1}{\eta'(r)h}}sinc(\frac{(\eta(r)-200h)}{h})$ (see Eq. (\ref{eq33}) in Section \ref{section2.2}).

\section{Choosing Fits for the DBS and WBS Tails } \label{appendix:e}

\subsubsection{DBS Fits}
As mentioned in Section \ref{section4.1}, in order to choose a suitable function with which to attach the HO and sinc tail, we must first identify $r_{\textrm{threshold}}$.  We identify this as the point beyond which the log of the absolute HO DBS deviates from linearity and the log of the absolute sinc DBS is still linear (see Fig. \ref{fig3}).  Next, we do several linear fits for points lying in the vicinity of $r_{\textrm{threshold}}$ using Mathematica's\textsuperscript{\textregistered} fitting features mentioned in \ref{appendix:b}.  In this region, the log of the absolute sinc DBS is almost a linear function.  Table \ref{tabE9} shows a sequence of slopes obtained by linearly fitting points of the log of the absolute sinc DBS in the vicinity of $r_{\textrm{threshold}}$.  We also quote the fitting error of the slope $\lambda'$.  $\delta r$ denotes the distance from $r_{\textrm{threshold}}$.  For example, at $\delta r=0$ (or 0 fm away from $r_{\textrm{threshold}}$), we plot 21 points ranging from $r=13.7 $ fm to $r=13.9 $ fm using the sinc WBS values and perform a linear fit to obtain a corresponding $\lambda'$.  While the $\lambda'$ values are not well converged, they are all within three percent of $\lambda_{\textrm{DBS}}=2.71$ $\textrm{fm}^{-1}$.  In Section \ref{section4.1}, we attach the exponential tail corresponding to the $\delta r=0$ fit to both the sinc and HO DBS.

\begin{table}[H]
 \renewcommand{\arraystretch}{2}
 \caption{The absolute slope of a sequence of fits that are based on sinc WBS points for $r$ values near $r_{\textrm{threshold}}$.  While the points do not seem to converge, all calculated $\lambda'$ are within three percent of $\lambda_{\textrm{DBS}}$, which is about $2.71$ $\textrm{fm}^{-1}$ (see Section \ref{results and comparisons}).   }
 
\vspace{3mm}

     \centering
    \begin{tabular} {|c|c|c|}
    \hline
    $\delta r (\textrm{fm})$ & $|\lambda'| (\textrm{fm}^{-1})$ & Standard Error of $\lambda' (\textrm{fm}^{-1}$) \\ \hline
     -0.2 & 2.66988 & $5.8\cdot10^{-4}$ \\ \hline
   -0.1 & 2.69572 & $3.0\cdot10^{-3}$ \\ \hline
    0 & 2.73465 & $1.5\cdot10^{-3}$ \\ \hline
   0.1 & 2.70977 & $4.3\cdot10^{-3}$\\ \hline
   0.2 & 2.63726 & $3.8\cdot10^{-3}$\\ \hline

    \end{tabular}
    \label{tabE9}
    \end{table}

\subsubsection{WBS Fits}

In order to verify that the sinc WBS is in fact exponential with the expected asymptotic behavior $e^{-\lambda_{\textrm{WBS}}r}$ and justify the methods we use in Section \ref{section4.2}, we perform exponential fits (using the values of the sinc WBS as data) at various distances from $r_{\textrm{threshold}}$.  Like in Table \ref{tabE9}, Table \ref{tabE10} shows the results of the fits.  $\delta r$ denotes the distance from $r_{\textrm{threshold}}$ and $\lambda'$ shows the decay value of the exponential.  For example, the $\lambda'$ corresponding to the same row as $\delta r=0$ is the result of fitting sinc WBS points in the vicinity of $r_{\textrm{threshold}}$.  The standard error of $\lambda'$ is the error using the NonlinearModelFit feature in Mathematica\textsuperscript{\textregistered}.  Like in the DBS, we fit 21 points within 0.1 fm of a given threshold point.

Note the increasing accuracy as we go further away from $r_{\textrm{threshold}}$.  Correspondingly, we find that $\lambda'$ converges to a fixed value.  For practical reasons, we cannot fit values too far from $r_{\textrm{threshold}}$ so we plot the values of Table \ref{tabE10} onto a graph and fit a function of the form $-ae^{-br}+\lambda'$.  The result is $\lambda'=0.4469$ $\textrm{fm}^{-1}$ with a standard error of $2.4\cdot10^{-4}$ $\textrm{fm}^{-1}$.  This is close, but not quite the same as $\lambda_{\textrm{WBS}}$, where the ratio $\frac{\lambda'}{\lambda_{\textrm{WBS}}}=1.012$.  

\begin{table}[H]
 \renewcommand{\arraystretch}{2}
 \caption{ The decay constants obtained by exponential fits of the tail using the sinc wave function points within 0.1 fm of $\delta r+r_{\textrm{threshold}}$.  Note how the standard error decreases and $\lambda'$ converges. }

\vspace{3mm}

     \centering
    \begin{tabular} {|c|c|c|}
    \hline
    $\delta r (\textrm{fm})$ & $\lambda' (\textrm{fm}^{-1})$ & Standard Error of $\lambda' (\textrm{fm}^{-1})$ \\ \hline
     0 & 0.42108 & $1.3\cdot10^{-4}$ \\ \hline
   1 & 0.43586 & $6.1\cdot10^{-5}$ \\ \hline
    2 & 0.44256 & $2.6\cdot10^{-5}$ \\ \hline
   3 & 0.44529 & $9.7\cdot10^{-6}$\\ \hline
   4 & 0.44618 & $2.3\cdot10^{-6}$\\ \hline
  5 & 0.44627 & $7.0\cdot10^{-7}$\\ \hline
   
    \end{tabular}
 \label{tabE10}
    \end{table}

\begin{figure}[H]
\centering
\includegraphics[width=12cm, height=5 cm]{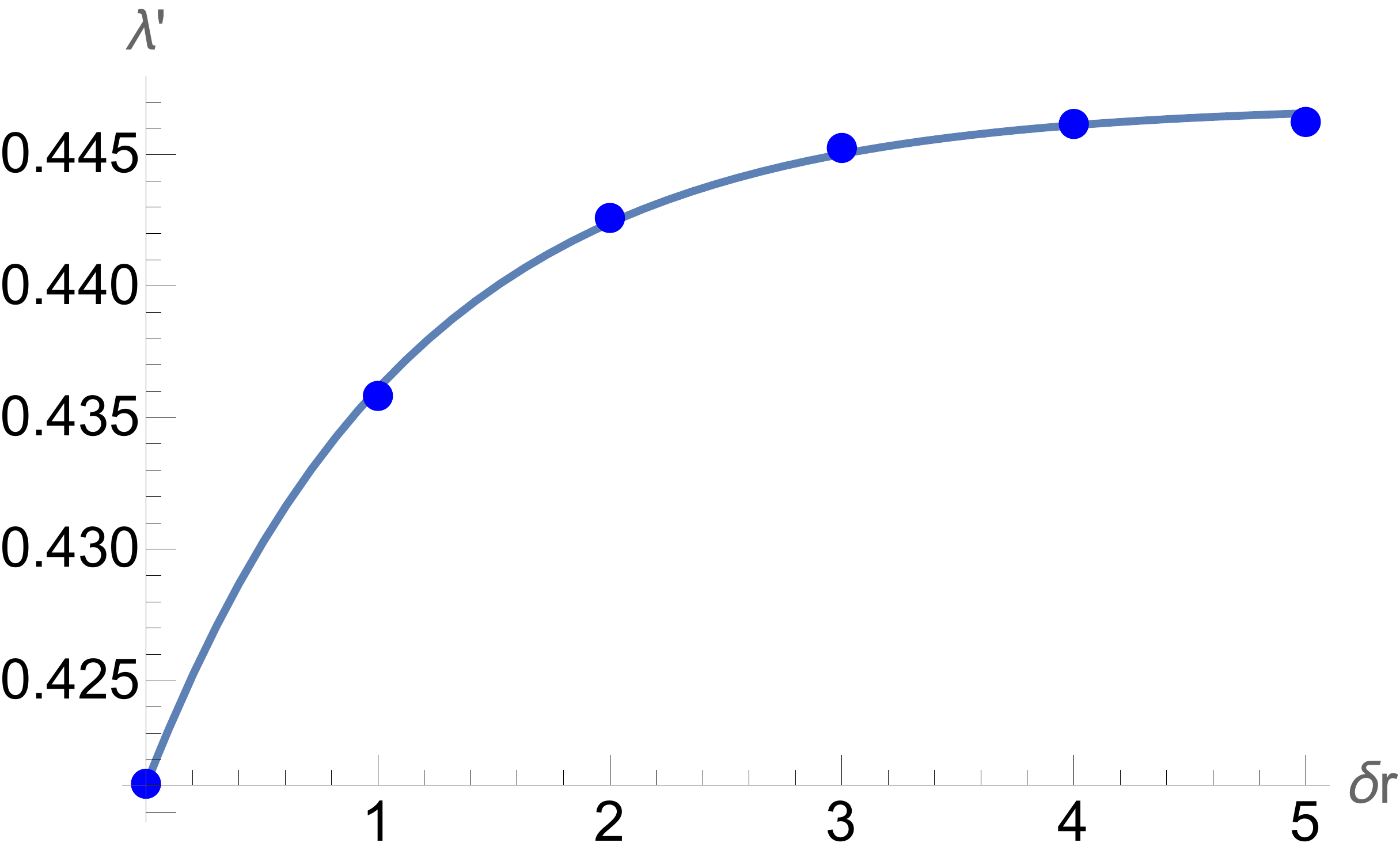}
\caption{ The plot of the points in Table \ref{tabE10} fitted with the function of the form $-ae^{-br}+\lambda'$ in order to get the asymptotic value of $\lambda'$.  The asymptotic value is in close agreement with the $\lambda'$ values in Table \ref{tabE10} but disagrees with $\lambda_{\textrm{WBS}}$ by about one percent. }
\label{figE11}
\end{figure}

\vspace{30mm}


 \begin{center}
\textbf{Acknowledgements}
\end{center}

 We thank the ISU Nuclear Structure research group members for helpful discussions and advice: Soham Pal, Shiplu Sarker, Dr. Pieter Maris, Dr. Andrey Shirkov, Dr. Peng Yin, Weijie Du, Mengyao Huang, and Matthew Lockner.  In addition, we thank Dr. Dossay Oryspayev at the ISU Electrical and Computer Engineering Department and Dr. Gerd Baumann at the German University in Cairo Math Department for helpful advice on the C programming and the sinc methodology aspect of the research, respectively.  This work was supported by the US Department of Energy (DOE) under Grant Nos. DE-SC0018223 (SciDAC-4/NUCLEI) and DE-FG02-87ER40371.

\end{document}